\documentclass[letterpaper,twocolumn,10pt]{article}
\usepackage{usenix}
\usepackage{times} %
\usepackage{amsmath,amssymb,amsfonts} %
\usepackage{graphicx} %
\usepackage{subcaption} %
\usepackage{xcolor} %
\usepackage{hyperref} %
\usepackage{listings}
\usepackage{stmaryrd}

\usepackage{booktabs}
\usepackage{tabularx}
\usepackage{multirow}
\usepackage{multicol}

\usepackage{fontawesome5}
\usepackage{wasysym}

\usepackage{url}
\makeatletter
\g@addto@macro{\UrlBreaks}{\UrlOrds}
\makeatother

\definecolor{figurecolor}{RGB}{22,90,220}
\definecolor{citecolor}{RGB}{198,81,19}
\hypersetup{ colorlinks=true, urlcolor=black, linkcolor=figurecolor, citecolor=citecolor, pdfstartview=FitH,
        pdftitle={SoK: Demystifying the multiverse of MPC protocols},
        pdfauthor={}
        }

\newcommand{\todo}[1]{}
\renewcommand{\todo}[1]{{\color{red} TODO: {#1}}}
\newcommand{\fakepara}[1]{\vspace{0mm}\noindent\textbf{#1}\quad}

\DeclareFixedFont{\ttb}{T1}{txtt}{bx}{n}{9} %
\DeclareFixedFont{\ttm}{T1}{txtt}{m}{n}{9}  %

\usepackage{color}
\definecolor{deepblue}{rgb}{0,0,0.5}
\definecolor{deepred}{rgb}{0.6,0,0}
\definecolor{deepgreen}{rgb}{0,0.5,0}

\newcommand\pythonstyle{\lstset{
language=Python,
basicstyle=\ttm,
otherkeywords={self},             %
keywordstyle=\ttb\color{deepblue},
emph={MyClass,__init__},          %
emphstyle=\ttb\color{deepred},    %
stringstyle=\color{deepgreen},
frame=single,                         %
showstringspaces=false,
breaklines=true, %
postbreak=\mbox{\textcolor{red}{$\hookrightarrow$}\space}, %
numbers=left,
numberstyle=\scriptsize\color{blue},
numbersep=4pt,
}}

\lstnewenvironment{python}[1][]
{
\pythonstyle
\lstset{#1}
}{}

\usepackage{cleveref}
\crefname{section}{\S\!}{\S\S}
\crefname{appendix}{\S\!}{\S\S}
\crefname{figure}{Figure}{Figures}
\crefname{table}{Table}{Tables}
\definecolor{aoe}{rgb}{0.0, 0.5, 0.0}
\newcommand{\sharing}[1]{\ensuremath{\llbracket #1 \rrbracket}}
\newcommand{\bitnumber}[1]{\ensuremath{\left( #1 \right)}}

\def\Snospace~{\S{}}

\newcommand{\colorref}[2]{%
{\small\hypersetup{linkcolor=citecolor}\hyperref[#1]{#2}(\autoref{#1})}%
}
\newcommand{\binary}{{\small $\mathbb{Z}_2$}}
\newcommand{\field}{{\small $\mathbb{Z}_{p}$}}
\newcommand{\ring}{{\small $\mathbb{Z}_{2^k}$}}
\newcommand{\malicious}{\faSkull}
\newcommand{\semihonest}{\faBalanceScale}
\newcommand{\honest}{{\Large\smiley}}
\newcommand{\dishonest}{\faUserSecret}
\newcommand{\additive}{\textbf{$\oplus$}}
\newcommand{\replicated}{\faCopy}

\newcommand{\evalgraphscale}{0.3}
\newcommand{\maingraphscale}{0.5}
\newcommand{\compgraphscale}{0.25}

\begin{document}
    
    \title{SoK: Demystifying the multiverse of MPC protocols}

    \author{Roberta De Viti, Vaastav Anand, Pierfrancesco Ingo, Deepak Garg\\Max Planck Institute for Software Systems}
    \date{}
    \maketitle

This paper systematizes knowledge on the performance of Multi-Party Computation (MPC) protocols. Despite strong privacy and correctness guarantees, MPC adoption in real-world applications remains limited by high costs (especially in the malicious setting) and lack of guidance on choosing suitable protocols for concrete workloads. We identify the theoretical and practical parameters that shape MPC efficiency and conduct an extensive experimental study across diverse benchmarks. Our analysis discusses the trade-offs between protocols, and highlights which techniques align best with different application scenarios and needs. By providing actionable guidance for developers and outlining open challenges for researchers, this work seeks to narrow the gap between MPC theory and practice.     %

\section{Introduction}
\label{sec:intro}

Secure multiparty computation (MPC) enables mutually distrusting parties to compute over private inputs with strong privacy and correctness guarantees~\cite{chaum1988multiparty}. Despite many proposed applications (e.g., secure analytics~\cite{volgushev2019conclave,Bater2017,poddar2021senate, de2022covault}, auctions~\cite{bogetoft2006practical}, and machine learning (ML)~\cite{mohassel2017secureml}), real-world adoption remains limited for two practical reasons. First, \emph{cost:} MPC protocols are expensive, especially if secure
against adversaries that can arbitrary deviate from the protocol. This cost can sometimes be offset with higher resource budgets (e.g., bandwidth/compute provisioning~\cite{de2022covault}), but whether the trade-off is worthwhile depends on the specific use case. Second, \emph{choice:} developers face a sprawling design space of threat models (semi-honest vs.\ malicious compromise of parties, honest vs.\ dishonest majority), domain in which data is represented (binary, field or ring), specific protocols, and protocol-specific optimizations.
There exist a myriad of protocols with different security models and underlying mathematical computation domains, each of which has a different performance for a given set of compute operations. At the same time, there is a lack of consolidated guidance that maps workloads and deployment constraints to protocol families. In practice, one would need to be closely familiar with \emph{many} protocols to make an informed choice.

\paragraph{Goal and audience.}
This paper is intended for two groups. \emph{Newcomers and student researchers entering the area}, who wish to get an overview of MPC protocols before diving deep into dozens of papers. \emph{Application developers}, who wish to pick an efficient protocol for given computations and operating conditions (threat model, available bandwidth, number of parties)
with minimal regret.

Prior work helps only partially. For instance, Cerebro~\cite{zheng2021cerebro} compiles ML pipelines to MPC, automatically selecting among protocol backends and planning the execution; it is a valuable tool when a workload fits its ML-centric operators and assumptions. However, its scope is ML-specific, thus its evaluation focuses on ML tasks. 
Outside ML, evidence is scattered across individual protocol papers. However, in most papers on MPC protocol design, the evaluation section reports \emph{theoretical} metrics (Big-O bounds, gate/AES/OT counts) that do not automatically translate into end-to-end costs for practical primitives (e.g., a 1024-way comparison or a sequence of matrix multiplications). Thus, one must read many papers end-to-end to infer the practical impact on concrete primitives. 
The MP\,-SPDZ paper~\cite{keller2020mp} provides the most comprehensive framework-level overview to date, and we use MP\,-SPDZ as our experimental framework. However, the paper's evaluation centers on a \emph{single} microbenchmark and does not systematize protocol choice across multiple primitives, number of parties, or available bandwidth.

We address this gap by comparing the latency and total data transmission of different protocols for a varied set of computational primitives, under controlled attributes (input size, integer bit width, number of parties, and available bandwidth).
To the best of our knowledge, this is the first paper to jointly systematize and empirically compare general-purpose MPC protocol families across primitives and threat models, and %
to provide \emph{empirically grounded} insights and open-source artifacts to extend the experiments.

\paragraph{Approach.}
We systematize the protocol landscape (\autoref{sec:theory}, \autoref{sec:domains}, \autoref{sec:primitives}, \autoref{sec:addingcorrectness}) and run a controlled empirical study (\S\ref{sec:proto}) across representative protocol families and adversary models (semi-honest/malicious, honest/dishonest majority). We benchmark four canonical primitives (integer \emph{comparison}, \emph{sort}, \emph{inner product}, and \emph{matrix multiplication}), which are very common across many applications. For each benchmark, we vary input size ($N$), integer bit width, number of parties, and available bandwidth (1–20\,Gbps). We report latency and total data transmitted, and develop insights about the relative performance of protocols on specific benchmarks under given conditions. %

\paragraph{Takeaways.}
(i) There is no universal winner: performance is driven by \emph{workload structure} (bit-centric vs.\ multiply-accumulate) and \emph{available bandwidth}. In particular, constraining network link capacity can change the relative order of the latency of different protocols. (ii) The relative performance of protocols is generally stable under changes in $N$ and integer bit width.
(iii) Honest-majority protocols tolerate growth in $N$ and number of parties more gracefully; malicious dishonest-majority protocols can become bandwidth-limited unless arithmetic dominates. %

\paragraph{Contributions.}
\begin{itemize}
	\item \textbf{Taxonomy} (\autoref{sec:theory}, \autoref{sec:domains}, \autoref{sec:offlineonline}, \autoref{sec:addingcorrectness}): a concise map of general-purpose MPC protocol families by threat model and domain of data representation.
	\item \textbf{Comprehensive measurements} (\autoref{sec:proto}): a standardized benchmark suite of four common primitives (comparison, sort, inner product, matrix multiplication) and multiple attributes (input size, integer bit width, number of parties, available bandwidth), reporting both latency and total bytes transmitted during protocol execution.
	\item \textbf{Empirical findings} (\autoref{sec:proto}): observations about the relative performance of protocols on individual benchmarks under specific conditions.
\end{itemize}

\section{Background and Terminology}
\label{sec:theory}

Multi-party computation (MPC)~\cite{chaum1988multiparty,Yao1982} is a class of cryptographic techniques that allow a set of parties to jointly compute a function over their private inputs while keeping those inputs hidden from one another. At a high level, MPC protocols aim to guarantee two fundamental properties:

\paragraph{Confidentiality.} Inputs provided by the parties (or external data providers) remain secret throughout the execution of the protocol. No party should learn any information about another party’s inputs beyond what is revealed by the final output.

\paragraph{Output integrity (correctness).} At the end of the computation, either (i) all honest parties obtain the correct output of the target function on the joint inputs, or (ii) the protocol aborts due to detected adversarial behavior. In particular, an adversary must not be able to undetectably alter the output.

\subsection{Threat models}
\label{threatmodels}
\label{sec:adversary}

The security guarantees of an MPC protocol depend on the adversarial model it assumes. We consider an adversary that may compromise up to $t$ out of $n$ parties. A $t$-out-of-$n$ protocol ensures confidentiality and integrity as long as at most $t$ parties are corrupted; the threshold $t$ is protocol-specific (see~\cref{numbercorrupted}). %
Parties that are not corrupted are considered \emph{honest}, meaning they follow the protocol as specified and do not attempt to deviate or leak information.
Next, we now describe two standard adversary behaviors: \emph{semi-honest} and \emph{malicious}.

\paragraph{Semi-honest (passive) adversary.}
A semi-honest, or honest-but-curious (HbC), adversary follows the protocol but attempts to infer information from observed messages or memory patterns. The main security goal in this model is input confidentiality, as integrity is guaranteed by assumption. Classical examples include Yao's Garbled Circuits~\cite{Yao1982}, which allows two parties to joinly compute any function encoded as a Boolean circuit, and the Goldreich-Micali-Wigderson (GMW) protocol~\cite{micali1987play}, which generalizes to the $n$-party setting using arithmetic circuits. Semi-honest protocols often serve as a foundation for constructing protocols secure against stronger adversaries.

\if 0
\paragraph{A passive, semi-honest, or honest-but-curious (HbC) adversary}
follows the protocol but attempts to infer secret information on the input of the honest parties. The HbC model assumes that the compromised parties are \textit{honest} in the sense that they adhere to the protocol, but are \textit{curious} about learning secret information. In this model, the main concern is information leakage: the primary goal is to maintain \textit{privacy} (or \textit{input confidentiality}) against eavesdropping and traffic or memory-pattern analysis. We note that correctness (or \textit{output integrity}) is guaranteed by assumption, as compromised parties execute the protocol as if they were honest. The best known example of a HbC MPC protocol is the original Yao's Garbled Circuits~\cite{Yao1982}, which allows two parties to jointly compute any function, encoded as a Boolean circuit, without revealing their inputs to each other. Another example is the Goldreich-Micali-Wigderson (GMW) protocol~\cite{micali1987play}, which encodes any function as an arithmetic circuit rather than Boolean, and generalizes to $n$-party settings. As discussed in the latter work,
MPC HbC protocols may serve as a starting point to design protocols that withstand an \textit{active} adversary.
\fi

\paragraph{Malicious (active) adversary.}
A malicious adversary may arbitrarily deviate from the protocol, tamper with messages, provide incorrect data, or collude with other corrupted parties. Its goal may be to compromise confidentiality, integrity, or availability. Protocols secure against malicious adversaries must ensure confidentiality and output integrity (\autoref{sec:theory}), even with up to $t$ corruptions.
\if 0
\paragraph{An active or malicious adversary} can replace the current MPC protocol with any polynomial time algorithm of its choosing. It can tamper with protocol messages, provide false information or data to honest parties or try to cause selective failures on honest users' data. The compromised parties can collude with each other. The adversary's goal is to compromise the confidentiality, integrity, or availability of the computation. A maliciously-secure protocol must guarantee the following requirements, even in the presence of at most $t$ out of $n$ compromised parties:
\begin{itemize}
	\item \textit{Confidentiality}: An active adversary might try to expose secret inputs from honest parties, e.g. causing a selective failure. Therefore, malicious MPC protocols have to prove that the confidentiality of the inputs provided by the parties is guaranteed even when the adversary can behave arbitrarilly.
\item \textit{Correctness} (or \textit{output integrity}): The output of the computation is correct. 
\item \textit{(Optional) Fairness}: Either all or no party learns the output of the computation at the same time. 
\item \textit{(Optional) Output delivery}: All parties always receive the output of the computation. 
\end{itemize}
\fi

Optionally, it may provide two additional properties: \emph{fairness}, either all parties learn the output, or none do; or \emph{output delivery}, all parties receive an output.
These optional properties trade off with efficiency and are often relaxed~\cite{lindell2017framework}. To guarantee output integrity, malicious-secure protocols extend semi-honest designs with mechanisms such as pre-compiled circuits, commitments, and zero-knowledge proofs (ZKPs)~\cite{micali1987play, composable}, forcing adversaries to prove that their behavior matches that of an honest party. For example, the "GMW compiler"~\cite{micali1987play} builds on GMW by requiring ZKPs for each message, though this incurs high overhead. To mitigate costs, subsequent work explores alternatives such as protocol randomization and MACs on circuit wires~\cite{amdcircuits, genkin2015efficient}.
%

\if 0 %
\paragraph{A covert adversary}
is an active adversary whose goal is not only to compromise confidentiality and/or integrity of the computation but also to remain undetected while doing it. Since most MPC protocols will abort when potentially malicious behaviour is detected, an openly malicious adversary accepts that, in exchange for the possibility of compromising the security of a subset of honest parties, there is almost certain detection. On the other hand a covert adversary is more cautious, trying to avoid detection while gaining every advantage it can.  Thus, unlike an arbitrarily malicious adversary, a covert adversary keeps a low profile by restricting their set of actions. The notion of covert adversary was introduced by Aumann and Lindell \cite{aumann2010security} with an eye on practical rather than ideal security settings. The key idea to trade off the security of maliciously-secure protocols for efficiency. In particular, covertly-secure protocols detect and penalize malicious behavior of up to $t$ compromised parties (while maintaining privacy and correctness) \textit{with a given probability} $\rho < 1$. This idea of lowering down the probability of detection allows to gain lower communication and computational cost. Hence, these protocols they are more suitable for resource-constrained environments or when the threat model of a specific application allows for a lower level of security.
\fi

\subsubsection{Number of corrupted parties}
\label{numbercorrupted}
As mentioned in~\cref{threatmodels}, $t$-out-of-$n$ protocol provides \emph{$t$-threshold security}: the computation remains secure as long as at most $t$ parties are corrupted. The value of $t$ is protocol-specific and determines whether the setting assumes an \emph{honest majority} ($t < n/2$) or a \emph{dishonest majority} ($t \geq n/2$).

\paragraph{Honest majority.} 
Here, more than half of the parties are honest. Thus, the honest parties can always outnumber the compromised ones.This assumption simplifies protocol design and analysis, and typically results in fewer communication rounds and lower computational overhead. The downside is limited resilience: only a minority of parties can be corrupted, making such protocols unsuitable in high-risk settings. In the two-party case, $t=1$ is the only meaningful threshold.\footnote{If $t=0$, no corruption is tolerated; if $t=2$, no honest party remains.} For $n>2$, examples include Araki et al.'s protocols in the semi-honest~\cite{araki2016} and malicious~\cite{araki2017optimized} settings. 

\paragraph{Dishonest majority.} 
Here, a majority of the parties may be corrupted, allowing resilience against up to $t=n-1$ corruptions. This setting is attractive when trust assumptions are weak or there is a higher risk of compromise. However, protocols are typically more complex and communication-heavy, especially under malicious adversaries. Generally, these protocols assume $t=n-1$. Hazay et al.~\cite{hazay2018concretely} show that requiring at least two honest parties ($n-t>1$) reduces authentication costs in Boolean circuits. Escudero et al.~\cite{escudero2023superpack} extend the idea to arithmetic circuits, achieving the first ``concretely efficient" dishonest-majority protocol: depending on the percentage of honest parties, their protocol is comparable to state-of-the-art honest-majority protocols. The SPDZ family~\cite{damgaard2012multiparty,spdz2k} represents another prominent line of dishonest-majority protocols.
\if 0
Generally, dishonest-majority protocols are secure up to $t=n-1$. Hazay et al.~\cite{hazay2018concretely} discuss the benefit of assuming $n-t>1$ in dishonest-majority ($t > n/2$) Boolean circuits: assuming a small minority of honest parties rather than a single honest party allows for shorter authentication keys, which reduces the key generation time. Escudero et al.~\cite{escudero2023superpack} explore the same idea in arithmetic circuits and propose the first ``concretely efficient" dishonest-majority protocol; depending on the percentage of honest parties, their protocol is comparable to state-of-the-art honest-majority protocols.
%
%
\fi

\subsubsection{Other considerations}
We highlight orthogonal dimensions relevant to MPC threat models, also common in other cryptographic settings.  

\paragraph{Information-theoretic vs computational security.} 
Protocols without assumptions on adversarial resources achieve \textit{information-theoretic} (perfect) security, while those relying on bounded adversaries (probabilistic polynomial time algorithms) achieve \textit{computational} security.

\paragraph{Static vs dynamic (or adaptive) adversaries.} 
A static adversary fixes corrupted parties before execution; a dynamic adversary can adaptively corrupt during execution. Some protocols (e.g., BGW~\cite{ben2019completeness}) tolerate dynamic corruption when the parties communicate over secure channels, but this does not hold in general~\cite{cramer1999efficient}.  

\paragraph{Denial of Service (DoS).} 
\if 0
A goal of the adversary might be to disrupt the execution of the MPC protocol. To this end, a malicious adversary can exploit the fact that, typically, MPC protocols provide \textit{security-with-abort}: if a compromised party deviates from the protocol, the honest parties will detect malicious behavior and abort the computation to prevent the adversary from manipulating the results. Thus, these protocols are susceptible to DoS attacks, as an adversary can force the continuous abortion of the protocol.
To counter DoS attacks, some protocols (e.g., ~\cite{ishai2014secure, baum2016efficient, brandt2020constructing}) extend the notion of security-with-abort to security with \textit{identifiable} abort (ID-MPC): upon abort, the honest parties discover the identity of at least one of the compromised parties. 
An alternative is to design an \textit{abort-free} protocol by guaranteeing \textit{output delivery}. However, this alternative requires a bound on the number of compromised parties $t$ (e.g., $t < n/2$ or $t < n/3$) ~\cite{honeybadger}, thus is only possible in honest-majority settings. (In general, it has been proven that dishonest-majority abort-free execution is not possible, as also reported in~\cite{ishai2014secure}).
\fi
Most MPC protocols provide \textit{security-with-abort}, making them vulnerable to DoS: an adversary can repeatedly trigger aborts. Countermeasures include \textit{identifiable abort} (ID-MPC)~\cite{ishai2014secure, baum2016efficient, brandt2020constructing}, which exposes at least one cheater, or \textit{abort-free} protocols with guaranteed output delivery. However, guaranteed output delivery requires an honest-majority (e.g., $t<n/2$ or $t<n/3$)~\cite{honeybadger}; in particular, guaranteed output delivery has been proven impossible under a dishonest majority ($t\ge n/2$)~\cite{ishai2014secure}.

\subsection{Secret-sharing schemes}

In MPC, secret values must be distributed across the parties such that no individual party can reconstruct them, while a sufficient subset of parties can recover the secret collectively. We refer to these techniques as \emph{secret-sharing schemes}. 

%

%
%

\if 0
%
%
\fi

\paragraph{Additive Secret Sharing.}
In additive secret sharing, a dealer (owner of the secret) splits a secret $x$ into $n$ shares whose sum equals $x$. Specifically, the dealer samples $n-1$ random values $\{a_1, \ldots, a_{n-1}\}$, assigns $a_j$ to party $P_j$, and keeps $x - \sum_{z=1}^{n-1} a_z$ as its own share. Each share looks random, yet the sum reconstructs $x$.

\paragraph{Boolean (XOR-based) Secret Sharing.}
XOR-based secret sharing is the Boolean analogue of the additive scheme. The dealer samples $n-1$ random bitstrings of the same length as the secret $x$, and sets the last share to $x \oplus \bigoplus_{z=1}^{n-1} a_z$. The secret is the XOR of all shares.

\paragraph{Shamir's Secret Sharing.} 
\if 0
Shamir's scheme \cite{shamir1979share} is a widely-used polynomial-based TSS scheme. Its key idea is to encode a secret value as the constant term of a random polynomial of degree $n-1$. Each party receives a share that corresponds to the evaluation of this polynomial at a distinct point. The original secret can be reconstructed via polynomial interpolation, given the knowledge of all the $n$ shares. To input a secret value $x$, the dealer, party $P_i$, chooses $n-1$ random coefficients $\left\{a_j\right\}_{j=1}^{t-1}$ and $n$ random evaluation points $\left\{z_i\right\}_{i =1}^{n}$, then it computes the polynomial
\begin{equation*}
	\left\{ f\left(z_i\right)\right\}_{i = 1}^n = x + \sum_{j=1}^{n - 1} a_j \cdot z^j.
\end{equation*}
Finally, the dealer sets the share of party $P_i$ to $\left[z_i, f\left(z_i\right)\right]$. 
\fi

Shamir’s scheme~\cite{shamir1979share} encodes a secret value as the constant term of a random polynomial of degree $n-1$. To share a secret $x$, the dealer samples $n-1$ random coefficients $\{a_j\}_{j=1}^{n-1}$ and defines
\[
f(z) = x + \sum_{j=1}^{n-1} a_j \cdot z^j.
\]
It then selects $n$ distinct, non-zero evaluation points $\{z_i\}_{i=1}^n$ and distributes to each party $P_i$ the share $[z_i, f(z_i)]$. The secret can be reconstructed by collecting all $n$ shares and interpolating the polynomial, which uniquely determines $f(0) = x$.

\if 0
\smallskip\noindent\textit{Blakley's Secret Sharing} \cite{blakley1979safeguarding} is a geometry-based TSS scheme. The secret is represented as the intersection point of $n$ hyperplanes in a $(t-1)$-dimensional space. Each party receives a share corresponding to one of these hyperplanes. The secret can be reconstructed by finding the intersection point of at least $t$ hyperplanes, while $t-1$ or fewer hyperplanes reveal no information about the secret.
\fi

\paragraph{Threshold Secret Sharing.}
\if 0
Threshold Secret Sharing~\cite{ito1989secret} is a technique used to adapt a secret-sharing scheme in such a way that to reconstruct a secret input a party would only need the shares of $1 < t < n$ parties instead of the shares of all the $n$ parties participating in the computation. \newline
\paragraph{Replicated secret sharing.} In \textit{replicated secret sharing} multiple shares are given to each party and, therefore, the quorum of computing parties needed to reconstruct a secret is reduced. Replicated secret sharing is often used in conjunction with the additive or XOR-based secret sharing schemes. \newline
\paragraph{Threshold Shamir secret sharing.} For Shamir secret sharing, the dealer decides the degree of the secret polynomial and, therefore, she decides the number of shares needed to recover the secret.
\fi

Threshold secret sharing~\cite{ito1989secret} generalizes the schemes above by requiring only $t$ out of $n$ shares to reconstruct the secret, rather than all $n$. The choice of $t$ is directly tied to the adversarial setting (see \autoref{numbercorrupted}): if at most $t$ parties can be corrupted, then $n-t$ honest parties suffice for reconstruction. For instance, in \emph{threshold Shamir secret sharing}, the dealer sets the polynomial degree $t-1$ to determine how many shares are required to recover the secret. Another variant is \emph{replicated secret sharing}, which lowers the quorum of parties required for reconstruction with additive or XOR-based schemes. Replicated secret sharing ensures that (i) no subset of fewer than $t$ parties can recover the secret, and (ii) any subset of $t$ parties collectively holds all the necessary shares. Concretely, the dealer generates $n$ additive shares of the secret and distributes multiple shares to each participant, hence the name ``replicated''.
These threshold variants form the backbone of many MPC protocols analyzed in this work.

\subsection{Choosing threat model, number of parties, and protocol family}
\label{subsec:howto}

\paragraph{Threat model.}
The threat model should be chosen to match the \emph{risk one must tolerate} and the \emph{latency/bandwidth one can afford}. Semi-honest protocols are faster but assume weaker adversaries. Malicious protocols add mechanisms to detect arbitrary deviations and therefore show a performance overhead. As mentioned in~\autoref{sec:intro}, this overhead can be partially offset with more resources (e.g., bandwidth/cores, over-provisioning), but whether the trade-off is worthwhile is application-specific. The corruption threshold (honest vs.\ dishonest majority) should reflect (i) the expected risk of compromise, (ii) latency requirements (more rounds in dishonest-majority protocols), and (iii) optional requirements such as fairness or guaranteed output delivery (given by dishonest-majority protocols).

\paragraph{Number of parties.}
\if 0
The choice of number of parties may depend on the use case. In some application scenarios (e.g., cooperative~\cite{poddar2021senate} or federated~\cite{roth2021mycelium,margolin2023arboretum} analytics), the number of parties is dictated by the number of data contributors that are directly involved in the MPC protocol and do not outsource computation (e.g., different stakeholders owning datasets of sensitive data or subset of data sources in federated committees).
In other application scenarios, the computation is outsourced to multiple servers engaging in MPC protocols (e.g., large-scale multi-party analytics~\cite{de2022covault}, ML computation~\cite{watson2022piranha}). In this setting, the choice of the number of servers is flexible, and may depend on the efficiency of the underlying MPC protocol and the estimated risk of compromise. So far, most of such prior work in systems research have selected $n=2$, $n=3$ and rarely $n=4$. The reason may be multi-fold: standardized MPC frameworks like EMP-toolkit~\cite{emp-toolkit} and MP-SPDZ~\cite{keller2020mp} mainly offer MPC protocol implementations for 2--3 parties; most protocols with $n>3$ might not be efficient enough for real-world settings; a few parties may be enough in scenarios with an estimated low risk of simultaneous compromise.
\fi
%
If computation is \emph{cooperative} or \emph{federated} (i.e., data sources participate directly) the number of parties $n$ follows the number of data sources~\cite{poddar2021senate,roth2021mycelium,margolin2023arboretum}. If computation is \emph{server-aided}, one can choose a small coalition of infrastructure parties to balance trust and cost~\cite{de2022covault,watson2022piranha}. In the latter setting, the number of servers is flexible and depends on the efficiency of the underlying MPC protocol and the estimated risk of compromise. In practice, latency and communication often scale superlinearly with the number of parties~\cite{poddar2021senate,wang2017globalscale}, and widely used frameworks (EMP-toolkit~\cite{emp-toolkit}, MP-SPDZ~\cite{keller2020mp}) offer mature, efficient stacks primarily for $n=2$ or $n=3$ (with limited $n=4$).

Thus, one should prefer the \emph{smallest} coalition that meets the threat model, and only scale $n$ upward when policy or trust constraints require it.

\paragraph{Protocol family.}
We assume that the designers of a given MPC protocol have already chosen the optimal secret-sharing scheme for their protocol, because the sharing scheme is typically coupled to the protocol domain (binary, field, or ring). 
Thus, the remaining practical choice is the \emph{protocol family} that best matches the application to deploy and its workload. This work aims to give guidelines for this choice, and provide empirical evidence at the basis of this choice: Sections~\autoref{sec:domains}, ~\autoref{sec:offlineonline}, and ~\autoref{sec:addingcorrectness} review domains and common paradigms, while~\autoref{sec:proto} reports empirical results where particular protocol families fail to scale for certain element bit widths or input sizes. 

A high-level guideline is the following: First, one should identify the dominant primitives in their application workload: bit-centric operations (comparisons, selections) tend to favor Boolean/GC-style protocols; multiply-accumulate workloads (inner product, matrix ops) tend to favor arithmetic sharing over fields/rings. %
Where supported, mixed computation can bridge domains for non-linear kernels, but its gains are workload- and protocol-dependent. Likewise, if a given deployment experiences idle periods, one can pre-process input-independent work ahead of time. Thus, choosing protocols that pre-process more computation during these low-load windows reduces the critical path at request time and helps reducing tail latency (we detail this in \autoref{sec:offlineonline}).

In summary, the practical rule is to \emph{match the protocol to the hot path}: choose the domain that keeps frequent operations cheap, and take into account that trade-offs may change at scale and depending on the actual resources at hand (compute, bandwidth).

\section{Domains}
\label{sec:domains}

MPC protocols operate over different \emph{domains} in which both public and secret values are represented. Common domains are $\mathbb{Z}_2$, $\mathbb{Z}_{2^k}$, and $\mathbb{Z}_p$. We defer the mathematical details to the appendix (\autoref{app:mathbasis}). The choice of domain directly impacts the efficiency of basic operations (addition, multiplication, comparison) and introduces domain-specific security considerations. In particular, additions and multiplications are cheaper in arithmetic domains, while comparisons are often more efficient in the binary domain. Next, we review these domains and the classes of protocols that operate on them.

\subsection{Binary $ \left( \mathbb{Z}_2 \right)$}
\label{subsec:binary_domain}

\if 0
MPC is said to operate in \textit{binary} domain when it operates on integers belonging to $\mathbb{Z}_2$. In this domain, MPC operates on binary values, effectively evaluating boolean circuits. The most popular family of protocols operating in $\mathbb{Z}_2$ are derivations from the original Yao's GCs protocol~\cite{Yao1982}. In GCs, the evaluation of AND operations is, usually, more costly than the evaluation of XOR operation and, therefore, those MPC protocols are usually best suited to evaluate comparisons, which are XOR reliant operations, rather than arithmetic operations, which rely on the evaluation of AND gates. Due to the success of the original Yao's GC protocol, there have been a lot of efficiency improvements in this protocol over the past decades~\cite{choi2012security,zahur2015two,huang2011faster,acharya2023new,songhori2015tinygarble,kolesnikov2014flexor}. In general, the binary domain offers fast evaluation of comparison operations, therefore, over the past years hybrid protocols in which there is a swap between different domains depending on the operations we need to evaluate have been proposed ~\cite{dabits}. We discuss those protocols in \ref{subsec:mix}.
\fi

The \emph{binary domain} encodes values as integers belonging to $\mathbb{Z}_2$, and computation proceeds via Boolean circuits. The canonical example is Yao’s Garbled Circuits (GCs)~\cite{Yao1982}, where one party (the generator) encodes a circuit by encrypting and permuting its truth tables, and the other (the evaluator) evaluates it using oblivious transfer (OT)~\cite{even1985randomized} to obtain only the output consistent with both parties' inputs. This technique guarantees input confidentiality while enabling arbitrary two-party computation, and later work generalizes GCs to multiparty settings. A long line of research has optimized GCs with respect to communication, computation, and circuit representation~\cite{choi2012security,zahur2015two,huang2011faster,acharya2023new,songhori2015tinygarble,kolesnikov2014flexor}. Another method to evaluate Boolean circuits does not require garbling but it requires distributing XOR-based secret shares of private inputs to all computing parties and the direct evaluation of Boolean gates on those secretly shared values ~\cite{GMW1987mentalgame}. Protocols that run in the \emph{binary} domain are typically optimized for XOR gates, making comparison-heavy workloads efficient, whereas arithmetic operations (requiring many AND gates) are comparatively expensive. As a result, hybrid protocols have emerged that combine domains, e.g., evaluating comparisons in $\mathbb{Z}_2$ and arithmetic in $\mathbb{Z}_{p}$ or $\mathbb{Z}_{2^k}$~\cite{dabits}, ~\cite{demmler2015aby}, ~\cite{escudero2020improved}. We briefly discuss these protocols in \autoref{subsec:mixing_domains_paper} and more thoroughly in \autoref{subsec:mixing_domains_appendix}.

\subsection{Arithmetic $(\mathbb{Z}_{p}, \mathbb{Z}_{2^k})$}
\label{subsec:arith_domain}
The \emph{arithmetic domain} represents values either as integers modulo a prime $p$ or modulo $2^k$. These settings are well-suited for additions and multiplications, while comparisons are typically more expensive than in the binary domain. We briefly discuss additions and multiplication in $\mathbb{Z}_{p}$ and $\mathbb{Z}_{2^k}$ in \autoref{subsubsec:ops}.

\paragraph{Prime Fields $\left(\mathbb{Z}_{p}\right)$.} 
Computation in a prime field domain uses integers modulo $p$ or polynomials over $\mathrm{GF}(p^k)$, where $p$ is prime and $k$ is a positive integer ($k \geq 1$).

\paragraph{Rings $\left(\mathbb{Z}_{2^k}\right)$.}
In a ring domain, values are represented on integers modulo $2^k$. A key advantage of ring-based protocols is that the representation of secret-shared values align with machine-word representations. Thus, local additions are fast and even more complex operations like comparisons, bit slicing and multiplications are more efficient. However, ring-based protocols also introduce security challenges (see \autoref{sec:macs}).

\subsubsection{Basic arithmetic operations}
\label{subsubsec:ops}

\paragraph{Addition.} Adding two secret-shared values in either $\mathbb{Z}_{p}$ or $\mathbb{Z}_{2^k}$ is a local operation: each party adds its shares to compute the new share.  For instance, let us denote a secret-shared value as \sharing{z}. Then,
to calculate $\sharing{z} = \sharing{x} + \sharing{y}$, each party locally evaluates $\sharing{z}_i = \sharing{x}_i + \sharing{y}_i$ where $\sharing{x}_i$ and $\sharing{y}_i$ represent the shares of $\sharing{x}$ and $\sharing{y}$ belonging to the $i$-th party.

\paragraph{Multiplication.} Multiplying two secret-shared values requires interaction among the computing parties. The standard technique is Beaver multiplication triples~\cite{beavertriple}, which allow secure multiplication with minimal communication rounds. A Beaver triple is a random correlated tuple $\left\{\sharing{a}, \sharing{b}, \sharing{c}\right\}$, where $\sharing{a}$ and $\sharing{b}$ are randomly chosen and $\sharing{c} = \sharing{a \cdot b}$. We detail their construction and usage in multiplications in \autoref{sec:beaver_multiplication}, and revisit their role in malicious security in \autoref{sec:addingcorrectness}. Generating and validating, i.e. checking that $\sharing{c} = \sharing{a \cdot b}$ and no malicious parties introduced any errors, has been a focus for many papers over the years, notably ~\cite{mascot} ~\cite{keller2018overdrive}. We provide details about triples' generation and validation in \autoref{subsubsec:beaver_triples_gen} and \autoref{subsec:primitives_app}, respectively. While Beaver triples are typically implemented in most protocols, different techniques (e.g.,~\cite{ben2019completeness}) have also been explored. 

\if 0
\fakepara{Multiplication}: Multiplying two secret shared values $\sharing{x}$ and $\sharing{y}$ requires one or more communication rounds. In most cases, multiplications are computed by sacrificing a Beaver multiplication triple. We will describe in details what Beaver multiplication triples are and how they are generated and used in \ref{sec:beaver_multiplication}. In general, while Beaver multiplication triples provide a fast way to evaluate multiplication between secretly shared values, different techniques are presented in the literature, such as \cite{goldwasser1988completeness}.
\fi

\subsection{Crossing domains}
\label{subsec:mixing_domains_paper}
\label{subsec:crossing}

\if 0
While arithmetic operations such as additions and multiplications are more efficiently computed in the arithmetic domain while comparisons, which are a fundamental part of most algoritms, are more efficiently computed in the Boolean domain.
The mixed circuit approach aims to achieve the best of both worlds by dynamically switching sub-protocols during computation. Clearly, for mixed circuits to be of practical use, the cost of switching back and forth to perform a given computation in another domain must be less than the cost of performing the same computation in the original domain without any switching. In this subsection, we briefly discuss the main techniques used to dinamically switch domains. 
\fi

Arithmetic domains offer efficient additions and multiplications, while Boolean domains are better suited for comparisons and bitwise operations. Since both types of operations are fundamental in most algorithms, some protocols~\cite{demmler2015aby, mohassel2017secureml} adopt a \emph{mixed} approach: dynamically switching representations so that each operation is executed in its most efficient domain. For this to be beneficial, the cost of switching must be lower than directly evaluating the operation in the original domain.  

Two main methods are known to dinamically switch computation domain: \emph{local share conversion} and the usage of a particular trype of correlated randomness namely \emph{daBits} and \emph{edaBits}. Local share conversion is a methodology that allows a lightweight and, mostly, local conversion of secret shares from the arithmetic to the boolean domain and viceversa. The downside of this technique is that it can only be used in conjunction with additive replicated secret sharing and only under a semi-honest threat model. We discuss \emph{local share conversion} in details in \autoref{subsubsec:local_share_conv_app}. To overcome the limitations of local share conversion and allow mixed computations under a malicious threat model, daBits (double-authenticated bits)~\cite{dabits} and edaBits (extended daBits)~\cite{escudero2020improved} have been proposed. \emph{daBits} (double-authenticated bits)~\cite{dabits} represent the same random bit shared in both arithmetic and Boolean domains, and \emph{edaBits} (extended daBits)~\cite{escudero2020improved} extend this idea by providing a random value $r \in \mathbb{Z}_p$ in the arithmetic domain together with its bit decomposition in the Boolean domain. These primitives enable efficient cross-domain conversions. This type of input-independent material is called \emph{correlated randomness}, as it consists of random values shared across domains or parties while preserving specific consistency relations. We provide further technical details on mixed circuits, daBits, and edaBits in \autoref{subsubsec:dabits_edabits_app}.
%
%

\if 0
\paragraph{daBits and edaBits.} daBits (double authenticated bits) and edaBits (extended double authenticated bits) are specifically designed to allow efficient conversion between arithmetic and binary secret sharing. Briefly, a daBit is an authenticated random bit $r$ secret shared among the computing parties both in binary and arithmetic domain. edaBits are a generalization of daBits, specifically, an edaBits is a tuple composed by a random number $r \in \mathbb{Z}_p$ shared in the arithmetic domain and its bit decomposition shared in the binary domain. More details about mixed circuits, daBits and edaBits can be found in \autoref{subsec:mixing_domains_appendix}.
\fi
%
%
%
     %
\section{Offline and Online computation}
\label{sec:primitives}
\label{sec:offlineonline}

Many MPC protocols split execution into two phases: (i) a \emph{pre-processing} or \emph{offline phase}, where parties perform input-independent operations, and (ii) an \emph{online phase}, where the actual function is evaluated on secret inputs. The offline phase is typically used to generate \emph{correlated randomness} -- values that do not depend on the parties' inputs but must satisfy specific consistency relations -- such as Beaver triples for secure multiplications (\autoref{subsubsec:ops}), or daBits and edaBits for cross-domain conversions (\autoref{subsec:crossing}). Because these primitives are expensive to generate securely yet independent of the inputs, they can be precomputed and validated in the offline phase, leaving the online phase to lightweight operations.

This separation offers practical advantages in real-world deployments: offline computation can be scheduled ahead of time or during idle periods, while the online phase remains short, reducing the latency of the critical-path. 

Some protocols~\cite{malrep} instead use a \emph{post-processing} paradigm: parties execute the computation is executed optimistically during the online phase, and correctness is verified afterward. Although different in structure, both offline and post-processing techniques aim to shift input-independent work outside the critical online path.  

Since the choice of pre-/post-processing is intrinsic to each protocol, our evaluation considers each protocol as a whole. A deeper study that disentangles online/offline costs and explores trade-offs under application-specific online constraints is left to future work (\autoref{sec:future}).

\section{Techniques for malicious security}
\label{sec:addingcorrectness}

So far, we have discussed protocols independently of their adversarial setting (see \autoref{sec:adversary}). Next, we give an overview of the fundamental techniques used to upgrade semi-honest protocols to malicious security. These techniques are important for understanding both the additional guarantees provided and the overheads observed in practice.

In this setting, protocols must ensure that adversarial parties cannot tamper with the computation undetected. The concrete mechanisms vary with the type of computational domain (\autoref{sec:domains}). We summarize the main approaches below.

\subsection{Binary protocols} 

While Yao's GCs (\autoref{subsec:binary_domain}) were originally designed for the semi-honest setting, they can be extended to malicious security. A widely used approach is \emph{cut-and-choose}~\cite{zhu2016cut}: the generator creates many garbled circuits, and the evaluator asks to \emph{open} (i.e., reveal and check) a random subset for correctness. If all opened circuits are correct, the remaining ones are evaluated. This makes it unlikely that the generator can cheat without being caught, but at the cost of constructing and transmitting many circuits. (In particular, to achieve statistical security $2^{-\rho}$ (with $\rho \geq 40$), $\rho$ independent GCs must be generated and transmitted between the parties.) The overhead grows quickly for large circuits. To reduce this cost, later work amortizes the cut-and-choose check across multiple executions. (If the same function is to be evaluated $\tau$ times, only $O(\rho / \log \tau)$ circuits are required.) Despite these optimizations, maliciously secure GCs remain substantially more expensive than their semi-honest counterparts.

\if 0
While Yao's GC protocol only provides semi-honest security, researchers have successfully extended the protocol to provide malicious security~\cite{}. A popular approach for developing such protocols is to apply the \emph{cut-and-choose} technique~\cite{zhu2016cut} on the Yao GC protocol. In this technique, multiple circuits are constructed and sent by one party, out of which some of them are opened and checked for correctness by the other party. With this technique, to achieve statistical security $2^{-\rho}, (\rho \geq 40)$, the optimal approach requires $\rho$ garbled circuits to be transmitted between the parties. For large circuits, this incurs large overheads. To overcome
these overheads, some approaches have proposed an amortization of the cost for functions that need to be evaluated multiple times. To amortize over $\tau$ executions, one only needs $O(\frac{\rho}{log \tau})$ garbled circuits.
\fi

%
\subsection{Arithmetic protocols} 
Arithmetic protocols typically rely on two techniques to achieve malicious security: (i) attaching \emph{message authentication codes (MACs)} to all secret-shared values, and (ii) validating preprocessed randomness (\autoref{sec:offlineonline}) through a method known as \emph{sacrifice}. These mechanisms ensure that corrupted parties cannot bias the computation without being detected.

\paragraph{MACs.} 
Each secret-shared value $\sharing{x}$ is accompanied by a MAC label tied to a global secret key $\alpha$. Concretely, for each secret-shared value $\sharing{x}$, the parties also hold a share of $\sharing{\alpha \cdot x}$. This allows any modification of  $x$ to be detected during reconstruction, when MACs are verified. While this approach is sound over fields, it requires statistical extensions to remain sound over rings. In particular, this approach fails with probability $1/2$ over rings; to reduce this failure probability to $2^{-s}$, all values (including $\alpha$) are extended by $s$ statistical security bits~\cite{keller2020mp}. We provide more details in \autoref{sec:macs}.

\if 0
Message authentication codes (MACs) are needed to guarantee that no malicious parties can corrupt in any way secret inputs or intermediate results. In general, MACs are a family of function $\mathcal{F} : \mathbb{Z}_{\star} \rightarrow \mathbb{Z}_{\star} $ such that given an error $\delta > 0$ and a  function $f \in \mathcal{F}$, it holds that $\mathbb{P} \left[ f\left(x\right) = f\left(x + \underline{\delta}\right) \right] \le p$ where $p$ is required to be small and less than or equal to $2^{-s}$. In pratice, most arithmetic MPC protocols use multiplicative MACs: a global MAC key $\alpha$ is generated and shared among the parties, then for each secret shared value $\sharing{x}$ a MAC label is computed as $\sharing{m} = \sharing{\alpha \cdot x}$ and shared among the parties. While this construction fulfills all the necessary requirements when used on elements belonging to a field, for elements belonging to a ring, it fails with probability $\frac{1}{2}$. To provide integrity in Ring-based protocols, all values, including the global MAC key $\alpha$, are extended by $s$ bits, where the desired failure probability is $2^{-s}$.  For a detailed mathematical explanation, please refer to \autoref{sec:macs}
\fi

\paragraph{Sacrifice.} 
Arithmetic protocols rely on correlated randomness generated in the offline phase, such as Beaver triples, to perform secure multiplications (\autoref{subsubsec:ops}). If a malicious adversary inputs incorrect triples, the results of subsequent multiplications would be corrupt. To prevent this, protocols use \emph{sacrifice}: some triples are randomly chosen and opened to check their consistency, while the rest are retained for use in the online phase. Although this discards part of the preprocessed material, it ensures with high probability that the remaining triples are correct.
\if 0
During the preprocessing phase, Beaver triples, i.e. triples $\left(\sharing{a}, \sharing{b}, \sharing{c}\right)$, such that $\sharing{c} = \sharing{a \cdot b}$, are generated by the parties. However, an active adversary can introduce an error $\delta$ such that $\sharing{c} \ne \sharing{a \cdot b}$; if a corrupt triple is used to compute a product, then the result of this operation will be inevitably corrupt. Therefore, all triples have to be checked before use. To do the checking, usually, one or more Beaver triples are opened and, therefore, \textit{sacrificed} to validate one.
\fi

\bigskip\noindent In summary, binary protocols typically achieve malicious security through cut-and-choose techniques, which rely on redundancy and probabilistic checking, while arithmetic protocols rely on algebraic consistency enforced by MACs and sacrifice checks on correlated randomness. Both approaches ``compile" semi-honest designs into maliciously secure ones, but at significant efficiency cost. 
\section{Evaluation}
\label{sec:proto}

Next, we discuss our empirical evaluation, beginning with our experimental setup and methodology before analyzing the results.

\begin{table}[t]
\centering
\begin{tabularx}{\linewidth}{Xrccc}
\toprule
\textbf{Protocol} &  \textbf{Domain} &  \textbf{Majority}  & \textbf{Security}  &  \textbf{Sharing}\\
\midrule
\colorref{subsec:mal:dh}{Tinier} & \binary & \dishonest & \malicious & \additive \\
\colorref{subsec:mal:sh:h}{Yao} & \binary & \honest & \semihonest & \additive \\
\colorref{subsec:mal:sh:h}{MalRepBin} & \binary & \honest & \malicious & \replicated \\
\colorref{subsec:mal:sh:h}{PsRepBin} & \binary & \honest & \malicious & \replicated \\
\colorref{subsec:mal:sh:h}{CCD} & \binary & \honest & \semihonest & {\small $Sh$} \\
\colorref{subsec:mal:sh:h}{MalCCD} & \binary & \honest & \malicious & {\small $Sh$} \\
\colorref{subsec:mal:dh}{Mascot} & \field & \dishonest & \malicious & \additive \\
\colorref{subsec:sh:dh}{Semi} & \field & \dishonest & \semihonest & \additive \\
\colorref{subsec:mal:dh}{LowGear} & \field & \dishonest & \malicious & \additive \\
\colorref{subsec:sh:dh}{Hemi} & \field & \dishonest & \semihonest & \additive \\
\colorref{subsec:sh:dh}{Temi} & \field & \dishonest & \semihonest & \additive \\
\colorref{subsec:sh:dh}{Soho} & \field & \dishonest & \semihonest & \additive \\
\colorref{subsec:mal:sh:h}{RepField} & \field & \honest & \semihonest & \replicated \\
\colorref{subsec:mal:sh:h}{PsRepField} & \field & \honest & \malicious & \replicated \\
\colorref{subsec:mal:sh:h}{SyRepField} & \field & \honest & \malicious & \replicated \\
\colorref{subsec:mal:sh:h}{ATLAS} & \field & \honest & \semihonest & {\small ATLAS}\\
\colorref{subsec:mal:sh:h}{Shamir} & \field & \honest & \semihonest & {\small $Sh$} \\
\colorref{subsec:mal:sh:h}{MalShamir} & \field & \honest & \malicious & {\small $Sh$} \\
\colorref{subsec:mal:sh:h}{SyShamir} & \field & \honest & \malicious & {\small SPDZ+$Sh$}\\
\colorref{subsec:mal:dh}{SPDZ2k} & \ring & \dishonest & \malicious & {\small SPDZ}\\
\colorref{subsec:sh:dh}{Semi2k} & \ring & \dishonest & \semihonest & \additive \\
\colorref{subsec:sh:dh}{Ring} & \ring & \honest & \semihonest & \replicated \\
\colorref{subsec:mal:sh:h}{PsRepRing} & \ring & \honest & \malicious & \replicated \\
\colorref{subsec:mal:sh:h}{SyRepRing} & \ring & \honest & \malicious & {\small SPDZ+}\replicated\\
\colorref{subsec:mal:sh:h}{MalRepRing} & \ring & \honest & \malicious & \replicated \\
\bottomrule
\end{tabularx}
\caption{Protocols used for analysis with their characteristics. Domain(\ring=Ring,\field=Field,\binary=Binary), Majority(\honest=Honest,\dishonest=Dishonest), Security(\semihonest=Semi-Honest,\malicious=Malicious), Sharing Scheme(\additive=Additive/XOR, \replicated=Replicated, $Sh$=Shamir).}.
\label{tab:protocols}
\end{table}

\subsection{Setup and Methodology}

We run all experiments on a cluster of 8 machines (Intel Xeon Gold 6244, 3.60\,GHz, 16 cores, 495\,GB RAM) running Debian GNU/Linux 12 and connected via two 1/10\,GbE Broadcom NICs to a Cisco Nexus 7000 switch. To reduce experimental variance, all experiments use only one core per machine. 

We implement our benchmarks in the MP-SPDZ framework~\cite{keller2020mp} to evaluate the protocols in \autoref{tab:protocols}. To understand the effect of network bandwidth limits, we cap link capacity at \{20, 10, 5, 1\}\,Gbps using Wondershaper 1.4.1~\cite{wondershaper} in some experiments. 

We benchmark four primitives: \emph{comparison}, which compares two
arrays of length \(N\) pointwise; \emph{sort}, which sorts an array of
length \(N\); \emph{matrix multiplication (matmul)}, which multiplies
two \(N\times N\) matrices; and \emph{inner product}, which computes
the inner product of two arrays of length \(N\). For sorting, we use a
variant of radix sort optimized for MPC~\cite{hamada2014oblradixsort};
for inner product and matrix multiplication, we use the optimized
native implementations provided by MP-SPDZ. We choose these four
primitives for their ubiquity in data analytics and because they use
different kinds of operations on integers: comparison uses only
logical operations on integers (also called combinatorial operations);
radix sort uses bit decomposition and arithmetic operations (integer
addition and multiplication); inner product and matmul use only
arithmetic operations.

We run each protocol in different configurations -- with daBits,
edaBits, local share conversion and without any of these
features. Unless otherwise specified, we report the results of the
protocol configuration that provides the lowest latency. We test every
protocol and benchmark combination with the integer bit-width set to
64 and 128 separately (blue and orange bars, respectively, in our
graphs).

Unless otherwise specified, all protocols are run with 3 parties,
using 3 of our 8 machines. The only exception is the Yao protocol,
which supports only 2 parties and, hence, is run on 2 parties only
(our use of Yao can be viewed as a degenerate 3-party MPC, where one
party remains idle, and at most one party may be compromised
semi-honestly).
We also experiment with varying number of parties (2--8) for protocols
that support this (\autoref{subsec:varyingn}).

We report two metrics: (i) end-to-end latency of computing the primitive, and (ii) global data sent --- the total amount of data transmitted by all parties. For both metrics, lower numbers indicate better performance. We enforce a 10-minute wall-clock timeout per run. Given the modest sizes of inputs we test, we consider 10 minutes a very permissive cut-off. If an execution times out, then we kill the execution without letting it finish. As a consequence, for configurations that time out we do not obtain and do not report numbers for global data sent. The \emph{timeout} threshold is indicated by a dashed line in our graphs.

\subsection{Performance analysis}
\label{sec:evaluation}

\begin{figure*}[h]%
	\centering%
		\begin{subfigure}{0.25\textwidth}\centering%
		\includegraphics[scale=\evalgraphscale]{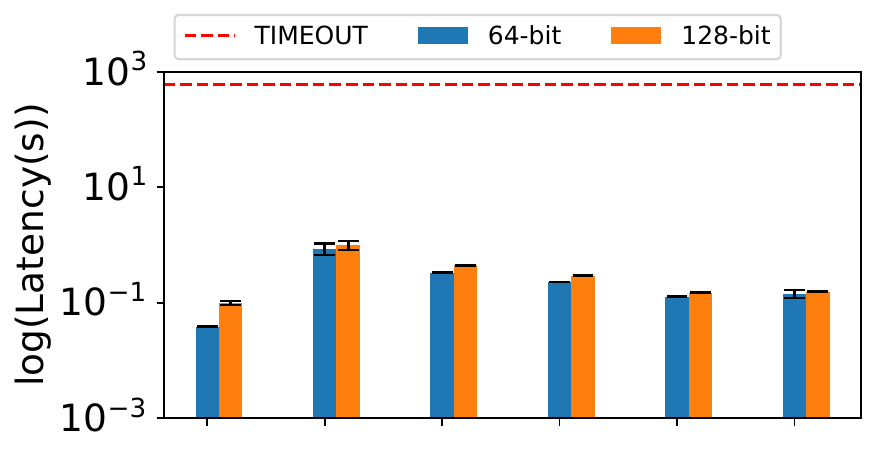}%
	\end{subfigure}%
	\begin{subfigure}{0.25\textwidth}\centering%
		\includegraphics[scale=\evalgraphscale]{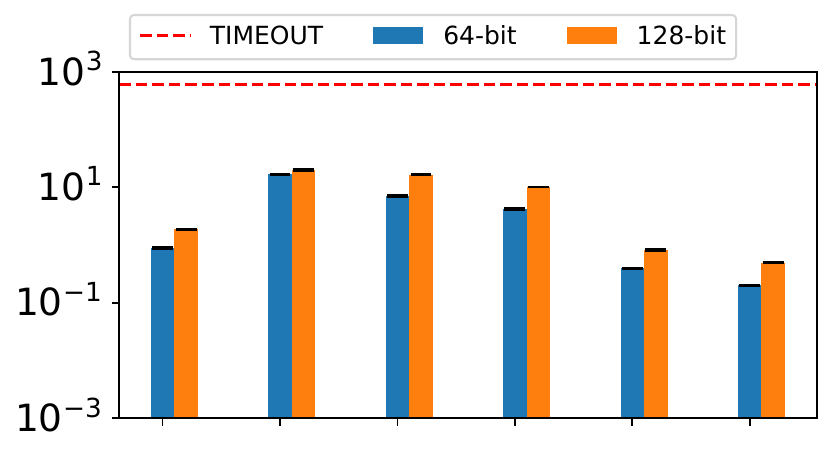}%
	\end{subfigure}%
	\begin{subfigure}{0.25\textwidth}\centering%
		\includegraphics[scale=\evalgraphscale]{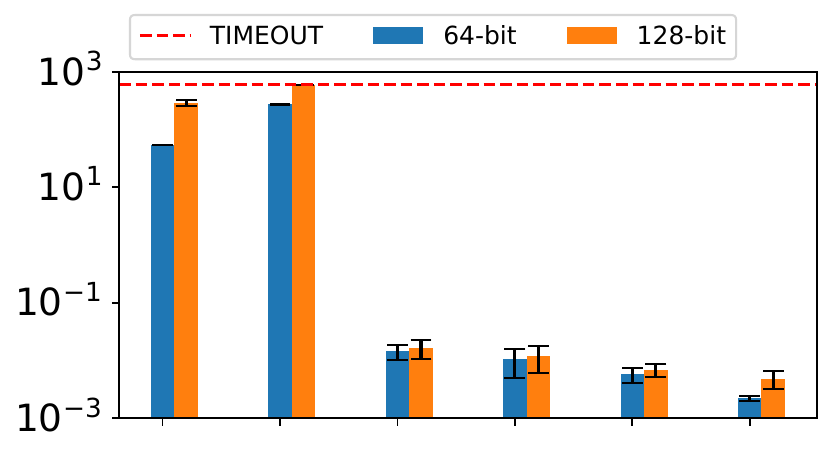}%
	\end{subfigure}%
	\begin{subfigure}{0.25\textwidth}\centering%
		\includegraphics[scale=\evalgraphscale]{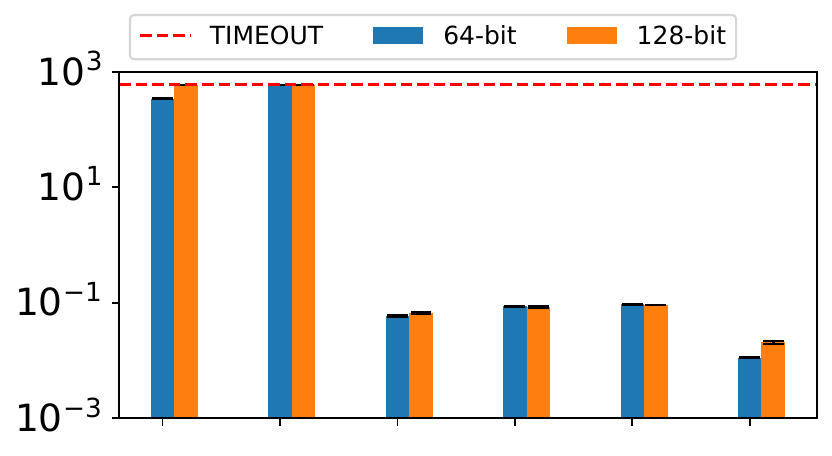}%
	\end{subfigure}%
	\centering%
	
	\begin{subfigure}{0.25\textwidth}\centering%
		\includegraphics[scale=\evalgraphscale]{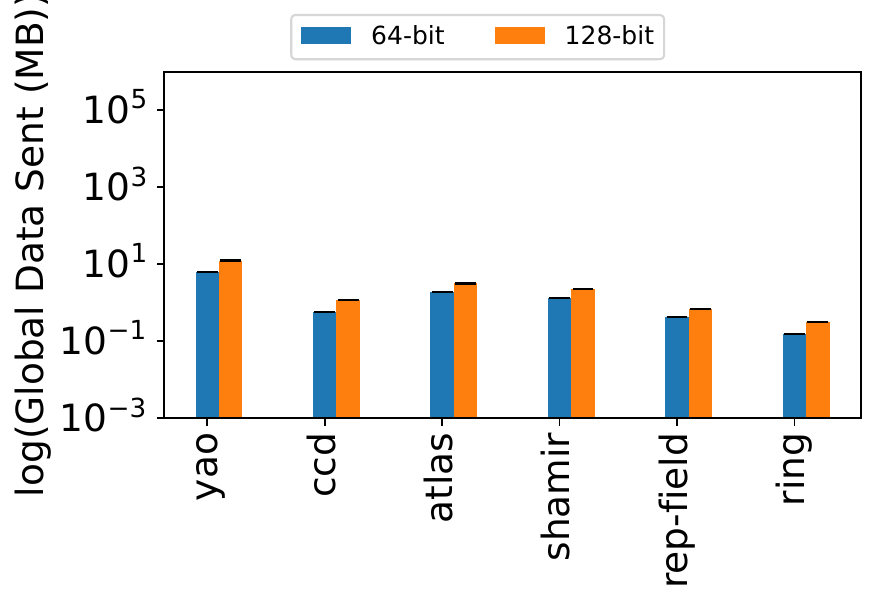}%
		\caption{Comparison, N=1024}%
		\label{fig:sh_h_proto_comp}%
	\end{subfigure}%
	\begin{subfigure}{0.25\textwidth}\centering%
		\includegraphics[scale=\evalgraphscale]{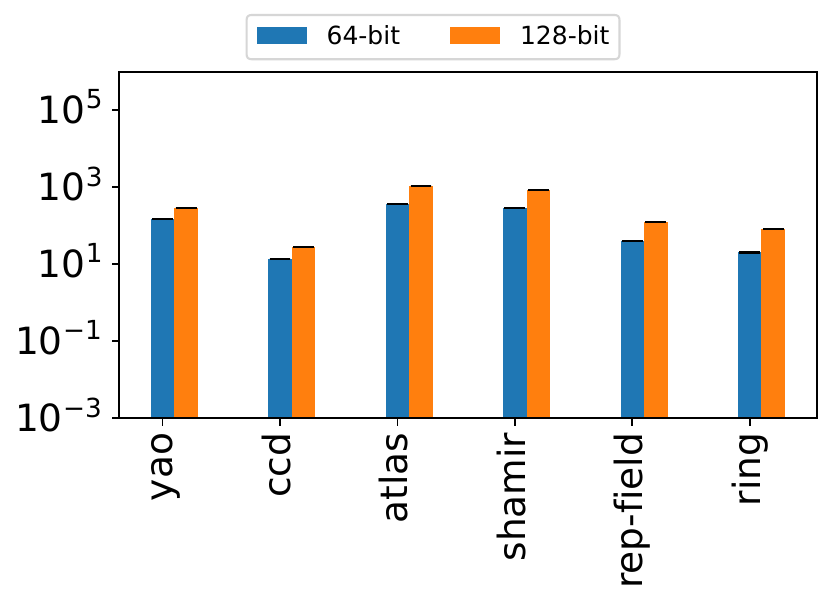}%
		\caption{Sort, N=1024}%
		\label{fig:sh_h_proto_sort}%
	\end{subfigure}%
	\begin{subfigure}{0.25\textwidth}\centering%
		\includegraphics[scale=\evalgraphscale]{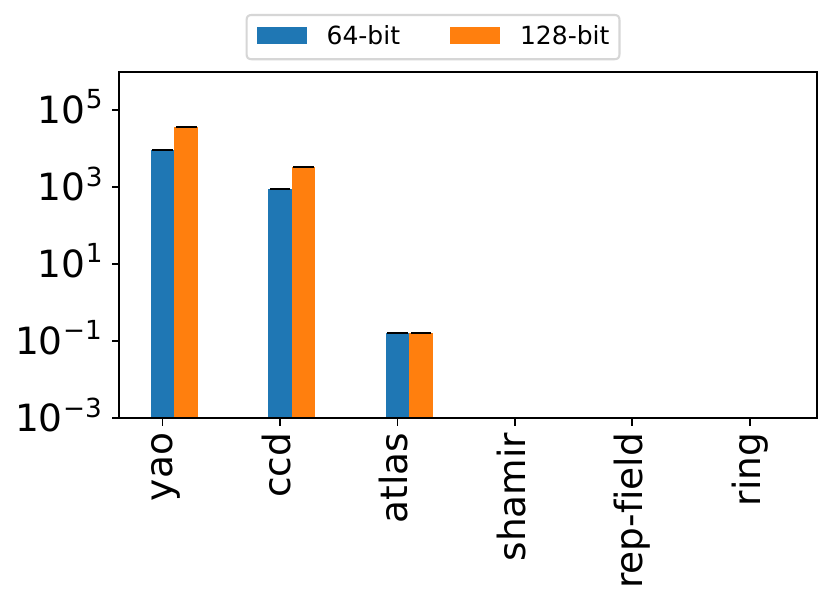}%
		\caption{Innerproduct, N=65536}%
		\label{fig:sh_h_proto_ip}%
	\end{subfigure}%
	\begin{subfigure}{0.25\textwidth}\centering%
		\includegraphics[scale=\evalgraphscale]{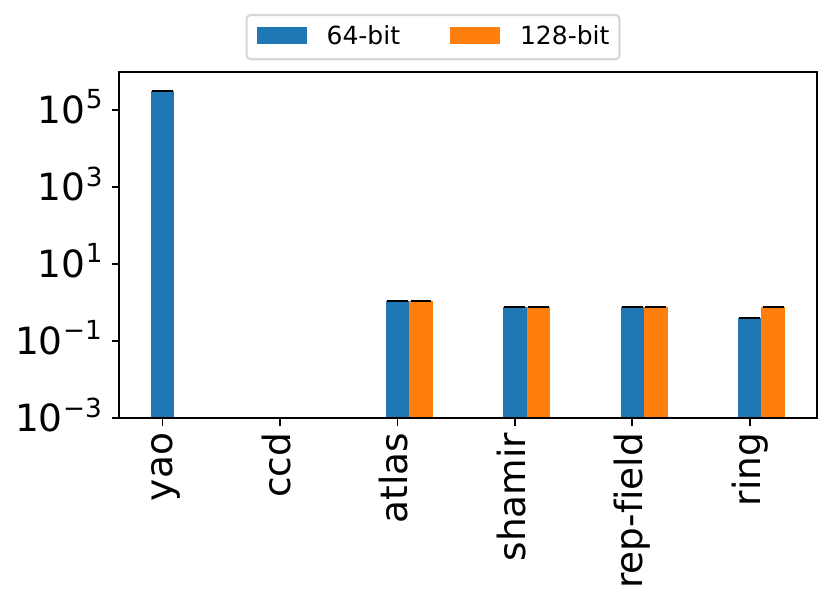}%
		\caption{Mat Mul, N=128}%
		\label{fig:sh_h_proto_mm}%
	\end{subfigure}%
	\caption{Performance of semi-honest protocols with honest majority (sh\_h) on our benchmarks}%
	\label{fig:sh_h_proto}%
\end{figure*}

\begin{figure*}[h]%
	\centering%
		\begin{subfigure}{0.25\textwidth}\centering%
		\includegraphics[scale=\evalgraphscale]{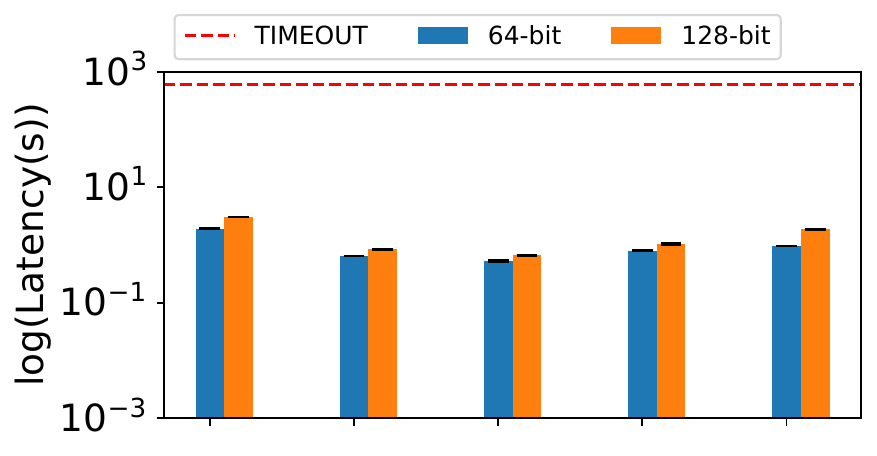}%
	\end{subfigure}%
	\begin{subfigure}{0.25\textwidth}\centering%
		\includegraphics[scale=\evalgraphscale]{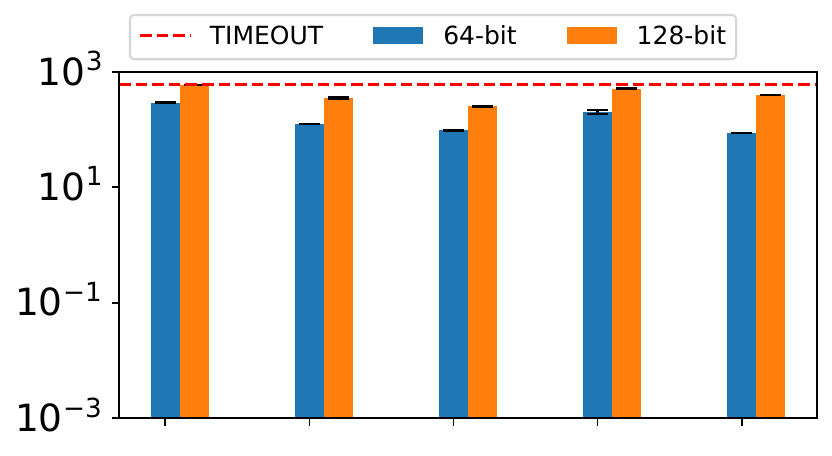}%
	\end{subfigure}%
	\begin{subfigure}{0.25\textwidth}\centering%
		\includegraphics[scale=\evalgraphscale]{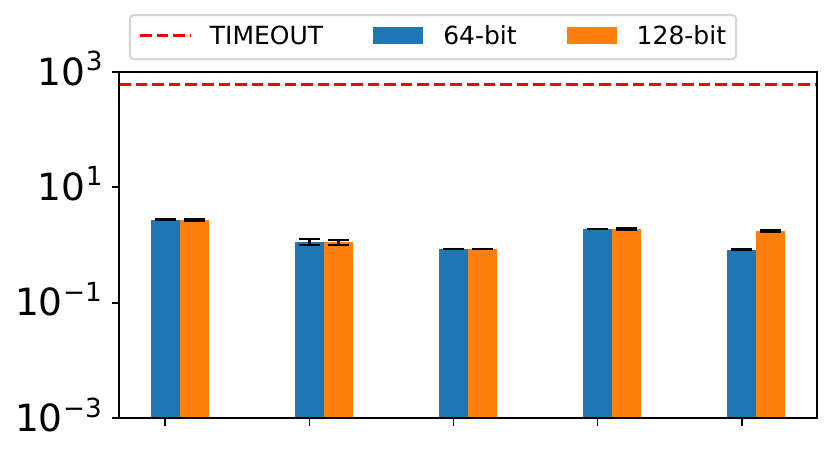}%
	\end{subfigure}%
	\begin{subfigure}{0.25\textwidth}\centering%
		\includegraphics[scale=\evalgraphscale]{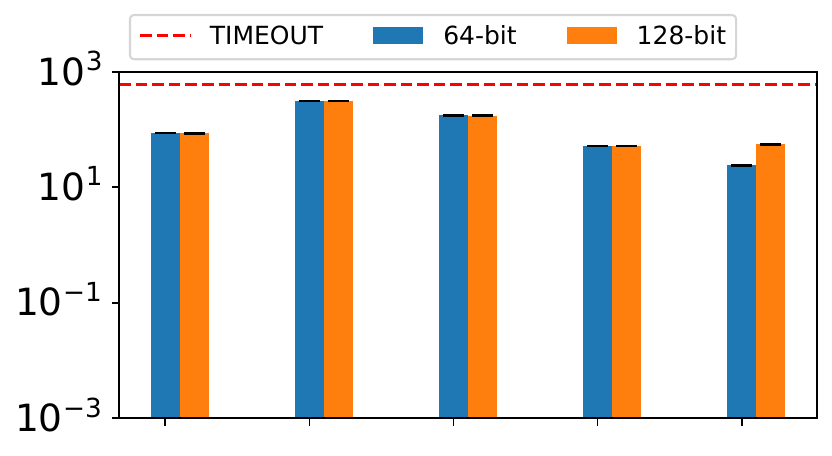}%
	\end{subfigure}%
	\centering%
	
		\begin{subfigure}{0.25\textwidth}\centering%
		\includegraphics[scale=\evalgraphscale]{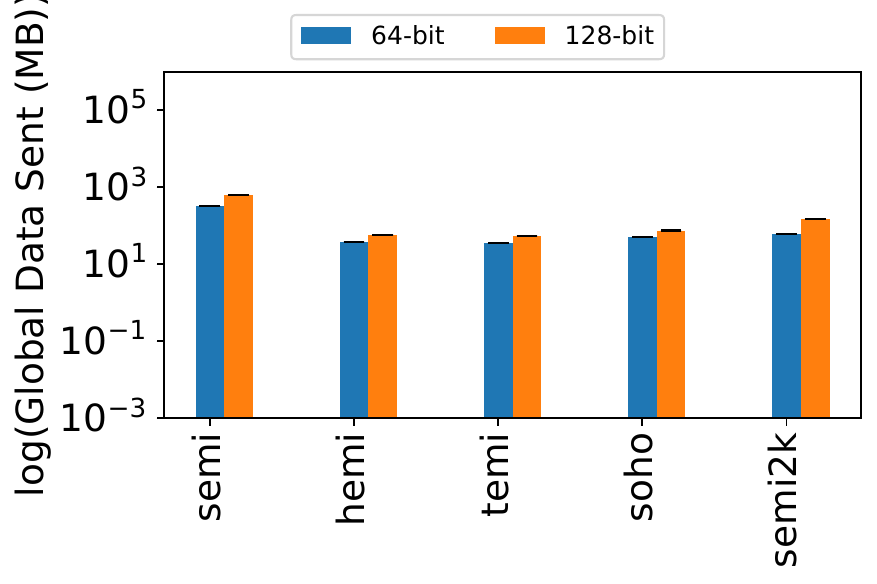}%
		\caption{Comparison, N=1024}%
		\label{fig:sh_dh_proto_comp}%
	\end{subfigure}%
	\begin{subfigure}{0.25\textwidth}\centering%
		\includegraphics[scale=\evalgraphscale]{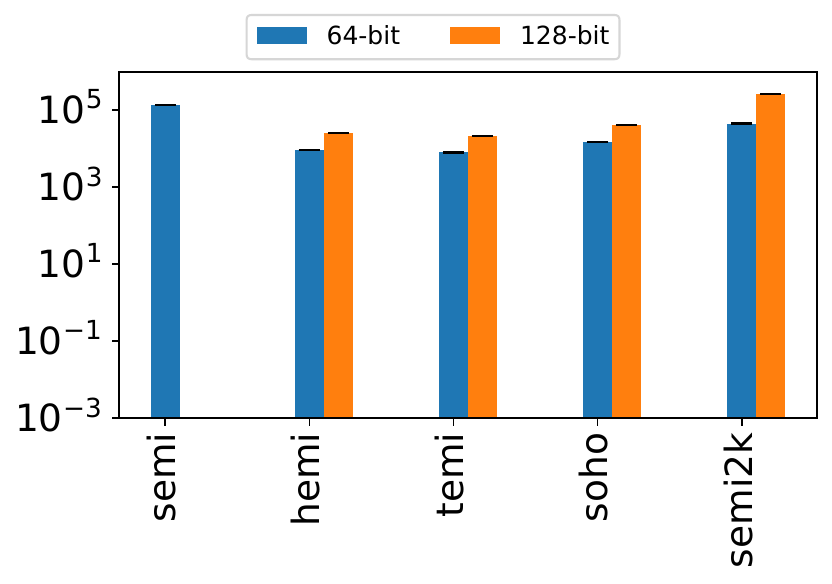}%
		\caption{Sort, N=1024}%
		\label{fig:sh_dh_proto_sort}%
	\end{subfigure}%
	\begin{subfigure}{0.25\textwidth}\centering%
		\includegraphics[scale=\evalgraphscale]{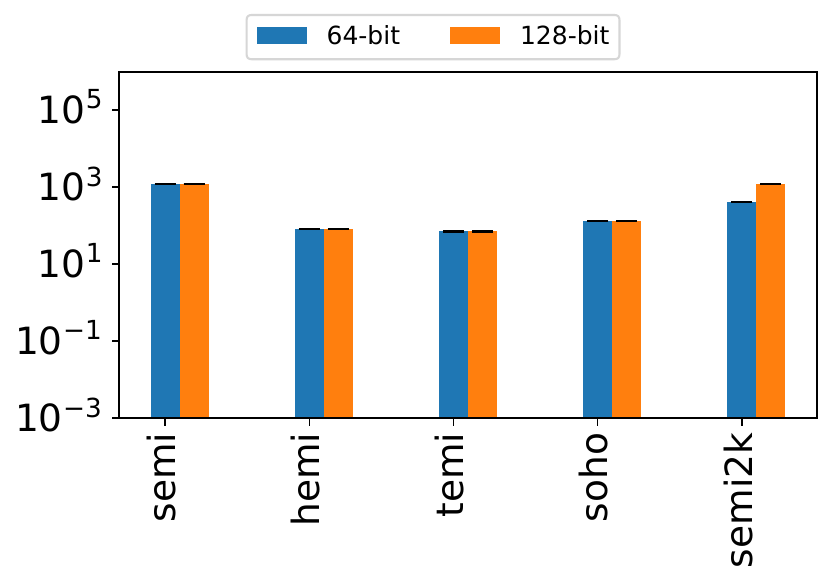}%
		\caption{Innerproduct, N=65536}%
		\label{fig:sh_dh_proto_ip}%
	\end{subfigure}%
	\begin{subfigure}{0.25\textwidth}\centering%
		\includegraphics[scale=\evalgraphscale]{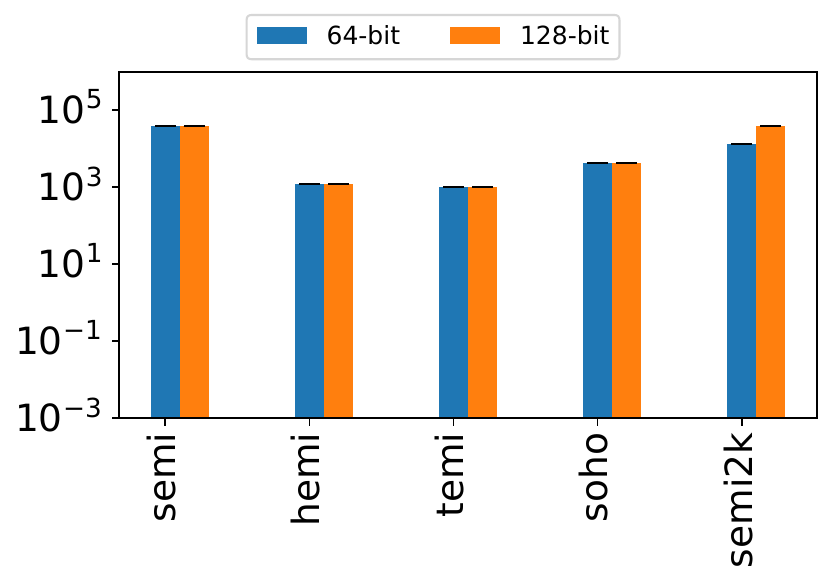}%
		\caption{Mat Mul, N=128}%
		\label{fig:sh_dh_proto_mm}%
	\end{subfigure}%
	\caption{Performance of semi-honest protocols with dishonest majority (sh\_dh) on our benchmarks}%
	\label{fig:sh_dh_proto}%
\end{figure*}

\begin{figure*}[h]%
	\centering%
		\begin{subfigure}{0.25\textwidth}\centering%
		\includegraphics[scale=\evalgraphscale]{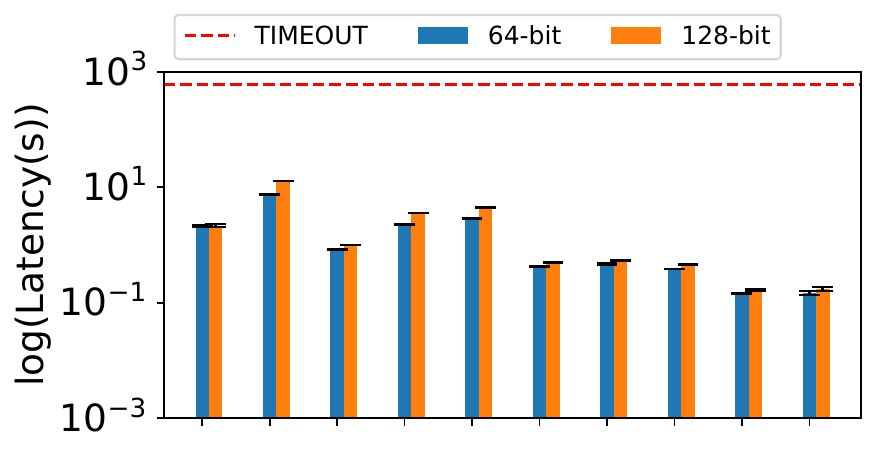}%
	\end{subfigure}%
	\begin{subfigure}{0.25\textwidth}\centering%
		\includegraphics[scale=\evalgraphscale]{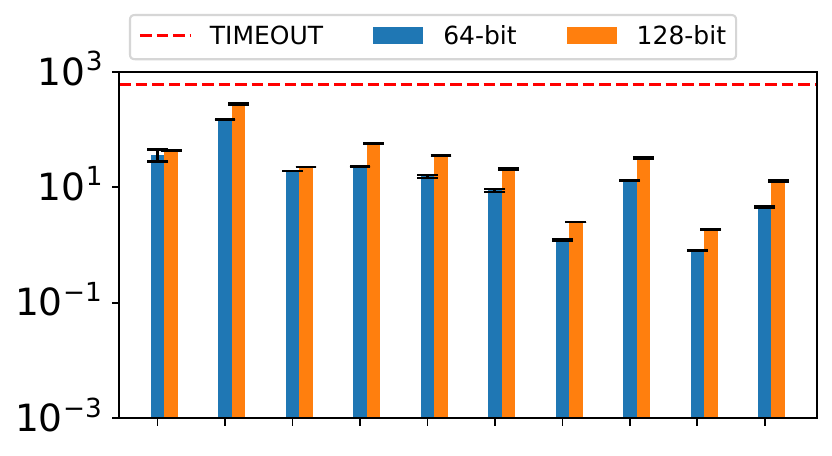}%
	\end{subfigure}%
	\begin{subfigure}{0.25\textwidth}\centering%
		\includegraphics[scale=\evalgraphscale]{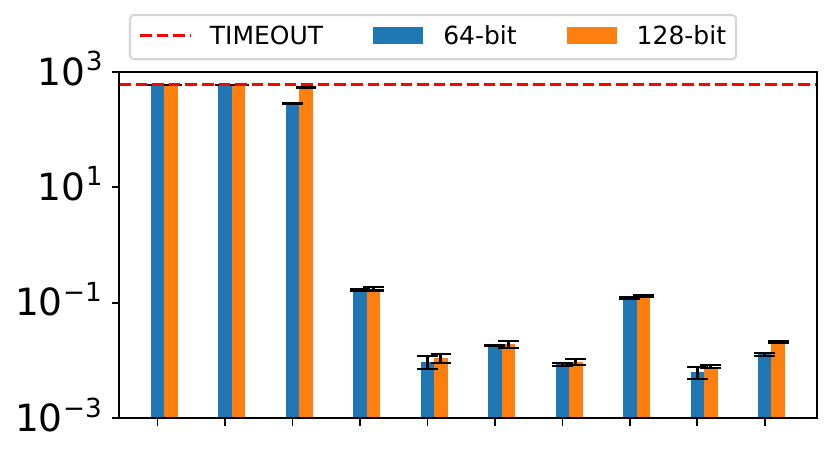}%
	\end{subfigure}%
	\begin{subfigure}{0.25\textwidth}\centering%
		\includegraphics[scale=\evalgraphscale]{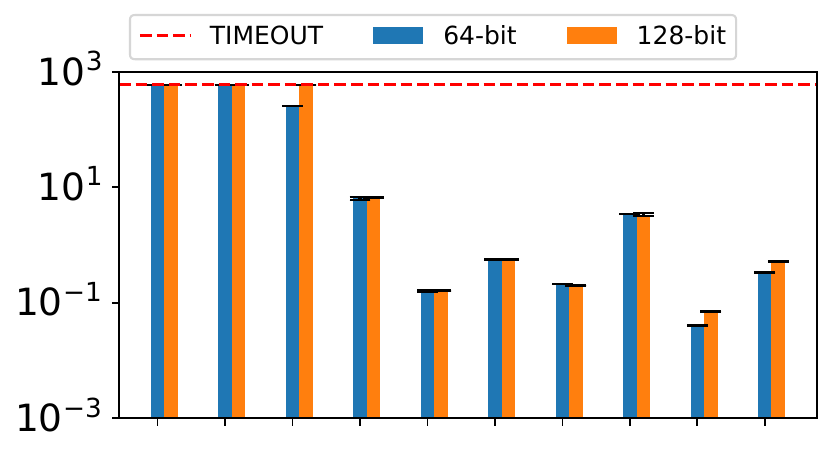}%
	\end{subfigure}%
	
	\centering%
		\begin{subfigure}{0.25\textwidth}\centering%
		\includegraphics[scale=\evalgraphscale]{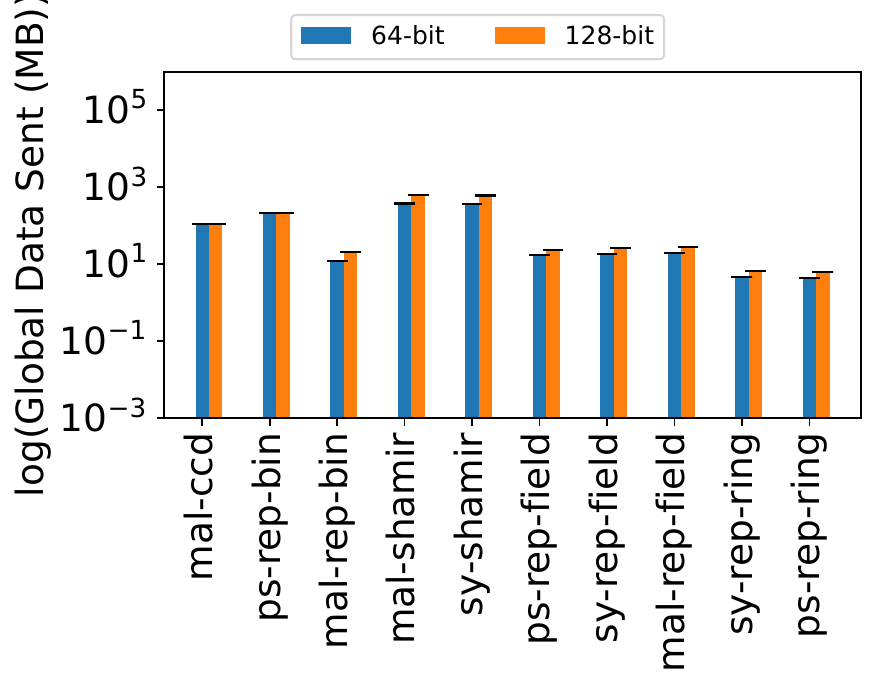}%
		\caption{Comparison, N=1024}%
		\label{fig:mal_h_proto_comp}%
	\end{subfigure}%
	\begin{subfigure}{0.25\textwidth}\centering%
		\includegraphics[scale=\evalgraphscale]{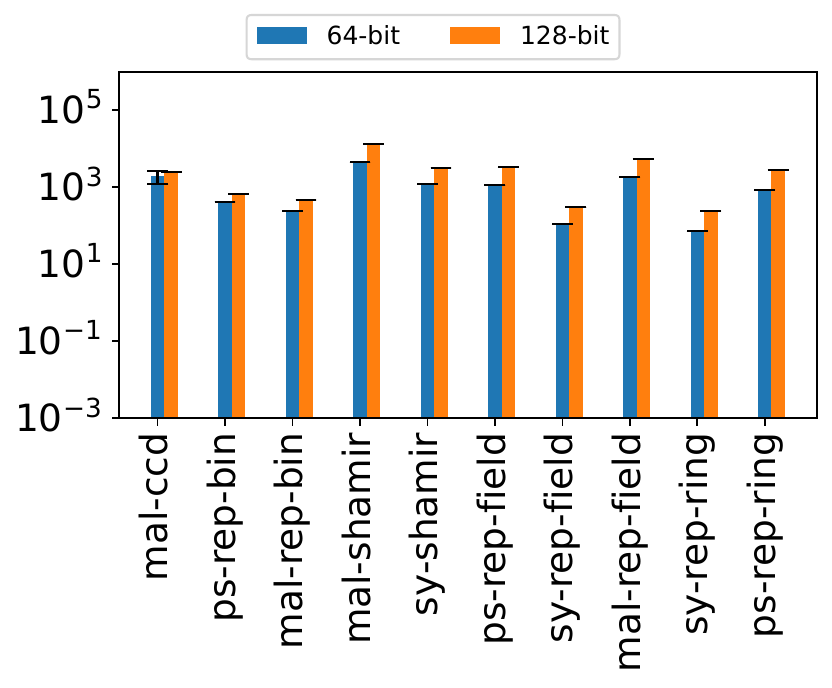}%
		\caption{Sort, N=1024}%
		\label{fig:mal_h_proto_sort}%
	\end{subfigure}%
	\begin{subfigure}{0.25\textwidth}\centering%
		\includegraphics[scale=\evalgraphscale]{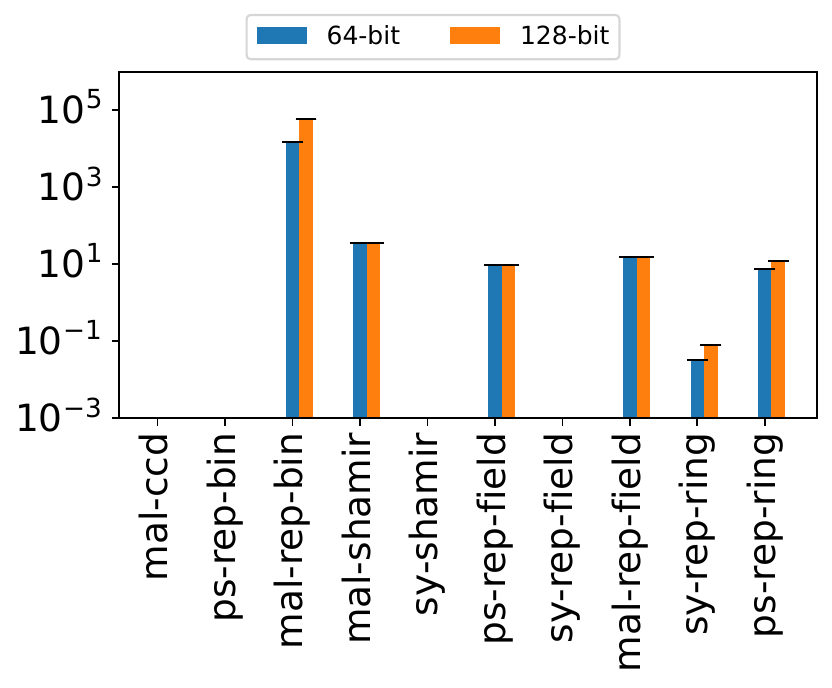}%
		\caption{Innerproduct, N=65536}%
		\label{fig:mal_h_proto_ip}%
	\end{subfigure}%
	\begin{subfigure}{0.25\textwidth}\centering%
		\includegraphics[scale=\evalgraphscale]{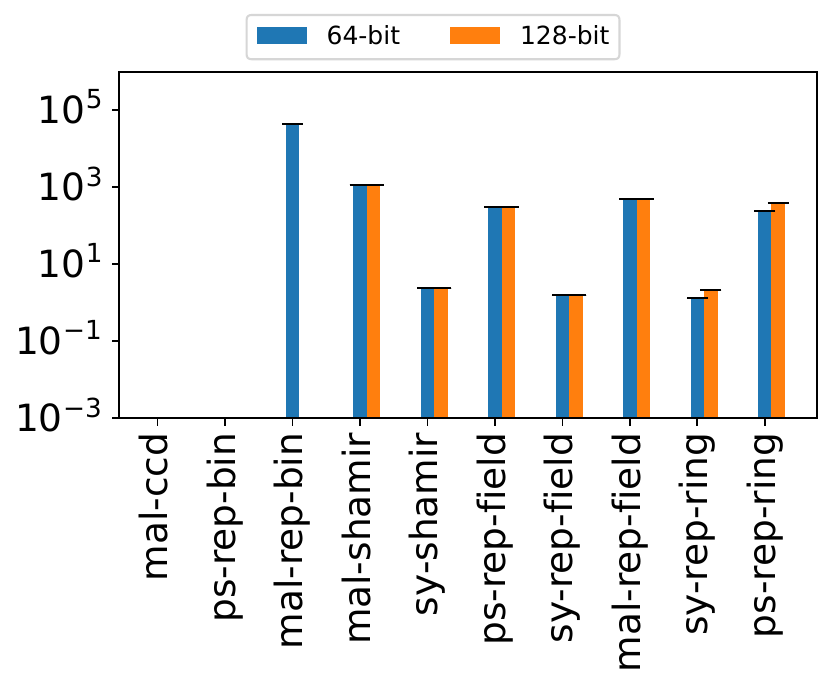}%
		\caption{Mat Mul, N=128}%
		\label{fig:mal_h_proto_mm}%
	\end{subfigure}%
	\caption{Performance of malicious protocols with honest majority (mal\_h) on our benchmarks}%
	\label{fig:mal_h_proto}%
\end{figure*}

\begin{figure*}[h]%
\centering%
\begin{subfigure}{0.25\textwidth}\centering%
	\includegraphics[scale=\evalgraphscale]{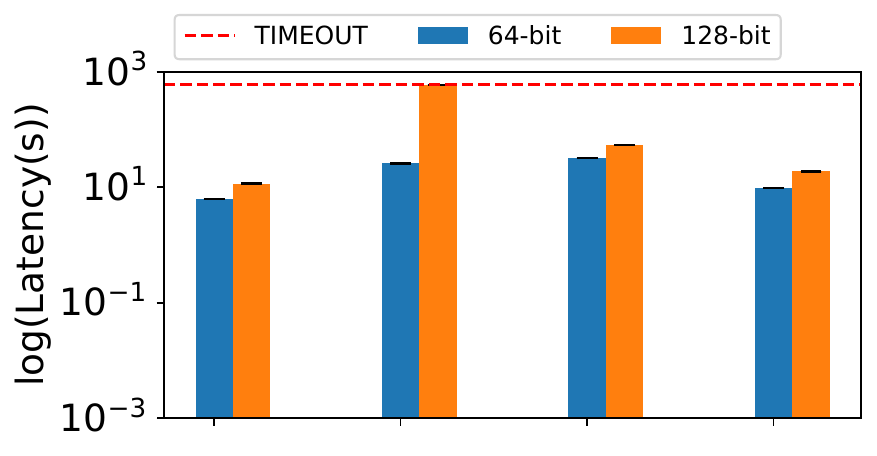}%
	\label{fig:mal_dh_proto_lat_comp}%
\end{subfigure}%
\begin{subfigure}{0.25\textwidth}\centering%
\includegraphics[scale=\evalgraphscale]{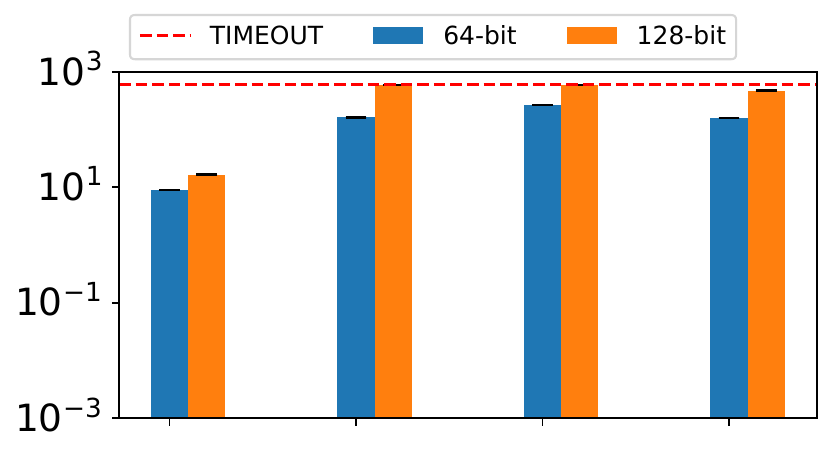}%
\label{fig:mal_dh_proto_lat_sort}%
\end{subfigure}%
\begin{subfigure}{0.25\textwidth}\centering%
\includegraphics[scale=\evalgraphscale]{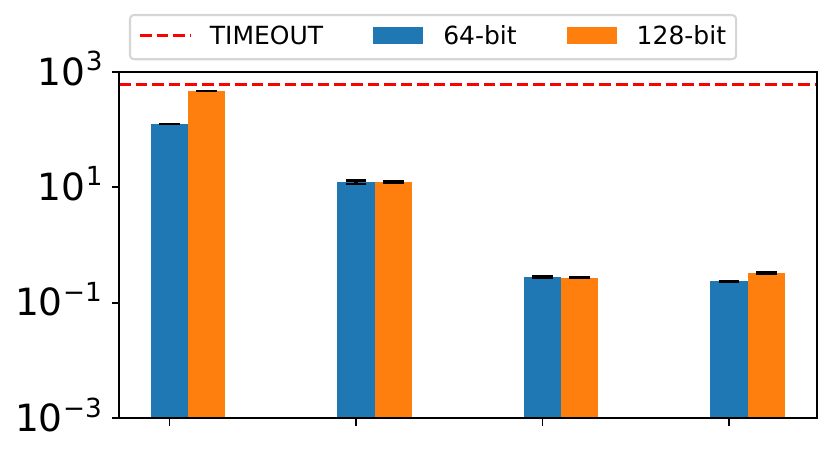}%
\end{subfigure}%
\begin{subfigure}{0.25\textwidth}\centering%
\includegraphics[scale=\evalgraphscale]{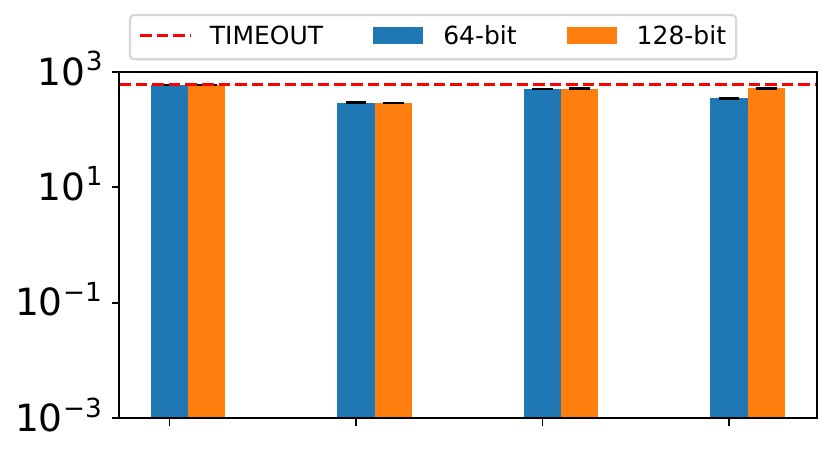}%
\end{subfigure}%
\centering%

\begin{subfigure}{0.25\textwidth}\centering%
	\includegraphics[scale=\evalgraphscale]{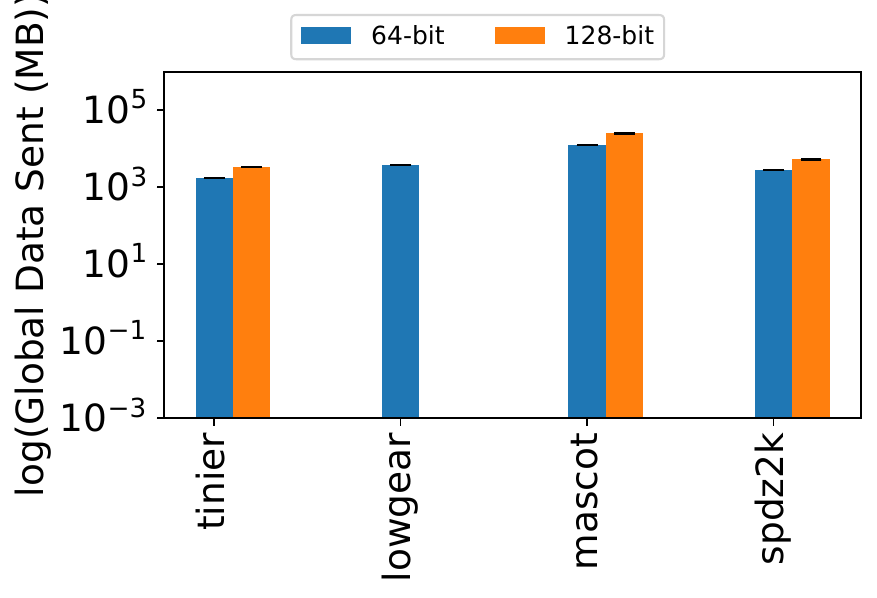}%
	\caption{Comparison, N=1024}%
	\label{fig:mal_dh_proto_comp}%
\end{subfigure}%
\begin{subfigure}{0.25\textwidth}\centering%
\includegraphics[scale=\evalgraphscale]{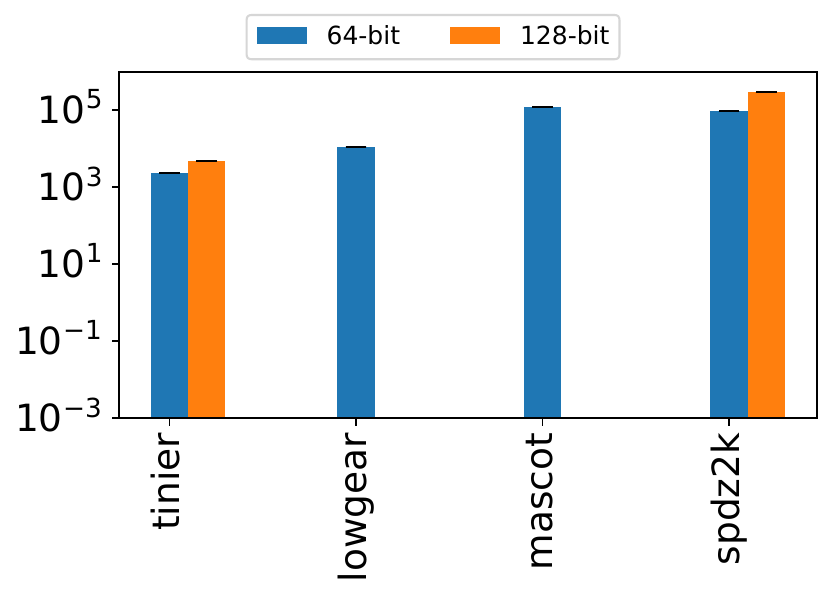}%
\caption{Sort, N=128}%
\label{fig:mal_dh_proto_sort}%
\end{subfigure}%
\begin{subfigure}{0.25\textwidth}\centering%
\includegraphics[scale=\evalgraphscale]{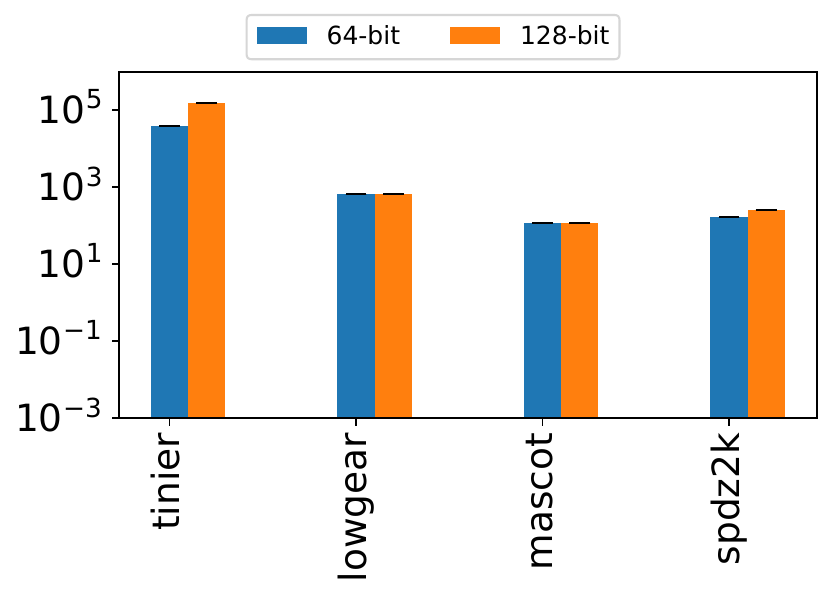}%
\caption{Innerproduct, N=1024}%
\label{fig:mal_dh_proto_ip}%
\end{subfigure}%
\begin{subfigure}{0.25\textwidth}\centering%
\includegraphics[scale=\evalgraphscale]{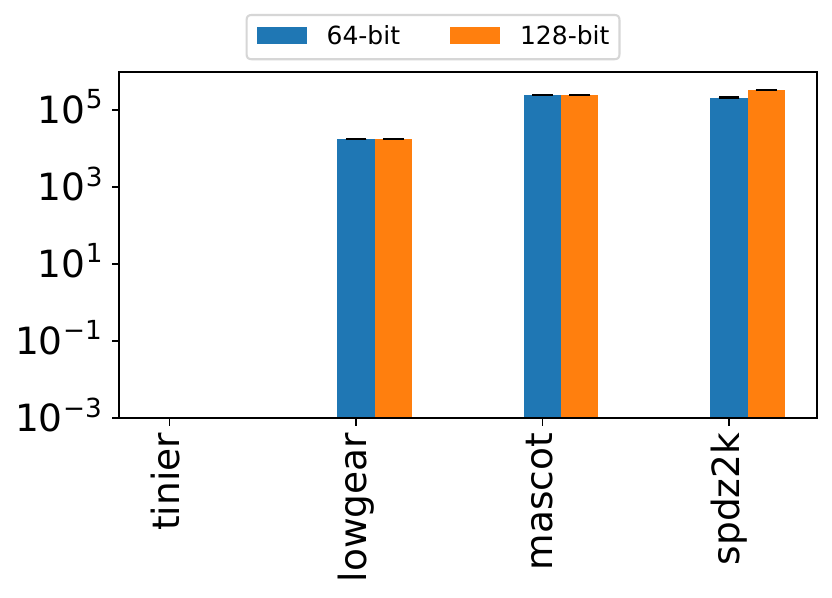}%
\caption{Mat Mul, N=128}%
\label{fig:mal_dh_proto_mm}%
\end{subfigure}%
\caption{Performance of malicious protocols with dishonest majority (mal\_dh) on our benchmarks}%
\label{fig:mal_dh_proto}%
\end{figure*}

\subsubsection{Relative Protocol Performance}

We first report the relative performances of different protocols on
our four primitives. Empirically, we found that the size of the input
\(N\) does not change the relative order of performance of
protocols. Hence, for each primitive we pick an \(N\) that brings out
the difference between the best and worst protocols most noticeably.

We divide protocols into four categories by their threat models
(\autoref{tab:protocols}): semi-honest with honest-majority
(sh\_h), semi-honest with dishonest-majority (sh\_dh), malicious with
honest-majority (mal\_h), and malicious with dishonest-majority
(mal\_dh) (\autoref{sec:adversary}). We report results for each
primitive and each category separately in
\autoref{fig:sh_h_proto}--\ref{fig:mal_dh_proto}.

\paragraph{\underline{Comparison, N = 1024 ($2^{32}$)}}

\paragraph{sh\_h} (\autoref{fig:sh_h_proto_comp}): The lowest latency
is obtained by the \textsc{yao} protocol, which relies on garbled
circuits, while \textsc{ring} transmits the least amount of data. This
is unsurprising: \textsc{yao} is primarily a streaming protocol, which
does not synchronize the parties often, accounting for its lower
latency. However, \textsc{yao} also represents each bit of the
plaintext computation as a 128-bit garbled value in our
implementation, so it transmits an enormous amount of data. 

\paragraph{sh\_dh} (\autoref{fig:sh_dh_proto_comp}): In this category,
\textsc{hemi} and \textsc{temi} are neck-to-neck in having the lowest
latency and the lowest data sent; \textsc{soho} is close behind, and
the remaining protocols are also close. The better performance of
\textsc{hemi}, \textsc{temi} and \textsc{soho} is due to specific
optimizations that are built into these protocols. Note that all
protocols in the sh\_dh category use arithmetic domains; we were
unable to compile the binary domain protocols in this category in
MP-SPDZ.

\paragraph{mal\_h} (\autoref{fig:mal_h_proto_comp}):
\textsc{ps-rep-ring}/\textsc{sy-rep-ring} have very similar and the
best performance on both latency and total data sent. Both use
arithmetic domains based on rings. The protocol \textsc{mal-rep-bin},
which uses the binary domain, is close behind on both metrics. The
slight outperformance of protocols based on the arithmetic domain on a
benchmark that relies on bit-level logical operations (comparisons)
may be surprising; we believe that this is largely due to advancement
in techniques for efficient bit decomposition (specifically, the use
of edaBits) and other optimizations that
\textsc{ps-rep-ring}/\textsc{sy-rep-ring} rely on.

\paragraph{mal\_dh} (\autoref{fig:mal_dh_proto_comp}):
\textsc{tinier}, a binary domain protocol, is the fastest and most
network efficient protocol in this category, while \textsc{spdz2k}, an
arithmetic ring-based protcol is close behind. Here, \textsc{spdz2k}'s
advantage in using machine integers directly is offset by its need for
bit decomposition.

\paragraph{Summary (comparisons)}
For logical operations like integer comparisons, the best-performing
protocols based on the ring domain outperform or are close behind the
best protocols using the binary domain. While one may expect that the
binary domain would align better with logical operations, the ability
to use CPU addition and multiplication directly in the ring domain,
and the advent of fast bit decomposition techniques compensates for
the difference in representation. The exception to this observation is
garbled circuits (\textsc{yao}), which when usable, offer much lower
latency but comparatively higher bandwidth.

\smallskip
\paragraph{\underline{Sort, N = 1024 ($2^{10}$)}}

\mbox{} \smallskip

\noindent The sorting protocol we use relies on both bit decomposition
and arithmetic operations, not just bit decomposition. As a result,
relative to the \emph{comparison} benchmark, ring-based protocols
begin to outshine binary-based protocols on \emph{sort}.

In sh\_h, \textsc{ring} has the lowest latency, while \textsc{ccd}, a
binary protocol, is transmits the least amount of data and
\textsc{ring} is close behind. The low latency of \textsc{ccd} is due
to its use of a binary domain; our sorting algorithm relies heavily on
bit representations, so the remaining protocols, which are either
garble circuit-based or arithmetic-domain, transmit more data. In
sh\_dh, \textsc{hemi}/\textsc{semi} continue to lead in both having
the lowest latency and the least data volume, with \textsc{soho} close
behind.

In mal\_h, \textsc{sy-rep-ring} and \textsc{sy-rep-field} are
neck-to-neck in having the lowest latency and the lowest data
volume. The protocol \textsc{ps-rep-ring}, which was very close to
\textsc{sy-rep-ring} on comparisons, is significantly behind
\textsc{sy-rep-ring} on sorting. This difference can be attributed to
\textsc{ps-rep-ring}'s use of postprocessing of Beaver triples, which
is less effecient than \textsc{sy-rep-ring}'s pre-processing, when the
number of triples is large. This effect increases increases with the
number of arithmetic operations (the differences between the two
protocols are even more pronounced in the \emph{matmul} benchmark,
which use multiplications extensively). In mal\_dh, \textsc{tinier}, a
binary-domain protocol, continues to dominate other protocols despite
arithmetic-domain protocols having a structural advantage on the
arithmetic part of our sorting algorithm (\textsc{tinier} falls behind
arithmetic protocols on the \emph{inner product} and \emph{matmul}
benchmarks, which use only arithmetic operations, as explained below).

The best protocol choice matches \textit{comparison} in all
settings except mal-h, where (i) \textsc{sy-rep-ring} replaces
\textsc{ps-rep-ring} as choice for lowest latency and best overall,
and (ii) \textsc{mal-ccd} and \textsc{ps/mal-rep-bin} show lower
bandwidth, but $\sim35\times$ higher latency than
\textsc{sy-rep-ring}.

\paragraph{Summary (sort)}
The best protocols in each category except mal\_dh are all
ring-based. Our chosen sorting protocol relies on both bit
decomposition and arithmetic operations protocols that optimize
arithmetic operations begin to outshine other protocols.

\smallskip

\paragraph{\underline{Inner Product, N = 65,536 ($2^{16}$)}}

\mbox{}

\smallskip

Inner product relies even more heavily on arithmetic operations than
does our sorting algorithm. The relative performance of protocols on
\emph{inner product} is similar to that on \emph{sort}, but the
advantages of arithemtic-optimized protocols become more
pronounced.

In sh\_h, the best protocol on both our metrics is \textsc{ring} and
it leads binary protocols like \textsc{yao} and \textsc{ccd} by a
larger margin. In sh\_dh, \textsc{hemi}/\textsc{temi} lead with
\textsc{soho} close behind.

In mal\_h, \textsc{sy-rep-ring} offers the best latency, while
\textsc{sy-rep-field} offers the lowest data transmitted. In mal\_dh,
two arithmetic domain protocols, \textsc{spdz2k} and \textsc{mascot},
are neck-to-neck in offering the lowest latency and the lowest data
volume. The binary domain protocol \textsc{tinier}, which led on
\emph{comparison} and \emph{sort} is significantly behind due to
arithmetic-heavy nature of the present benchmark.

\paragraph{Summary (inner product)}
Since inner product relies on arithmetic operations, in all
categories, the best protocols are from the arithmetic domain. In
general ring-based protocols outperform field-based protcols, but
there are exceptions for data volume due to primitive-specific
optimizations, e.g., \textsc{sy-rep-field} has significantly lower
data volume than \textsc{sy-rep-ring}.

\smallskip
\paragraph{\underline{Matmul, N = 128 ($2^{7}$)}}

\mbox{} \smallskip

The relative order of performance of protocols on matrix
multiplication is almost the same as that on inner product, with a few
noteworthy exceptions. First, in mal\_h, the significant difference in
data transmitted by \textsc{sy-rep-field} and \textsc{sy-rep-ring}
vanishes; \textsc{sy-rep-ring} is at par with \textsc{sy-rep-field} on
the amount of data transmitted (and \textsc{sy-rep-ring} has lower
latency). Second, in mal\_dh, the best performing protocol is
\textsc{lowgear}, which is also field-based, but specifically
optimizes the offline process of triple generation, and excels when
the benchmark requires a very large number of multiplications, as in
this benchmark. \textsc{mascot} and \textsc{spdz2k}, which dominated
\emph{inner product}, do not use this optimization, and are,
therefore, slower than \textsc{lowgear} on this benchmark. 

\paragraph{Summary (matmul)}
Matrix multiplication is dominated by a specific arithmetic operation
--- multiplications. Consequently, arithmetic protcols outperform all
others on this primitive and, importantly, arithmetic protocols that
are optimized on Beaver triples (like \textsc{lowgear}) outshine
others, while those that scale poorly with the number of Beaver
triples fall behind others (e.g., like \textsc{ps-rep-ring} performs
significantly worse than \textsc{sy-rep-ring} on \emph{matmul}, as
compared to their relative performance on \emph{sort} or
\emph{comparison}).

\subsubsection{Impact of integer bit-width increase (64$\rightarrow$128)}

Across all primitives and threat model categories, moving from 64- to
128-bit integers shifts the Pareto frontier towards higher latency and
higher data volume, but the relative performance of protocols is
largely unchanged. Specifically, 128-bit integers raise costs
primarily by increasing the sizes of individual data elements, their
encodings and MACs (in maliciously-secure protocols), but the round
complexity does not change.

\subsubsection{Impact of local share conversion on ring protocols}

\begin{figure}
	\centering%
	\begin{subfigure}{0.5\linewidth}%
		\includegraphics[scale=\compgraphscale]{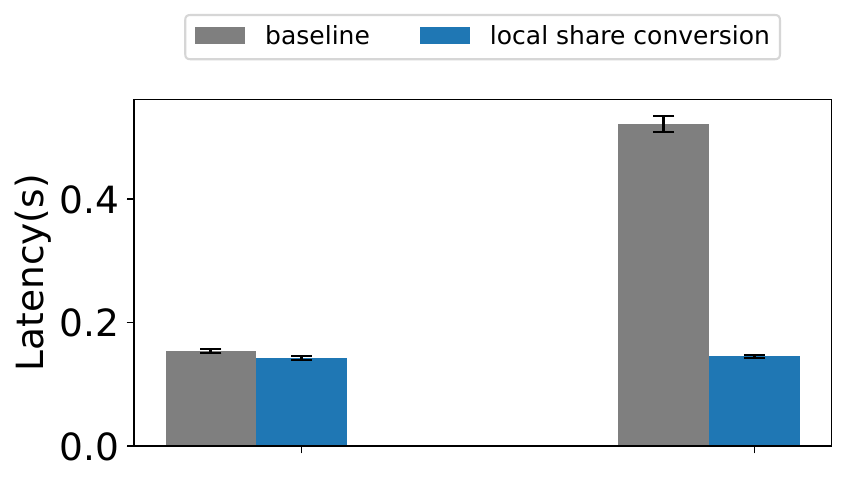}%
	\end{subfigure}%
	\begin{subfigure}{0.5\linewidth}%
		\includegraphics[scale=\compgraphscale]{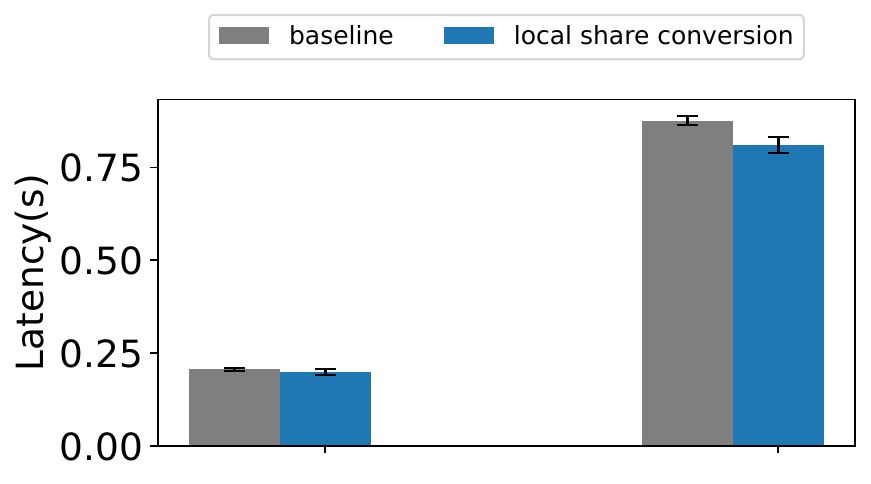}%
	\end{subfigure}%
	
	\begin{subfigure}{0.5\linewidth}%
		\includegraphics[scale=\compgraphscale]{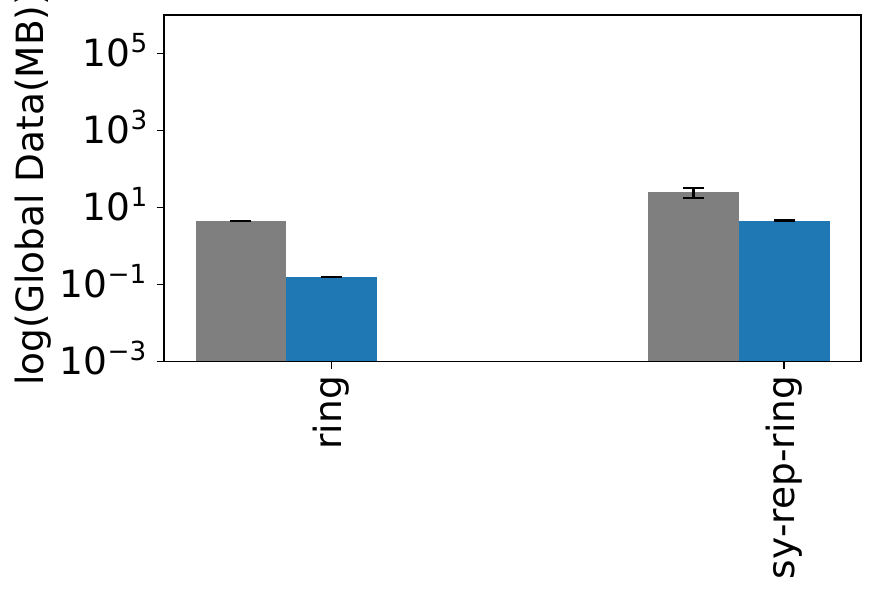}%
		\caption{Comparison, N=1024}%
		\label{fig:mc_comp}%
	\end{subfigure}%
	\begin{subfigure}{0.5\linewidth}%
		\includegraphics[scale=\compgraphscale]{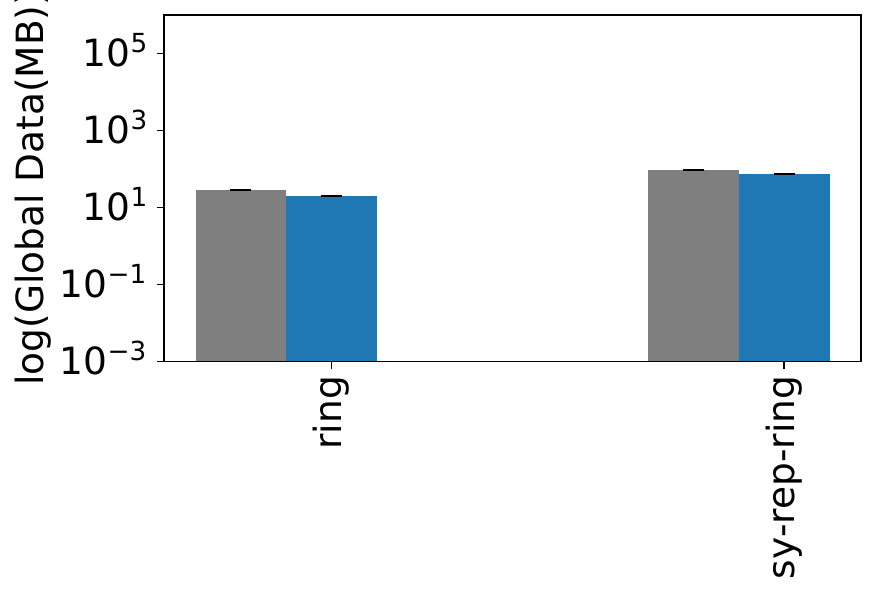}%
		\caption{Sort, N=1024}%
		\label{fig:mc_sort}%
	\end{subfigure}%
	\caption{Impact of enabling local share conversion on ring-based protocols}
	\label{fig:mixedcomp}
\end{figure}

In this section, we quantify the effect of local share conversion,
which is a technique to decompose a shared integer into shares of its
bit representation efficiently. This technique benefits computations
that rely on logical operations on integers. However, the technique
applies only to ring-based protocols that replicate shares. In
MP\,-SPDZ~\cite{keller2020mp}, this local share conversion is enabled
available via the \texttt{-Z} compilation flag, which is enabled by
default; to quantify the impact of enabling the flag, we override the
default to explicitly turn it off.

Among our primitives, \emph{comparison} and \emph{sort} are the right
stress tests for local share conversion as they both need bit
decompositions. \autoref{fig:mixedcomp} shows the effect of local
share conversion for two ring-based protocols, one semi-honest
(\textsc{ring}) and the other malicious (\textsc{sy-rep-ring}). The
gray lines are metrics without local share conversion, while the blue
lines are with local share conversion. As can be seen, with both
protocols, local share conversion reduces the latency and data volume
significantly on \emph{comparison} and to a lesser extent on
\emph{sort}. This is understandable because \emph{sort} has a large
component of arithmetic operations that do not benefit from local
share conversion, while \emph{comparison} benefits in its entirety.

We expect that other compound non-linear primitives (e.g., top-$k$,
joins, argmin/argmax, range filters, fixed/floating-point kernels)
will see similar benefit from local share conversion. We leave an
evaluation on these operators to future work (\autoref{sec:future}).

\subsubsection{Impact of input size N}
\label{subsec:varyingn}

\begin{figure*}[t]
\centering%
\begin{subfigure}{0.25\linewidth}%
\includegraphics[scale=\evalgraphscale]{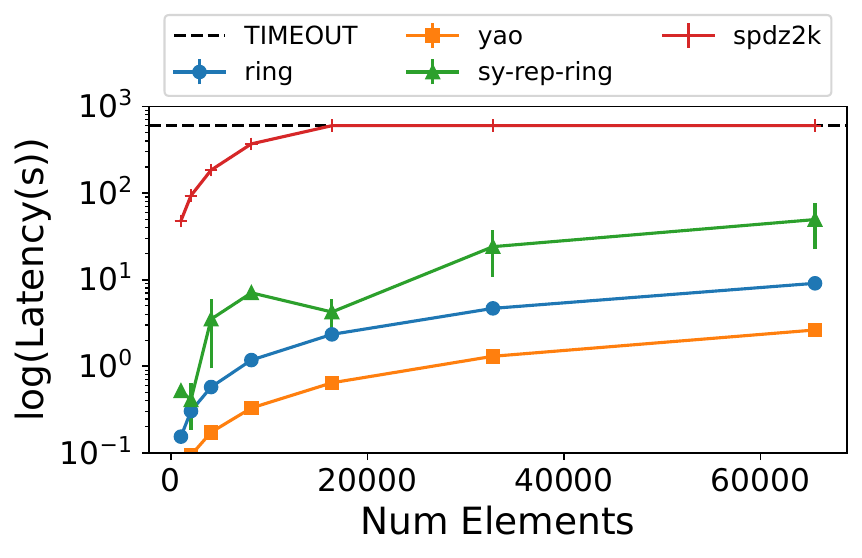}%
\caption{Comparison, $N=2\rightarrow65536$}%
\end{subfigure}%
\begin{subfigure}{0.25\linewidth}%
\includegraphics[scale=\evalgraphscale]{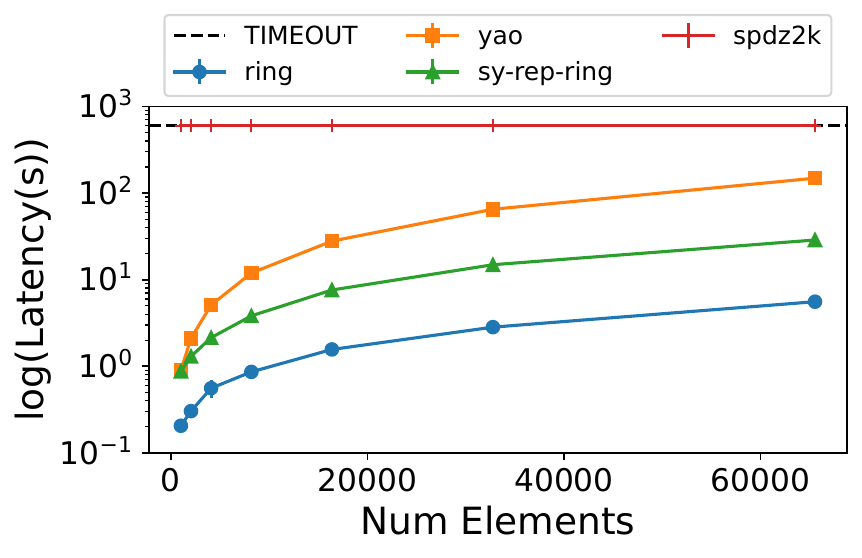}%
\caption{Sort, $N=2\rightarrow65536$}%
\end{subfigure}%
\begin{subfigure}{0.25\linewidth}%
\includegraphics[scale=\evalgraphscale]{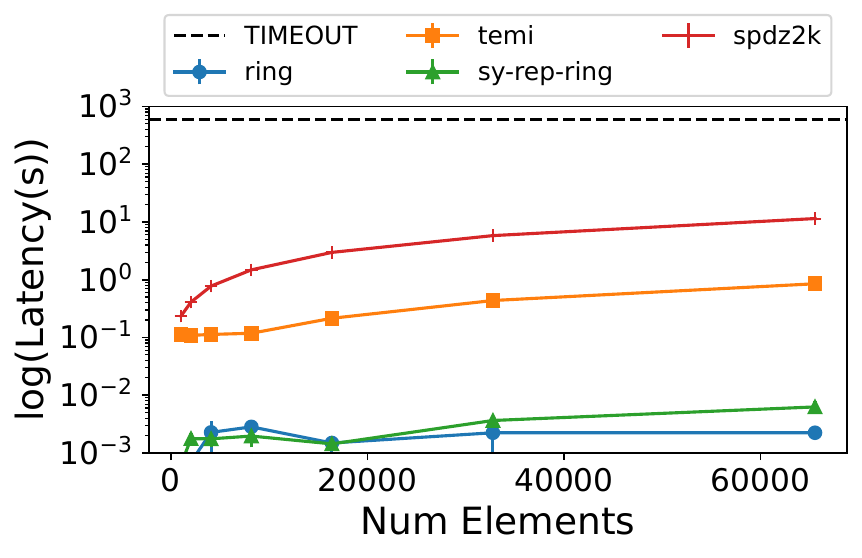}%
\caption{Inner Product, $N=2\rightarrow65536$}%
\end{subfigure}%
\begin{subfigure}{0.25\linewidth}%
\includegraphics[scale=\evalgraphscale]{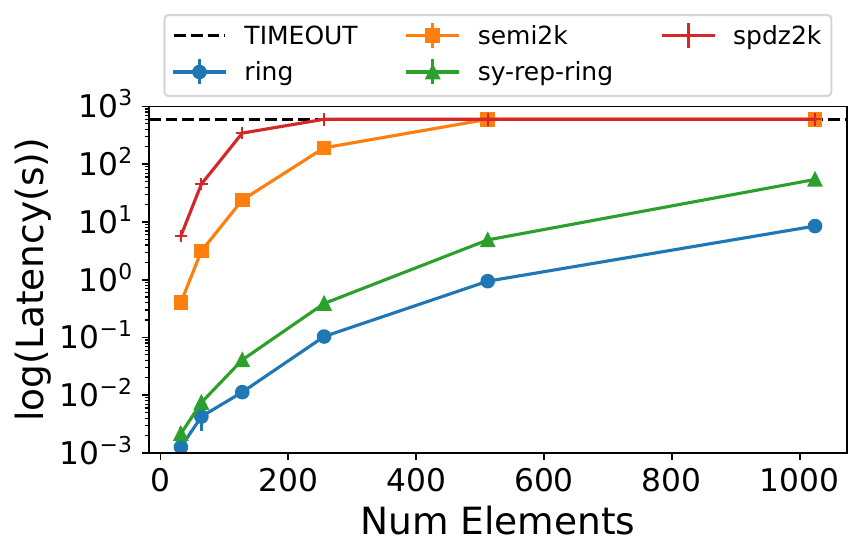}%
\caption{Matmul, $N=2\rightarrow1024$}%
\end{subfigure}%
\caption{Latency plots increasing input size $N$}%
\label{fig:varyingn}%
\end{figure*}

\autoref{fig:varyingn} shows how latency varies with changes in the
input size \(N\) for a subset of protocols. In general, latency
increases with the size of the input. The rate of increase depends on
\emph{both} the complexity of the underlying algorithm and the MPC
protocol. \emph{Comparison} and \emph{inner product} are linear $O(N)$
algorithms; our sorting algorithm is slightly non-linear
($O(N \log(N))$) and \emph{matmul} is $O(N^3)$. Accordingly, the lines
in the \emph{matmul} figure rise much faster than those in the other
figures.

Importantly, there are also protocol-specific differences. For
example, the lines for \textsc{ring} and \textsc{sy-rep-ring} are
parallel for \emph{sort}, indicating proportional scaling between the
two protocols on sorting, while the lines for the two protocols
diverge on \emph{matmul}, indicating that \textsc{sy-rep-ring} scales
less well on matrix multiplication than does \textsc{ring}.

\subsubsection{Performance under limited available bandwidth}

\begin{figure}
\centering%
\begin{subfigure}{\linewidth}%
\includegraphics[scale=\maingraphscale]{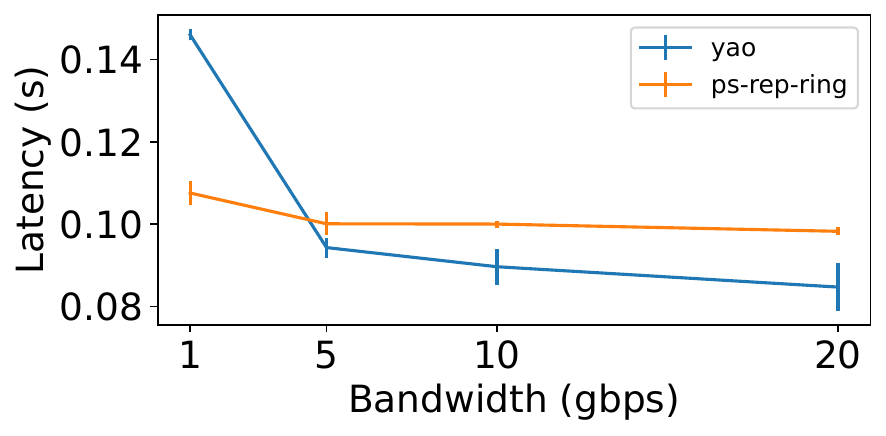}%
\caption{Comparison, $N = 250$, bit-width = 512 bit}%
\end{subfigure}

\begin{subfigure}{\linewidth}%
\includegraphics[scale=\maingraphscale]{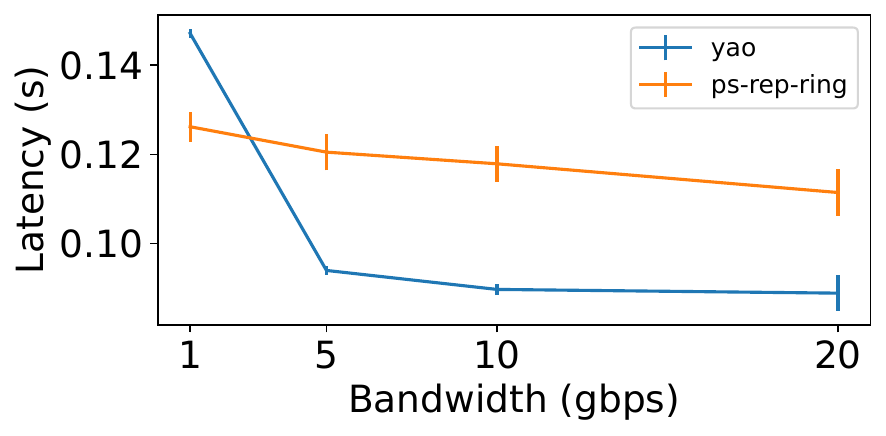}%
\caption{Comparison $N = 500$, bit-width = 256 bit}%
\end{subfigure}
\caption{Cross-over points between ps-rep-ring and yao based on bandwidth constraints}
\label{fig:constrained}
\end{figure}

We examined bandwidth-sensitivity of MPC by throttling link capacity
to 1/5/10/20\,Gbps and benchmarking \emph{comparison} on two
protocols: \textsc{yao} (sh\_h) and \textsc{ps-rep-ring}
(mal\_h). \autoref{fig:constrained} shows the results.

Note that \textsc{yao}'s latency drops steeply as bandwidth increases,
consistent with \textsc{yao} transmitting a lot of data. In contrast,
\textsc{ps-rep-ring} changes less sharply, reflecting its few online
rounds and comparatively compact messages.

Interestingly, when available bandwidth is low, \textsc{ps-rep-ring}
is faster \emph{despite} running in the malicious model with MAC
checks; this highlights that even a pick as efficient as \textsc{yao}
(on this primitive) can be suboptimal when link capacity is the
bottleneck. In summary, the best protocol for a task can depend on the
available network bandwidth, particularly when there is flexibility in
picking the threat model.

\subsubsection{Scaling with the number of parties (fixed $N{=}1024$)}

\begin{figure}
\centering%
\begin{subfigure}{\linewidth}%
\includegraphics[scale=\maingraphscale]{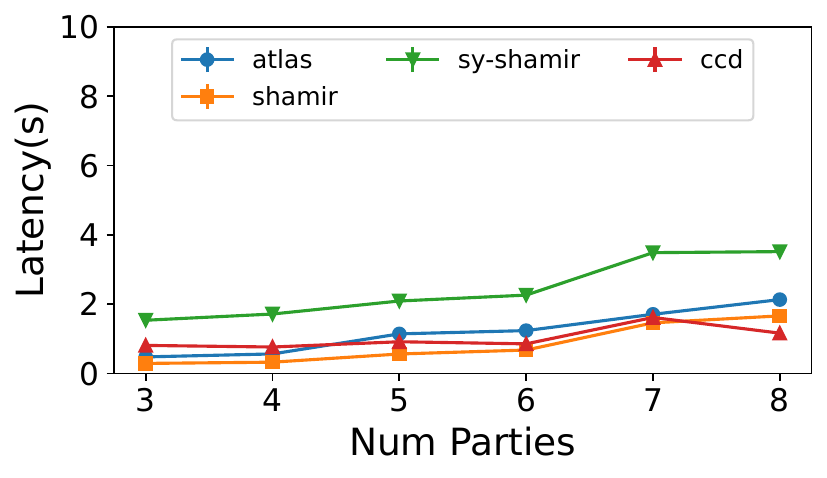}%
\caption{Comparison, N=1024}%
\label{fig:parties_scale_comparison}%
\end{subfigure}
\centering%
\begin{subfigure}{\linewidth}%
\includegraphics[scale=\maingraphscale]{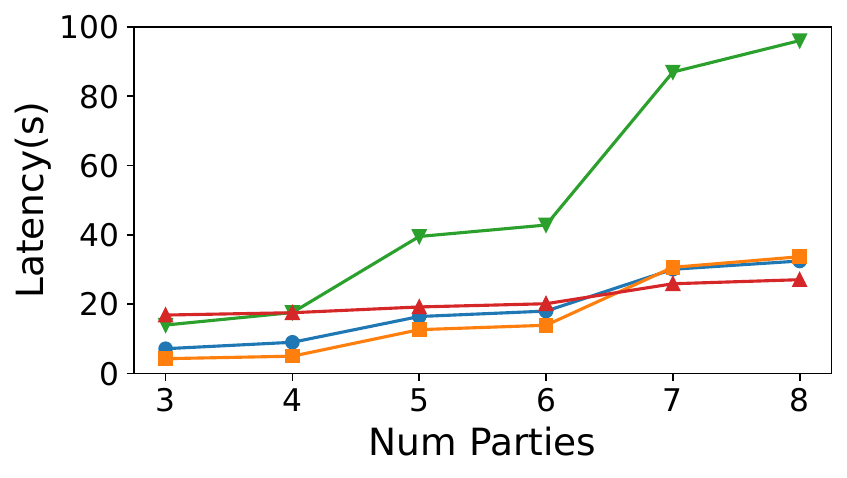}%
\caption{Sort, N=1024}%
\label{fig:parties_scale_sort}%
\end{subfigure}
\caption{Effect on latency for scaling the number of parties}
\label{fig:parties_scale}
\end{figure}

\autoref{fig:parties_scale} shows how latency of two primitives --
\emph{comparison} and \emph{sort} -- varies as a function of the
number of MPC parties on a subset of protocols that support a variable
number of parties. As expected, the latency increases with the number
of parties on both primitives and for all protocols, but the rate of
increase varies across protocols. For example, we notice that
\textsc{sy-shamir} scales poorly compared to \textsc{shamir}. This is
because \textsc{sy-shamir}'s SPDZ-like MAC protocol requires
additional pairwise communication between parties.

%
%
%
%
%
%
%

%

%

\if 0
Report results:
1. Protocols and max size they work with; \\
2. For each program, avg and std deviation for protocol; \\
3. Distinguish between computational costs and communication costs; \\
4. Nice table https://oaklandsok.github.io/papers/ladisa2023.pdf (Table I); \\
5. Report stacked costs (i.e., the avg and std dev above) distinguishing offline phase, online phase, the time spent for sacrificing, the time spent for encryption/decryption, etc. Basically more details on where the overheads are. \\

Now, focus on the overheads: where is the challenge? Where can we improve? Are there some new papers that are addressing the challenge and propose extensions or improvements in what seems to be the right direction?
\fi
     %
\section{Limitations and Future Work}
\label{sec:future}

Our analysis centers on MP-SPDZ~\cite{keller2020mp}, chosen for its broad protocol coverage (secret sharing, garbled circuits, HE hybrids) within a single open-source framework that enables \textit{apples-to-apples} comparisons: a shared benchmark and largely shared testing code reduce implementation noise from differing and codebases. Also, its wide adoption and active maintenance support reproducibility and reuse by the community. The trade-off is coverage: protocols and optimizations that have not yet been ported to MP-SPDZ fall beyond our evaluation. Extending coverage by implementing missing schemes in MP-SPDZ to enable further comparisons is left for future work.

Our study prioritizes core primitives because they are ubiquitous and foundational, providing a necessary baseline. As future work, we plan to analyze macro building blocks (e.g., filters/selections, joins, top-$k$, private set operations, softmax functions), which may exhibit different trade-offs.

Finally, an interesting direction for future work is a systematic study of the trade-offs between offline and online computation. While our evaluation considers each protocol as a whole, many applications place strict constraints on online latency but not on offline pre-processing. A dedicated analysis that disentangles these phases and explores such application-driven trade-offs would complement our study.  
\section{Related work}
\label{sec:related}

Hastings et al.~\cite{hastings2019sok} survey general-purpose MPC compilers, comparing languages, protocol back-ends, and developer UX, and provide runnable artifacts. Their work is orthogonal to ours: their focus is usability and architecture, rather than end-to-end latency/byte costs on concrete primitives across adversary models.

MP-SPDZ~\cite{keller2020mp} is a versatile implementation framework that unifies many protocol variants behind a common interface, spanning semi-honest/malicious security and honest/dishonest majority across binary, field, and ring computation.
The paper’s evaluation compares only a selected few protocols on a single microbenchmark, the inner product of two arrays with 100k 64-bit integers.
Thus, MP-SPDZ provides the most comprehensive framework-level overview to date, but it does not systematize protocol choice across different primitives. In contrast, this SoK compares protocol families across comparison, sort, inner product, and matrix multiply, normalizing to latency/bytes under controlled variations in input size, bit width, parties, and bandwidth, and providing selection rules for practitioners.

Contemporary to our work, Meisingseth and Rechberger~\cite{meisingseth2025sok} systematize differential privacy (DP) definitions for MPC, ordering the space by distribution model and computational perspective. They survey direct-implication and expressiveness relations, separations, and open problems. Their SoK clarifies when DP guarantees can be imposed in MPC and under what assumptions. This is orthogonal to our work, which provides end-to-end latency/communication measurements and selection guidance across primitives and threat models.

\if 0
Prior work partially addressed the second issue. Cerebro~\cite{zheng2021cerebro} is a tool that automatically selects a suitable MPC protocol for a given Machine Learning (ML) application. However, Cerebro is a tool designed for ML based applications, thus it may not meet the performance requirements for non-ML applications. Furthermore, Cerebro is not designed to compete against \textit{specialized} solutions for a given application; in order to minimize the performance cost, applications working in a malicious threat model and operating with large scale data are in need of such specialized solutions.
Thus, there still remains a lack of systematic understanding of the security performance tradeoffs of the various
MPC techniques and how these techniques compare with each other in different real-world application settings.
\fi     %
\section{Conclusions.}

This paper systematize the design space of general-purpose MPC and provides an empirical, cross-primitive comparison of protocol families under uniform, realistic conditions. By comparing latency and total data transmitted; varying input size, integer bit width, number of parties, and available bandwidth; and analyzing scaling behavior, we distilled practical selection rules rather than a single \emph{best} protocol -- showing how workload structure and network link capacity drive performance, why the relative performance of protocols is generally stable as inputs grow, when mixed computation helps, and how bandwidth caps impact protocol selection. 
Looking ahead, we see value in standardizing benchmark suites, extending measurements to hybrid designs, and integrating our rules into autotuners that choose protocols per primitive and deployment resources (compute, bandwidth).

	\bibliographystyle{plain}
    \bibliography{paper}

\begin{thebibliography}{10}

\bibitem{spdzwisepow2}
Mark Abspoel, Anders Dalskov, Daniel Escudero, and Ariel Nof.
\newblock An efficient passive-to-active compiler for honest-majority mpc over
  rings.
\newblock Cryptology ePrint Archive, Paper 2019/1298, 2019.
\newblock \url{https://eprint.iacr.org/2019/1298}.

\bibitem{acharya2023new}
Anasuya Acharya, Tomer Ashur, Efrat Cohen, Carmit Hazay, and Avishay Yanai.
\newblock A new approach to garbled circuits.
\newblock In {\em International Conference on Applied Cryptography and Network
  Security}, pages 611--641. Springer, 2023.

\bibitem{araki2018}
Toshinori Araki, Assi Barak, Jun Furukawa, Marcel Keller, Yehuda Lindell,
  Kazuma Ohara, and Hikaru Tsuchida.
\newblock Generalizing the spdz compiler for other protocols, 2018.
\newblock \url{https://eprint.iacr.org/2018/762}.

\bibitem{araki2017optimized}
Toshinori Araki, Assi Barak, Jun Furukawa, Tamar Lichter, Yehuda Lindell, Ariel
  Nof, Kazuma Ohara, Adi Watzman, and Or~Weinstein.
\newblock Optimized honest-majority mpc for malicious adversaries—breaking
  the 1 billion-gate per second barrier.
\newblock In {\em 2017 IEEE Symposium on Security and Privacy (SP)}, pages
  843--862. IEEE, 2017.

\bibitem{araki2016}
Toshinori Araki, Jun Furukawa, Yehuda Lindell, Ariel Nof, and Kazuma Ohara.
\newblock High-throughput semi-honest secure three-party computation with an
  honest majority.
\newblock Cryptology ePrint Archive, Paper 2016/768, 2016.
\newblock \url{https://eprint.iacr.org/2016/768}.

\bibitem{wondershaper}
S.~Séhier B.~Hubert, J.~Geul.
\newblock The wonder shaper 1.4.1.
\newblock \url{https://github.com/magnific0/wondershaper}.
\newblock Accessed: 08.08.2024.

\bibitem{Bater2017}
Johes Bater, Gregory Elliott, Craig Eggen, Satyender Goel, Abel Kho, and Jennie
  Rogers.
\newblock {SMCQL: Secure Querying for Federated Databases}.
\newblock {\em Proceedings of the VLDB Endowment}, 10(6):673--684, February
  2017.

\bibitem{baum2016efficient}
Carsten Baum, Emmanuela Orsini, and Peter Scholl.
\newblock Efficient secure multiparty computation with identifiable abort.
\newblock In {\em Theory of Cryptography: 14th International Conference, TCC
  2016-B, Beijing, China, October 31-November 3, 2016, Proceedings, Part I 14},
  pages 461--490. Springer, 2016.

\bibitem{beavertriple}
Donald Beaver.
\newblock Efficient multiparty protocols using circuit randomization.
\newblock In Joan Feigenbaum, editor, {\em Advances in Cryptology --- CRYPTO
  '91}, pages 420--432, Berlin, Heidelberg, 1992. Springer Berlin Heidelberg.

\bibitem{ben2019completeness}
Michael Ben-Or, Shafi Goldwasser, and Avi Wigderson.
\newblock Completeness theorems for non-cryptographic fault-tolerant
  distributed computation.
\newblock In {\em Providing sound foundations for cryptography: on the work of
  Shafi Goldwasser and Silvio Micali}, pages 351--371. 2019.

\bibitem{bogetoft2006practical}
Peter Bogetoft, Ivan Damg{\aa}rd, Thomas Jakobsen, Kurt Nielsen, Jakob Pagter,
  and Tomas Toft.
\newblock A practical implementation of secure auctions based on multiparty
  integer computation.
\newblock In {\em Financial Cryptography and Data Security: 10th International
  Conference, FC 2006 Anguilla, British West Indies, February 27-March 2, 2006
  Revised Selected Papers 10}, pages 142--147. Springer, 2006.

\bibitem{brandt2020constructing}
Nicholas Brandt, Sven Maier, Tobias M{\"u}ller, and J{\"o}rn M{\"u}ller-Quade.
\newblock Constructing secure multi-party computation with identifiable abort.
\newblock {\em Cryptology ePrint Archive}, 2020.

\bibitem{composable}
Ran Canetti, Yehuda Lindell, Rafail Ostrovsky, and Amit Sahai.
\newblock Universally composable two-party and multi-party secure computation.
\newblock Cryptology ePrint Archive, Paper 2002/140, 2002.
\newblock \url{https://eprint.iacr.org/2002/140}.

\bibitem{chaum1988multiparty}
David Chaum, Claude Cr{\'e}peau, and Ivan Damgard.
\newblock Multiparty unconditionally secure protocols.
\newblock In {\em Proceedings of the twentieth annual ACM symposium on Theory
  of computing}, pages 11--19, 1988.

\bibitem{ccd}
David Chaum, Claude Cr\'{e}peau, and Ivan Damgard.
\newblock Multiparty unconditionally secure protocols.
\newblock In {\em Proceedings of the Twentieth Annual ACM Symposium on Theory
  of Computing}, STOC '88, page 11–19, New York, NY, USA, 1988. Association
  for Computing Machinery.

\bibitem{chida2018}
Koji Chida, Daniel Genkin, Koki Hamada, Dai Ikarashi, Ryo Kikuchi, Yehuda
  Lindell, and Ariel Nof.
\newblock Fast large scale honest-majority mpc for malicious adversaries, 2018.
\newblock \url{https://eprint.iacr.org/2018/570}.

\bibitem{choi2012security}
Seung~Geol Choi, Jonathan Katz, Ranjit Kumaresan, and Hong-Sheng Zhou.
\newblock On the security of the “free-xor” technique.
\newblock In {\em Theory of Cryptography Conference}, pages 39--53. Springer,
  2012.

\bibitem{cramer1999efficient}
Ronald Cramer, Ivan Damg{\aa}rd, Stefan Dziembowski, Martin Hirt, and Tal
  Rabin.
\newblock Efficient multiparty computations secure against an adaptive
  adversary.
\newblock In {\em International conference on the Theory and Applications of
  Cryptographic Techniques}, pages 311--326. Springer, 1999.

\bibitem{spdz2k}
Ronald Cramer, Ivan Damgård, Daniel Escudero, Peter Scholl, and Chaoping Xing.
\newblock Spdz2k: Efficient mpc mod $2^k$ for dishonest majority.
\newblock Cryptology ePrint Archive, Paper 2018/482, 2018.
\newblock \url{https://eprint.iacr.org/2018/482}.

\bibitem{temi}
Ronald Cramer, Ivan Damgård, and Jesper~Buus Nielsen.
\newblock Multiparty computation from threshold homomorphic encryption.
\newblock Cryptology ePrint Archive, Paper 2000/055, 2000.
\newblock \url{https://eprint.iacr.org/2000/055}.

\bibitem{dalskov2012fantastic}
Anders Dalskov, Daniel Escudero, and Marcel Keller.
\newblock Fantastic four: Honest-majority four-party secure computation with
  malicious security.
\newblock Cryptology ePrint Archive, Paper 2020/1330, 2020.
\newblock \url{https://eprint.iacr.org/2020/1330}.

\bibitem{damgaard2012multiparty}
Ivan Damg{\aa}rd, Valerio Pastro, Nigel Smart, and Sarah Zakarias.
\newblock Multiparty computation from somewhat homomorphic encryption.
\newblock In {\em Annual Cryptology Conference}, pages 643--662. Springer,
  2012.

\bibitem{de2022covault}
Roberta De~Viti, Isaac Sheff, Noemi Glaeser, Baltasar Dinis, Rodrigo Rodrigues,
  Jonathan Katz, Bobby Bhattacharjee, Anwar Hithnawi, Deepak Garg, et~al.
\newblock Covault: A secure analytics platform.
\newblock {\em arXiv preprint arXiv:2208.03784}, 2022.

\bibitem{demmler2015aby}
Daniel Demmler, Thomas Schneider, and Michael Zohner.
\newblock Aby-a framework for efficient mixed-protocol secure two-party
  computation.
\newblock In {\em NDSS}, 2015.

\bibitem{malrep}
Hendrik Eerikson, Marcel Keller, Claudio Orlandi, Pille Pullonen, Joonas Puura,
  and Mark Simkin.
\newblock Use your brain! arithmetic 3pc for any modulus with active security.
\newblock Cryptology ePrint Archive, Paper 2019/164, 2019.
\newblock \url{https://eprint.iacr.org/2019/164}.

\bibitem{escudero2020improved}
Daniel Escudero, Satrajit Ghosh, Marcel Keller, Rahul Rachuri, and Peter
  Scholl.
\newblock Improved primitives for mpc over mixed arithmetic-binary circuits.
\newblock In {\em Advances in Cryptology--CRYPTO 2020: 40th Annual
  International Cryptology Conference, CRYPTO 2020, Santa Barbara, CA, USA,
  August 17--21, 2020, Proceedings, Part II 40}, pages 823--852. Springer,
  2020.

\bibitem{escudero2023superpack}
Daniel Escudero, Vipul Goyal, Antigoni Polychroniadou, Yifan Song, and Chenkai
  Weng.
\newblock Superpack: Dishonest majority mpc with constant online communication.
\newblock In {\em Annual International Conference on the Theory and
  Applications of Cryptographic Techniques}, pages 220--250. Springer, 2023.

\bibitem{even1985randomized}
Shimon Even, Oded Goldreich, and Abraham Lempel.
\newblock A randomized protocol for signing contracts.
\newblock {\em Communications of the ACM}, 28(6):637--647, 1985.

\bibitem{tinier}
Tore~Kasper Frederiksen, Marcel Keller, Emmanuela Orsini, and Peter Scholl.
\newblock A unified approach to mpc with preprocessing using ot.
\newblock Cryptology ePrint Archive, Paper 2015/901, 2015.
\newblock \url{https://eprint.iacr.org/2015/901}.

\bibitem{tiny}
Jun Furukawa, Yehuda Lindell, Ariel Nof, and Or~Weinstein.
\newblock High-throughput secure three-party computation for malicious
  adversaries and an honest majority.
\newblock Cryptology ePrint Archive, Paper 2016/944, 2016.
\newblock \url{https://eprint.iacr.org/2016/944}.

\bibitem{genkin2015efficient}
Daniel Genkin, Yuval Ishai, and Antigoni Polychroniadou.
\newblock Efficient multi-party computation: from passive to active security
  via secure simd circuits.
\newblock In {\em Annual Cryptology Conference}, pages 721--741. Springer,
  2015.

\bibitem{amdcircuits}
Daniel Genkin, Yuval Ishai, Manoj~M. Prabhakaran, Amit Sahai, and Eran Tromer.
\newblock Circuits resilient to additive attacks with applications to secure
  computation.
\newblock Cryptology ePrint Archive, Paper 2015/154, 2015.
\newblock \url{https://eprint.iacr.org/2015/154}.

\bibitem{GMW1987mentalgame}
Oded Goldreich, Silvio Micali, and Avi Wigderson.
\newblock How to play any mental game or a completeness theorem for protocols
  with honest majority.
\newblock In {\em STOC '87: Proceedings of the nineteenth annual ACM symposium
  on Theory of computing}, pages 218--229, 1987.

\bibitem{atlas}
Vipul Goyal, Hanjun Li, Rafail Ostrovsky, Antigoni Polychroniadou, and Yifan
  Song.
\newblock Atlas: Efficient and scalable mpc in the honest majority setting.
\newblock Cryptology ePrint Archive, Paper 2021/833, 2021.
\newblock \url{https://eprint.iacr.org/2021/833}.

\bibitem{hamada2014oblradixsort}
Koki Hamada, Dai Ikarashi, Koji Chida, , and Katsumi Takahashi.
\newblock Oblivious radix sort: An efficient sorting algorithmfor practical
  secure multi-party computation.
\newblock Cryptology ePrint Archive, Paper 2014/121, 2014.
\newblock \url{https://eprint.iacr.org/2014/121}.

\bibitem{hastings2019sok}
Marcella Hastings, Brett Hemenway, Daniel Noble, and Steve Zdancewic.
\newblock Sok: General purpose compilers for secure multi-party computation.
\newblock In {\em 2019 {IEEE} Symposium on Security and Privacy, {SP} 2019, San
  Francisco, CA, USA, May 19-23, 2019}, pages 1220--1237. {IEEE}, 2019.

\bibitem{hazay2018concretely}
Carmit Hazay, Emmanuela Orsini, Peter Scholl, and Eduardo Soria-Vazquez.
\newblock Concretely efficient large-scale mpc with active security (or,
  tinykeys for tinyot).
\newblock In {\em International Conference on the Theory and Application of
  Cryptology and Information Security}, pages 86--117. Springer, 2018.

\bibitem{huang2011faster}
Yan Huang, David Evans, Jonathan Katz, and Lior Malka.
\newblock Faster secure $\{$Two-Party$\}$ computation using garbled circuits.
\newblock In {\em 20th USENIX Security Symposium (USENIX Security 11)}, 2011.

\bibitem{ishai2014secure}
Yuval Ishai, Rafail Ostrovsky, and Vassilis Zikas.
\newblock Secure multi-party computation with identifiable abort.
\newblock In {\em Advances in Cryptology--CRYPTO 2014: 34th Annual Cryptology
  Conference, Santa Barbara, CA, USA, August 17-21, 2014, Proceedings, Part II
  34}, pages 369--386. Springer, 2014.

\bibitem{ito1989secret}
Mitsuru Ito, Akira Saito, and Takao Nishizeki.
\newblock Secret sharing scheme realizing general access structure.
\newblock {\em Electronics and Communications in Japan (Part III: Fundamental
  Electronic Science)}, 72(9):56--64, 1989.

\bibitem{keller2020mp}
Marcel Keller.
\newblock Mp-spdz: A versatile framework for multi-party computation.
\newblock In {\em Proceedings of the 2020 ACM SIGSAC conference on computer and
  communications security}, pages 1575--1590, 2020.

\bibitem{mascot}
Marcel Keller, Emmanuela Orsini, and Peter Scholl.
\newblock Mascot: Faster malicious arithmetic secure computation with oblivious
  transfer.
\newblock Cryptology ePrint Archive, Paper 2016/505, 2016.
\newblock \url{https://eprint.iacr.org/2016/505}.

\bibitem{keller2016mascot}
Marcel Keller, Emmanuela Orsini, and Peter Scholl.
\newblock Mascot: faster malicious arithmetic secure computation with oblivious
  transfer.
\newblock In {\em Proceedings of the 2016 ACM SIGSAC Conference on Computer and
  Communications Security}, pages 830--842, 2016.

\bibitem{lowhighgear}
Marcel Keller, Valerio Pastro, and Dragos Rotaru.
\newblock Overdrive: Making spdz great again.
\newblock Cryptology ePrint Archive, Paper 2017/1230, 2017.
\newblock \url{https://eprint.iacr.org/2017/1230}.

\bibitem{keller2018overdrive}
Marcel Keller, Valerio Pastro, and Dragos Rotaru.
\newblock Overdrive: Making spdz great again.
\newblock In {\em Annual International Conference on the Theory and
  Applications of Cryptographic Techniques}, pages 158--189. Springer, 2018.

\bibitem{kolesnikov2014flexor}
Vladimir Kolesnikov, Payman Mohassel, and Mike Rosulek.
\newblock Flexor: Flexible garbling for xor gates that beats free-xor.
\newblock In {\em Advances in Cryptology--CRYPTO 2014: 34th Annual Cryptology
  Conference, Santa Barbara, CA, USA, August 17-21, 2014, Proceedings, Part II
  34}, pages 440--457. Springer, 2014.

\bibitem{lindell2017framework}
Yehuda Lindell and Ariel Nof.
\newblock A framework for constructing fast mpc over arithmetic circuits with
  malicious adversaries and an honest-majority.
\newblock In {\em Proceedings of the 2017 ACM SIGSAC Conference on Computer and
  Communications Security}, pages 259--276, 2017.

\bibitem{honeybadger}
Donghang Lu, Thomas Yurek, Samarth Kulshreshtha, Rahul Govind, Rahul Mahadev,
  Aniket Kate, and Andrew Miller.
\newblock Honeybadgermpc and asynchromix: Practical asynchronousmpc and its
  application to anonymous communication.
\newblock Cryptology ePrint Archive, Paper 2019/883, 2019.
\newblock \url{https://eprint.iacr.org/2019/883}.

\bibitem{margolin2023arboretum}
Elizabeth Margolin, Karan Newatia, Tao Luo, Edo Roth, and Andreas Haeberlen.
\newblock Arboretum: A planner for large-scale federated analytics with
  differential privacy.
\newblock In {\em Proceedings of the 29th Symposium on Operating Systems
  Principles}, pages 451--465, 2023.

\bibitem{meisingseth2025sok}
Fredrik Meisingseth and Christian Rechberger.
\newblock Sok: Computational and distributed differential privacy for {MPC}.
\newblock {\em Proc. Priv. Enhancing Technol.}, 2025(1):420--439, 2025.

\bibitem{micali1987play}
Silvio Micali, Oded Goldreich, and Avi Wigderson.
\newblock How to play any mental game.
\newblock In {\em Proceedings of the Nineteenth ACM Symp. on Theory of
  Computing, STOC}, pages 218--229. ACM New York, NY, USA, 1987.

\bibitem{mohassel2018aby3}
Payman Mohassel and Peter Rindal.
\newblock Aby3: A mixed protocol framework for machine learning.
\newblock In {\em Proceedings of the 2018 ACM SIGSAC conference on computer and
  communications security}, pages 35--52, 2018.

\bibitem{mohassel2017secureml}
Payman Mohassel and Yupeng Zhang.
\newblock Secureml: A system for scalable privacy-preserving machine learning.
\newblock In {\em 2017 IEEE symposium on security and privacy (SP)}, pages
  19--38. IEEE, 2017.

\bibitem{damgard2011spdz}
Valerio Pastro, Ivan Damgård, Nigel Smart, and Sarah Zakarias.
\newblock Multiparty computation from somewhat homomorphic encryption.
\newblock Cryptology ePrint Archive, Paper 2011/535, 2011.
\newblock \url{https://eprint.iacr.org/2011/535}.

\bibitem{poddar2021senate}
Rishabh Poddar, Sukrit Kalra, Avishay Yanai, Ryan Deng, Raluca~Ada Popa, and
  Joseph~M Hellerstein.
\newblock Senate: a $\{$Maliciously-Secure$\}$$\{$MPC$\}$ platform for
  collaborative analytics.
\newblock In {\em 30th USENIX Security Symposium (USENIX Security 21)}, pages
  2129--2146, 2021.

\bibitem{dabits}
Dragos Rotaru and Tim Wood.
\newblock Marbled circuits: Mixing arithmetic and boolean circuits with active
  security.
\newblock Cryptology ePrint Archive, Paper 2019/207, 2019.
\newblock \url{https://eprint.iacr.org/2019/207}.

\bibitem{roth2021mycelium}
Edo Roth, Karan Newatia, Yiping Ma, Ke~Zhong, Sebastian Angel, and Andreas
  Haeberlen.
\newblock Mycelium: Large-scale distributed graph queries with differential
  privacy.
\newblock In {\em Proceedings of the ACM SIGOPS 28th Symposium on Operating
  Systems Principles}, pages 327--343, 2021.

\bibitem{shamir1979share}
Adi Shamir.
\newblock How to share a secret.
\newblock {\em Communications of the ACM}, 22(11):612--613, 1979.

\bibitem{songhori2015tinygarble}
Ebrahim~M Songhori, Siam~U Hussain, Ahmad-Reza Sadeghi, Thomas Schneider, and
  Farinaz Koushanfar.
\newblock Tinygarble: Highly compressed and scalable sequential garbled
  circuits.
\newblock In {\em 2015 IEEE Symposium on Security and Privacy}, pages 411--428.
  IEEE, 2015.

\bibitem{volgushev2019conclave}
Nikolaj Volgushev, Malte Schwarzkopf, Ben Getchell, Mayank Varia, Andrei
  Lapets, and Azer Bestavros.
\newblock {Conclave: Secure Multi-Party Computation on Big Data}.
\newblock In {\em Proceedings of the Fourteenth EuroSys Conference 2019},
  EuroSys '19, New York, NY, USA, 2019. Association for Computing Machinery.

\bibitem{emp-toolkit}
X.~Wang, A.~J. Malozemoff, and J.~Katz.
\newblock {EMP-toolkit: Efficient MultiParty computation toolkit}.
\newblock \url{https://github.com/emp-toolkit}, 2016.
\newblock Accessed: 2023-12-07.

\bibitem{wang2017globalscale}
X.~Wang, S.~Ranellucci, and J.~Katz.
\newblock {Global-Scale Secure Multiparty Computation}.
\newblock In B.~Thuraisingham, D.~Evans, T.~Malkin, and D.~Xu, editors, {\em
  Proceedings of the 2017 {ACM} {SIGSAC} Conference on Computer and
  Communications Security, {CCS} 2017}, pages 39--56. {ACM}, 2017.

\bibitem{watson2022piranha}
Jean-Luc Watson, Sameer Wagh, and Raluca~Ada Popa.
\newblock Piranha: A $\{$GPU$\}$ platform for secure computation.
\newblock In {\em 31st USENIX Security Symposium (USENIX Security 22)}, pages
  827--844, 2022.

\bibitem{Yao1982}
A.~C. Yao.
\newblock {Protocols for secure computations}.
\newblock In {\em 23rd Annual Symposium on Foundations of Computer Science
  (sfcs 1982)}, pages 160--164, 1982.

\bibitem{zahur2015two}
Samee Zahur, Mike Rosulek, and David Evans.
\newblock Two halves make a whole: Reducing data transfer in garbled circuits
  using half gates.
\newblock In {\em Advances in Cryptology-EUROCRYPT 2015: 34th Annual
  International Conference on the Theory and Applications of Cryptographic
  Techniques, Sofia, Bulgaria, April 26-30, 2015, Proceedings, Part II 34},
  pages 220--250. Springer, 2015.

\bibitem{zheng2021cerebro}
Wenting Zheng, Ryan Deng, Weikeng Chen, Raluca~Ada Popa, Aurojit Panda, and Ion
  Stoica.
\newblock Cerebro: A platform for multi-party cryptographic collaborative
  learning.
\newblock In {\em USENIX Security Symposium}, 2021.

\bibitem{zhu2016cut}
Ruiyu Zhu, Yan Huang, Jonathan Katz, and Abhi Shelat.
\newblock The $\{$Cut-and-Choose$\}$ game and its application to cryptographic
  protocols.
\newblock In {\em 25th USENIX Security Symposium (USENIX Security 16)}, pages
  1085--1100, 2016.

\end{thebibliography}

    \appendix
\section{Mathematical Basics}

\subsection{Groups, Rings, \& Fields}
\label{app:mathbasis}

\subsubsection{Groups}

A \textbf{group} is a non-empty set $G$ with a binary operation ($*$), such that it satisfies the following four properties:
\begin{itemize}
	\item \textit{Closure:} if $a \in G$ and $b \in G$ then $a*b \in G$.
    \item \textit{Associativity:} $\forall a, b, c \in G$, $a * (b * c) = (a * b) * c$.
    \item \textit{Identity:} $\forall a \in G$, $\exists e \in G$ such that $e * a = a * e = a$. The element $e$ is called identity element or neutral element.
    \item \textit{Inverse:} $\forall a \in G$, $\exists b \in G$, called the \textit{inverse} of $a$, such that $a * b = b * a = e$ (where $e$ is the identity element).
\end{itemize}

In addition, a group is said to be \textit{abelian} if it also satisfies:
\begin{itemize}
	\item \textit{Commutativity:} $\forall a, b, c \in G$, $a * b = b * a$.
\end{itemize}

A group is denoted as of finite order (or simply \textit{finite}) if it has a finite number of elements. In this case, the number of elements in $G$ is called the \textit{order} of $G$ and is denoted by $|G|$. A group with infinitely many elements is denoted as of infinite order (or infinite). 
\\

\subsubsection{Rings \& Fields}

With the definition of groups in mind, we can now formally describe rings and fields. Both rings and fields are sets equipped with two binary operations: \textbf{addition} $(+)$ and \textbf{multiplication} $(\cdot)$. A ring is a group under $(+)$ but not under $(\cdot)$, as it only partially satisfies the above properties.  \\

Formally, a \textbf{ring} is a set $R$ which is closed under $(+)$ and $(\cdot)$. Furthermore, $R$ is an abelian group under $(+)$ and satisfies the following properties under $(\cdot)$:
\begin{itemize}
	\item \textit{Associativity:} $\forall a, b, c \in R$, $a \cdot (b \cdot c) = (a \cdot b) \cdot c$.
	\item \textit{Identity:} $\forall a \in R$, $\exists e \in R$ such that $e \cdot a = a \cdot e = a$.
	\item \textit{Distributive:} $R$ satisfies the distributive properties of multiplication over addition. Formally, $\forall a, b, c \in R$, $a \cdot (b + c) = (a \cdot b) + (a \cdot c)$ and $(b + c) \cdot a = (b \cdot a) + (c \cdot a)$.
\end{itemize}

Note that (i) we do \textit{not} require multiplicative inverses and (ii) multiplication need not be commutative. When multiplication is commutative, we refer to \textit{commutative rings}. The simplest type of commutative ring is a field.
\\

Formally, a \textbf{field} is a set $F$ which is closed under $(+)$ and $(\cdot)$ . $F$ is a commutative ring, thus all properties above hold plus commutativity for multiplication. In particular, $F$ is an abelian group under $(+)$ with $0$ as the additive identity. Furthermore, $F - {0}$ is an abelian group under $(\cdot)$ with $1$ as the multiplicative identity; we subtract the additive identity $0$ because only non-zero elements are invertible under multiplication.
\\

A finite field (i.e., a field with finite number of elements) is also called a \textbf{Galois Field (GF)}. As mentioned above in the case of finite groups, the number of elements in $F$ is called the order of $F$ and it is denoted by $|F|$. In finite fields, $|F|$ is either a prime number or a prime power. Thus, every finite field has \textit{prime power order}. In other terms, for every prime power there is a finite field of that order. Formally, given a prime $p$ and an integer $k$, the prime power $q = p^k$ is the order of a (unique) field denoted as $F_q$ or $GF(q)$.
\\

The most common examples of fields are rational numbers $\mathbb{Q}$, real numbers $\mathbb{R}$, and complex numbers $\mathbb{C}$, while the set of integers $\mathbb{Z}$ is only a commutative ring (dividing two integers does not always result in an integer). The most common example of GF is the set of integers modulo a prime.

\subsection{Beaver's Triples} \label{sec:beaver_multiplication}

In the context of field or ring based MPC, i.e. when values are represented as elements belonging to either $\mathbb{Z}_{2^k}$ or $\mathbb{Z}_p$, multiplication can be efficiently computed using Beaver or multiplication triples ~\cite{beavertriple}. A Beaver triple is a triple of secret shared values $\sharing{a}$, $\sharing{b}$ and $\sharing{c}$, under the constraint that $\sharing{a \cdot b} = \sharing{c}$. The main challenge that comes with Beaver triples is how to generate and validate them efficiently: a malicious adversary might try to introduce errors in the triples, i.e. generate a triple in which $\sharing{a \cdot b} \ne \sharing{c}$, so that when this incorrect triple is used to efficiently compute a multiplication, the result would be incorrect. In the next paragraph, we will discuss the generation of Beaver triples, then we will explain how are they used during the execution of a MPC protocol. \newline

\subsubsection{Generation.}
\label{subsubsec:beaver_triples_gen}

Beaver triples generation relies either on \emph{Oblivious Transfer}(OT) ~\cite{keller2016mascot} or on \emph{Homomorphic Encryption}(HE) ~\cite{keller2018overdrive}. \newline

\paragraph{OT-based method ~\cite{keller2016mascot}.} To generate a Beaver triple $\left\{ \sharing{a}, \sharing{b}, \sharing{a \cdot b} \right\}$, every party samples $a \in \mathbb{Z}_p$ and bit-decomposes it, we denote the bit decomposition of $a^i$ as $\left(a^i_{\bitnumber{0}}, a^i_{\bitnumber{1}}, a^i_{\bitnumber{2}}, ..., a^i_{\bitnumber{\tau}}\right)$ and $b^i$. Then, every ordered pair of parties engage in a OT protocol, where party $i$ receives $q^{i,j}_{\bitnumber{0}}, q^{i,j}_{\bitnumber{1}}$ and party $j$ receives $s^{i,j}_h = q^{i,j}_{a_{\bitnumber{h}}}$, $\forall h \in \left[0, \tau\right]$. Now, party $i$ sends $d_{\bitnumber{h}}^{i,j} = q^{i,j}_{\bitnumber{0}} - q^{i,j}_{\bitnumber{1}} + b^i$, $\forall h \in \left[0, \tau\right]$. Party $j$ sets $t_{\bitnumber{h}}^{i,j} = s_{\bitnumber{h}}^{i,j} + a_{\bitnumber{h}}^j d_{\bitnumber{h}}^{i,j} = q_{0, \bitnumber{h}}^{i,j} + a_{\bitnumber{h}}^j b^i$. Note that, for every $h$ it holds
\begin{equation*}
	t_{\bitnumber{h}}^{i,j} = q_{0, \bitnumber{h}}^{i,j} + a_{\bitnumber{h}}^j b^i.
\end{equation*}
Finally, party $i$ sets $c^{i,j}_i = - \left[q_{0, \bitnumber{0}}^{i,j},q_{0, \bitnumber{1}}^{i,j}, ..., q_{0, \bitnumber{\tau}}^{i,j}\right]$ and party $j$ sets $c^{i,j}_j = \left[t_{\bitnumber{0}}^{i,j}, t_{\bitnumber{1}}^{i,j}, \dots, t_{\bitnumber{\tau}}^{i,j}\right]$. Since $c^{i,j}_i$ and $c^{i,j}_j$ are additive shares of $a^j \cdot b^i$, now the parties can receive their shares of $\sharing{a \cdot b} = \sharing{c}$ as
\begin{equation*}
	\sharing{c}_i = a^i \cdot b^i + \sum_{j \ne i} c_{i}^{i,j} + c^{j,i}_i.
\end{equation*}

\paragraph{HE-based method ~\cite{keller2018overdrive}.} Let $\text{Enc}_{PK}$ denote the encryption operation of a semi-homomorphic encryption scheme, with public key $PK$. Then, to generate a Beaver triple $\left\{ \sharing{a}, \sharing{b}, \sharing{a \cdot b} \right\}$, party $i$ unformly samples $a$ at random and send to party $j$ $\text{Enc}_{PK}\left(a\right)$. Upon receiving $\text{Enc}_{PK}\left(a\right)$, party $j$ replies with $K = b \cdot \text{Enc}_{PK}\left(a\right) + \text{Enc}_{PK}\left(c_j\right)$. Since, semi-homomorphic encryption schemes allow one multiplication of an encryption with a cleartext value, the decryption of $K$ would yield $c_i = b \cdot a - c_j$, which makes $\left(c_i, c_j\right)$ a valid additive secret sharing of $a \cdot b$.

\paragraph{Validation.} 
We discuss in details Beaver triples' validation methods for fields \autoref{subsubsec:lindell_and_nof}, for fields and rings and during both pre and post processing in \autoref{subsubsec:Eerikson_et_al} and with an arbitrary level of security in \autoref{subsubsec:araki_bucket_candc}.

\paragraph{Multiplication \cite{beavertriple}.}
In the \textit{online phase}, to multiply two secret shared values $\sharing{x}$ and $\sharing{y}$, we, first, open the blinded sums $\varepsilon = a + x$ and  $\delta = b + y$, i.e. the parties broadcast their shares of $\sharing{a + x}$ and $\sharing{b + y}$. Then, each party locally computes its own share of $z$ as
\begin{equation*}
	\sharing{z} = \varepsilon \cdot \sharing{y} - \delta \cdot \sharing{a} + \sharing{c}.
\end{equation*}
Note that the correctness of $\sharing{z}$ follows from the Beaver triples construction
\begin{align*}
	& z = \sum_{i = 1}^{n} \sharing{z} = \\
	&= \varepsilon \cdot \sum_{i = 1}^{n} \sharing{y}_i - \delta \sum_{i = 1}^{n} \sharing{a} + \sum_{i = 1}^{n} \sharing{c} = \\
	&= ay + xy - ab - ay + ab = xy,
\end{align*}
since $c = a \cdot b$. It is important to note that the shares of $z$, $\sharing{z}_i$ are locally computed by each party as its correctness relies on the homomorphic properties of the underlying secret sharing scheme, which may be either additive or Shamir. Using Beaver triples allows to offload most of the computation to a preprocessing phase, whose goal is to pre-compute intermediate results that can speed up the multiplication of secret values in the subsequent online phase. 

\subsection{Details: Crossing domains}
\label{subsec:mixing_domains_appendix}

Arithmetic operations such as additions and multiplications are more efficiently computed in the arithmetic domain while comparisons, which are a fundamental part of most algoritms, are more efficiently computed in the Boolean domain. The mixed circuit approach aims to achieve the best of both worlds by dynamically switching sub-protocols during computation. Clearly, for mixed circuits to be of practical use, the cost of switching back and forth to perform a given computation in another domain must be less than the cost of performing the same computation in the original domain without any switching. In this subsection, we briefly introduce the techniques used to achieve that.

\subsubsection{Local share conversion}
\label{subsubsec:local_share_conv_app}

\emph{Local share conversion} is a technique that allows parties to locally convert the arithmetic shares of a secret value $x$ to boolean shares of the same number, and viceversa. In the next two paragraphs we present the method proposed by Araki et al in \cite{araki2018}. Note that this technique can only be used with replicated additive secret sharing and in the absence of any MACs. \newline

\paragraph{From arithmetic to binary \cite{araki2018}. } Let the $\sharing{x}$ be with values $\left(x_1, x_2\right)$, $\left(x_2, x_3\right)$ and $\left(x_1, x_3\right)$ belonging to parties $1$, $2$ and $3$, respectively, where $x = x_1 + x_2 + x_3$. Now, each party locally converts each bit of their respective share into a new sharing $\left(x_{\bitnumber{j}}^1, 0\right)$, $\left(0, x_{\bitnumber{j}}^1 \right)$ and $\left(0, 0\right)$, which constitutes a valid $2$-out-of-$3$ additive sharing of $x_{\bitnumber{j}}^1$. Now, the parties hold a bitwise sharing of $x_1$, $x_2$ and $x_3$ but not a bitwise sharing of $x$. To obtain a bitwise sharing of $x$, we need to add those shares and since we have to consider the bit carries, local addition of shares is not sufficient. Now, the sum of three binary values can be computed as $\alpha_{\bitnumber{h}} = x_{\bitnumber{h}}^1 \oplus x_{\bitnumber{h}}^2 \oplus x_{\bitnumber{h}}^3$ but for the carry we have to consider that now we have to compute the sum of $4$ binary values and therefore there might be $2$-bits carries. In order to compute those, an helper function is defined $M \left(a, b, c\right) = a \cdot b \oplus a \cdot c \oplus b \cdot c$, this function computes whether the $1$'s or the $0$'s are the majority in $a$, $b$ and $c$ and returns a value accordingly. Finally, the sum of bitwise shares operation is reduced to
\begin{equation*}
	\sharing{x^1}_{\bitnumber{h}} = \alpha_{\bitnumber{h}}^1 \cdot c_{\bitnumber{h - 1}}^1 \cdot cc_{\bitnumber{h - 2}}^1 ,
\end{equation*}
where
\begin{align*}
	& \alpha_{\bitnumber{h}}^1 = x_{\bitnumber{h - 1}}^1 \oplus x_{\bitnumber{h - 1}}^2 \oplus x_{\bitnumber{h - 1}}^3 \\
	& \beta_{\bitnumber{h}}^1 = M\left( x_{\bitnumber{h - 1}}^1,x_{\bitnumber{h - 1}}^1, x_{\bitnumber{h - 1}}^1 \right) \\
	& \gamma_{\bitnumber{h}}^1 = M \left( \alpha_{\bitnumber{h}}^1, c_{\bitnumber{h- 1}}^1, cc_{\bitnumber{h- 2}}^1  \right) \\
	& c_{\bitnumber{h- 1}}^1 = \beta_{\bitnumber{h - 1}}^1 \oplus \gamma_{\bitnumber{h - 1}}^1\\
	& cc_{\bitnumber{h- 2}}^1 = \beta_{\bitnumber{h - 2}}^1 \oplus \gamma_{\bitnumber{h - 2}}^1.\\
\end{align*}

\paragraph{From binary to arithmetic \cite{araki2018}.} In order to convert binary shares of $x$ into arithmetic ones, the procedure is similar to the one described in the previous paragraph with the difference that now we need to cancel out the carries, since the sharing in the boolean domain is XOR based and, therefore, bitwise. Therefore, the arithmethic shares of the $h$-th bit of $x$ has to be computed as $x_{\bitnumber{h}} = \text{bit}\left(x_{\bitnumber{h}}\right) + 2 \cdot \text{carry} \left(x_{\bitnumber{h}}\right) - \text{carry}\left(x_{\bitnumber{h - 1}}\right)$. 

\subsubsection{daBits and edaBits}
\label{subsubsec:dabits_edabits_app}

When the threat model does not allow for local share conversion to be applied to secret shared values, then to evaluate hybrid circuits we need a different methods, namely \emph{daBits} and \emph{edaBits}.

\paragraph{daBits}~\cite{dabits}
is a method to facilitate conversion between arithmetic and binary secret sharing, assuming a malicious adversary and a dishonest majority. Let $x$ be a secret value, we will refer with $\sharing{x}$ and $\sharing{x}_2$ as the representations of $x$ secret-shared in the arithmetic and binary domain, respectively.\footnote{daBits is agnostic to any underlying MPC protocol, so the computation may take place over fields $\mathbb{Z}_{p}$  or rings $\mathbb{Z}_{2^k}$; in fact, this technique can also be used to switch between fields and rings and vice versa.} 
A daBit (\textit{double-shared authenticated bit}) is defined as:
\begin{center} daBit := $\left\{ \sharing{r}, \sharing{r}_2 \right\}$ \end{center}

\noindent where $r$ is a secret shared uniformly sampled random bit.\footnote{The parties are given one share each. The shares look random, and their composition results in the original random bit $r$ --- this is why the parties are said to be given \textit{correlated randomness}.}
Thus, each party is given two shares of $r$, one in the arithmetic domain, one in the binary domain. The value of the bit $r$ must be the same in both domains for the daBit to be \textit{correct}. We refer~\cite{dabits} for details on daBits generation and checking procedure. 
Next, we present a concrete instance of domain conversion between arithmetic secret sharing over $\mathbb{Z}_p$ and GCs over $\mathbb{Z}_2$ using daBits.
Suppose the parties hold $\sharing{x}_p$ and want to convert it to binary representation $\sharing{x}_2$. First, the parties generate a daBit $(\sharing{r_i}_p, \sharing{r_i}_2)$  for each bit of $\sharing{x}_p$.
Then, the parties evaluate $\sharing{x - r}_p$ in $\mathbb{Z}_p$, where $r$ is 
\begin{equation*}
	\sum_{i=0}^{\log p - 1} 2^i \cdot \sharing{r_i}_p.
\end{equation*}
The parties open $x-r$ and compute $\left( x - r \right) + r$ mod $p$ in the boolean domain, which results in $\sharing{x}_2$, this computation is facilitated by feeding the garbled circuit with the shares $\sharing{r_i}_2$. daBits can be used to convert an arithmetic representation of a secret value $\sharing{x}$ into a binary in a similar way, by replacing the arithmetic sum in the last step with the bitwise XOR operation.

While daBits facilitates conversion between domains, this technique is expensive. In particular, generating a random bit in $\mathbb{Z}_p$ with malicious security involves multiplication (or squaring) over $\mathbb{Z}_p$, which is an expensive operation. For example, when working in with a large modulus $p$, the cost of generating the daBits required to switch to $\mathbb{Z}_2$ to evaluate a comparison is close to the cost of evaluating that comparison directly in $\mathbb{F}_p$~\cite{escudero2020improved}. The high cost of daBits limits its adoption in practice.

\paragraph{edaBits (extended daBits)}~\cite{escudero2020improved}
is a similar technique that aims to overcome the limitations of daBits. In particular, edaBits are tuples consisting of random values in ${\{0,1}\}$ shared and authenticated in $\mathbb{Z}_p$ and the bit decomposition of the same values in $\mathbb{Z}_2$:
\begin{center} edaBit := $\left(\sharing{r}_p, \sharing{r_0}_2, \sharing{r_1}_2, ... \sharing{r_{m-1}}_2 \right)$ \end{center}

\noindent where $\sharing{r}_p \in \mathbb{Z}_p$ and $m = \log_2(p)$. Thus, a daBit is an edaBit with $m=1$.

The approach to switch domains is analogous to the daBits technique. 
\if 0
For instance, to convert a secret value $x \in \mathbb{F}_p$ from SS to GCs, the parties compute and open $x - r$ in $\mathbb{F}_p$, locally bit-decompose it into $(x - r)_0, (x - r)_1, ... (x - r)_{m-1}$, and use these bits in input to the GC. Within the GC, the parties reconstruct the original secret $x$ in $\mathbb{F}_2$ computing $(x - r) + r$ mod $p$. To switch from GCs to SS, the parties compute and open $x \oplus r$ in $\mathbb{F}_2$, which can then be used to reconstruct $x$ in $\mathbb{F}_p$.
\fi
In addition, edaBits can be exploited to evaluate comparison, truncation, and probabilistic truncation operations (on both signed and unsigned data types) without an explicit domain switch. We refer to~\cite{escudero2020improved} for details on the constructions of these operations using edaBits. The main take-away is that using edaBits significantly speeds-up most applications relying on these operations.

\subsubsection{Notable works on hybrid computation}

\paragraph{ABY}~\cite{demmler2015aby}
focuses on 2-party computation with a HbC threat model. This technique allows to switch between three types of secret-sharing (SS) schemes: Arithmetic, Boolean, and Yao. 
\begin{enumerate}
	\item In the arithmetic domain, ABY uses additive SS in the ring $\mathbb{Z}_{2^k}$: a secret value $x$ is represented as the sum of two shares, which are integers modulo $2^k$. The operations are addition and multiplication modulo $2^k$. Thus, ABY supports the protocols that rely on additive SS (e.g., BGW, SPDZ).
	\item In the Boolean domain, ABY uses XOR-based secret sharing. The operations are XOR, AND, and multiplexer operations, which are run bit-wise (in parallel). To run these operations, ABY implements GMW.
	\item In Yao's GCs, the function to be computed is represented as a Boolean circuit with encrypted gates. The operations are XOR and AND. To run these operations, ABY uses Yao's GCs protocol with optimizations (e.g., Free XOR). 
\end{enumerate}
ABY proves that switching between domains pays off when an application needs multiplications, comparisons, and multiplexer operations, which are faster with arithmetic, Yao, and Boolean sharings, respectively.
ABY has also been extended to the 3-party setting in a malicious adversary model (ABY3~\cite{mohassel2018aby3}).

\subsection{MACs}
\label{sec:macs}

In this section we present the most common MAC constructions: multiplicative MACs for fields, in \autoref{subsubsec:MAC_fields_app}, and for rings, in \autoref{subsubsec:Ring_fields_app}. Those two constructions, in the context of MPC, were first proposed by Damgard et al in ~\cite{damgard2011spdz} for values in $\mathbb{Z}_{p}$ and by Cramer et al in ~\cite{spdz2k} for values in $\mathbb{Z}_{2^k}$.

\subsubsection{MACs for Fields}
\label{subsubsec:MAC_fields_app}

In the SPDZ protocol~\cite{damgard2011spdz}, a global key $\sharing{\alpha} \in \mathbb{Z}_p$ is jointly sampled and its shares are distributed to the parties. Then, every secret input $\sharing{x}$ is associated with a MAC tag $\sharing{m} = \sharing{x \cdot \alpha}$. Therefore, the share of $\sharing{x}$ given to party $i$ is a tuple $\left[\sharing{x}_i, \sharing{m}_i, \sharing{\alpha}_i \right]$. At the end of the computation, the integrity of secret inputs and outputs is guaranteed by checking that
\begin{equation*}
	\sum_{i = 1}^{n} \sharing{m}_i - \sum_{i = 1}^{n} \sharing{\alpha \cdot x}_i = 0.
\end{equation*}
If the adversary tries to alter $\sharing{x}$ by introducing an error $\delta$ on $\sharing{x}$ and an error $\varepsilon$ on $\sharing{m}$, the following equation has to hold,
\begin{equation*}
	\sum_{i = 1}^{n} \sharing{m}_i + \varepsilon = \sum_{i = 1}^{n} \sharing{\alpha \cdot x}_i + \alpha \cdot \delta,
\end{equation*}
this happens if and only if
\begin{equation*}
	\varepsilon = \alpha \cdot \delta.
\end{equation*}
But since $\alpha$ is unknown to the adversary, passing the MAC check means correctly guessing $\alpha$, which may happen with a probability of $\frac{1}{|\mathbb{Z}_p|}$.

\subsubsection{MACs for Rings}
\label{subsubsec:Ring_fields_app}

The MAC construction described in the previous subsection relies on the fact that every element belonging to a field, besides the zero element, has a multiplicative inverse. In a ring, some elements do not have multiplicative inverses, more specifically, in the $\mathbb{Z}_{2^k}$ ring half the elements have no inverses. Hence, the MAC construction described in \autoref{subsubsec:MAC_fields_app} would fail with probability $\frac{1}{2}$. To combat this issue, the SPDZ2k protocol~\cite{spdz2k}, introduces a modification of the construction described in \autoref{subsubsec:MAC_fields_app}. First,a security parameter $s$ is chosen and then and then the global MAC key $\sharing{\alpha}$ is sampled uniformly at random in $\mathbb{Z}_{2^s}$.
The MAC tags are then computed as $\sharing{m} = \sharing{x \cdot \alpha } \mod{2^{k + s}} $. MAC tags, shares and the global key are shared among the computing parties as in \autoref{subsubsec:MAC_fields_app}. At the end of the computation, the integrity of secret inputs and outputs is guaranteed by checking that
\begin{equation*}
	\sum_{i = 1}^{n} \sharing{m}_i - \sum_{i = 1}^{n} \sharing{\alpha \cdot x}_i = 0 \mod{2^{k+s}}.
\end{equation*}
If the adversary tries to alter $\sharing{x}$ by introducing an error $\delta$ on $\sharing{x}$ and an error $\varepsilon$ on $\sharing{m}$, the following equation has to hold,
\begin{equation*}
	\sum_{i = 1}^{n} \sharing{m}_i + \varepsilon = \sum_{i = 1}^{n} \sharing{\alpha \cdot x}_i + \alpha \cdot \delta \mod{2^{k+s}},
\end{equation*}
this holds if and only if
\begin{equation*}
	\varepsilon = \alpha \cdot \delta \mod{2^{k+s}}.
\end{equation*}
Let $2^v$ be the largest power of $2$ that divides $\delta$,
given that $\delta \ne 0 \mod{2^k}$, since the adversary wants to introduce a non-zero error on $\sharing{x}$, $v < k$. Now,  
\begin{equation*}
	\frac{\varepsilon}{2^v} = \alpha \cdot \frac{\delta}{2^v}\delta \mod{2^{k+s - v}},
\end{equation*}
since $\frac{\delta}{2^v}$ is an odd integer by definition of $v$, it is invertible in $\mathbb{Z}_{2^{k+s}}$ and therefore the MAC check passes if and only if
\begin{equation*}
	\alpha = \frac{\varepsilon}{2^v} \cdot \left(\frac{\delta}{2^v}\right)^{-1} \mod{2^{k+s-v}},
\end{equation*}
which happens with probability $\frac{1}{2^{k+s-v}} \le \frac{1}{2^s}$. This proof was provided originally by Cramer et al in ~\cite{spdz2k}.

\section{MPC protocols in MP-SPDZ}
 \label{sec:MPSPDZ_impl}

\subsection{Malicious setting, dishonest majority (mal\_dh) \cite{keller2020mp}}
\label{subsec:mal:dh}
\begin{itemize}
	\item \textsc{mascot} and \textsc{spdz2k} denote the protocols by Keller et al \cite{mascot} and Cramer et al \cite{spdz2k}, respectively.
	\item \textsc{lowgear} and \textsc{highgear} denote the two protocols described by Keller et al \cite{keller2018overdrive}.
	\item \textsc{mama} denotes MASCOT \cite{mascot} augmented with multiple MACs per value to increase the security parameter to a multiple the original one.
	\item \textsc{tiny} denotes the adaption of SPDZ2k\cite{spdz2k} to the binary domain: the SPDZ2k\cite{spdz2k} sacrifice does not work for bits, so it was replaced by cut-and-choose as described in \ref{subsubsec:tiny_furukawa}.
	\item \textsc{tinier} denotes the protocol by Keller et al \cite{tinier} also using the cut-and-choose sacrifice by \ref{subsubsec:tiny_furukawa}.
	\item \textsc{mascot} denotes the protocol by Keller et al \cite{mascot}.
	\item \textsc{lowgear} denotes the protocol by Keller et al \cite{keller2018overdrive}.
	\item \textsc{spdz2k} denotes the protocol by Cramer et al \cite{spdz2k}.
\end{itemize} 

\subsection{Semi-honest setting, dishonest majority (sh\_dh) \cite{keller2020mp}} 
\label{subsec:sh:dh}
\begin{itemize}
	\item \textsc{semi} is an adaptation of MASCOT\cite{mascot} to semi honest security: all steps necessary to provide malicious security have been removed. This protocol generates Beaver triples using OT as the original MASCOT.
	\item \textsc{semi2k} is an adaptation of SPDZ2k\cite{spdz2k} to semi-honest security.
	\item \textsc{semibin} is a protocol that generates bit-wise multiplication triples using OT. This protocol does not associate secret values with MACs and, therefore, provides security only against SH adversaries.
	\item \textsc{hemi} and \textsc{soho} are adaptation of LowGear and HighGear \cite{lowhighgear} to semi honest security, respectively.
	\item \textsc{temi} denotes the adaption of the protocol by Cramer et al \cite{temi} to the semi honest threat model.
\end{itemize} 

\subsection{Honest majority (mal\_h, sh\_h)}
\label{subsec:mal:sh:h}
All protocols denotes by \emph{ps} execute multiplication optimistically and then check the results at the end of the computation, as described in \ref{subsubsec:lindell_and_nof}. Every other protocol generates triples and then uses the sacrifice methodology, described in \ref{subsubsec:lindell_and_nof}, to achieve malicious security.

\begin{itemize}
	\item \textsc{malicious-rep-ring} is a ring based, maliciously secure, honest-majority protocol that uses the \emph{pre-processing on rings} method, described in \autoref{subsubsec:Eerikson_et_al}. 
	\item \textsc{ps-rep-ring} corresponds to a ring based, maliciously secure, honest-majority protocol that uses the \emph{post-processing} method, described in \autoref{subsubsec:Eerikson_et_al}.
	\item \textsc{ccd} denotes the protocol by Chaum et al \cite{ccd}. \textsc{ccd}, \textsc{malccd} and \textsc{shamir} replicate secret shares according to the optimized approach by Araki et al \cite{araki2016}, described in \ref{subsubsec:araki_et_al1}.
	\item \textsc{shamir} and \textsc{malshamir} are optimized for a honest majority threat model with the method proposed by Cramer et al \cite{cramer1999efficient} and the method from Araki et al \cite{araki2016}, described in \autoref{subsubsec:araki_et_al1}.
	\item \textsc{syshamir}, \textsc{syrepring} and \textsc{syrepfield} run computation in $\mathbb{Z}_p$ according to the technique by Chida et al \cite{chida2018} and computation in $\mathbb{Z}_{2^k}$ according to the technique proposed by Abspoel et al \cite{spdzwisepow2}. Both techniques are described in \autoref{subsubsec:araki_et_al1} and \ref{subsubsec:abspoel_et_al}, respectively.
	\item \textsc{rep4} refers to the $4$-party protocol by Dalskov et al \cite{dalskov2012fantastic}, described in \ref{subsubsec:dalskov_ff}.
	\item \textsc{malicious-rep-bin} generates and verifies Beaver triples according to the method by Furukawa et al \cite{tiny}, described in \ref{subsubsec:tiny_furukawa}.
	\item \textsc{ps-rep-bin} verifies multiplication triples according to the post-sacrifice approach by Araki et al \cite{araki2017optimized}, described in \ref{subsubsec:araki_bucket_candc}.
	\item \textsc{ring} denotes a MPC protocol in $\mathbb{Z}_{2^k}$, secure against a semi-honest adversary and that supports additive secret sharing.
	\item \textsc{atlas} denotes the protocol by Goyal et al \cite{atlas}.
	\item \textsc{yao} denotes the original Yao's garbled circuit 2-party protocol by Yao et al ~\cite{Yao1982}.
\end{itemize}

\subsection{Primitives in honest-majority protocols}
\label{subsec:primitives_app}

When not explictly specified, it is assumed that all values generate by a protocol are consistent with the domain of the protocol, i.e. if a protocol runs in $\mathbb{Z}_p$ it is assumed that all values generated during the execution of that protocol are in $\mathbb{Z}_p$.

\subsubsection{Sacrifice \cite{lindell2017framework}} \label{subsubsec:lindell_and_nof} 
Let $\left\{\sharing{x_i}, \sharing{y_i}, \sharing{z_i}\right\}_{i = 1}^{N} \in \mathbb{Z}_p$ be a list of triples to verify held by the computing parties. Additionally, let  $\left\{\sharing{a_i}, \sharing{b_i}, \sharing{c_i}\right\}_{i = 1}^{N}$ be a list of additional multiplication triples. The parties generate a random sharing $\sharing{\alpha}$. For each of the triples and the additional triples, the parties:
\begin{itemize}
	\item Multiply $\sharing{x_i}$ and $\sharing{\alpha}$ to obtain $\sharing{\alpha \cdot x_i}$.
	\item Compute $\sharing{\rho_i} = \sharing{\alpha \cdot x_i} + \sharing{a_i}$ and $\sharing{\sigma_i} = \sharing{y_i} + \sharing{b_i}$.
	\item The parties compute the following multiplications: $\sharing{z_i} \cdot \sharing{\alpha_i} = \sharing{\alpha_i \cdot z_i}$, $\sharing{a_i} \cdot \sharing{\sigma_i} = \sharing{a_i \cdot \sigma_i}$ and $\sharing{\rho_i} \cdot \sharing{y_i} = \sharing{\rho_i \cdot y_i}$.
	\item The parties generate a random value $\phi_i$ and $\alpha$ is opened.
\end{itemize}
Once all the previously described steps are complete, the parties compute
\begin{align*}
	&\sharing{v_i} = \left[ \sharing{\alpha_i \cdot z_i} + \alpha_i \phi_i \cdot \sharing{x_i} \right] - \sharing{c_i} + \\
	&  \left[\sharing{a_i \cdot \sigma_i} + \phi_i \cdot \sharing{a_i }\right] - \left[ \sharing{\rho_i \cdot y_i} + \phi_i \sharing{\rho_i} \right],
\end{align*}
then $N$ additional random values are jointly generated by the parties, let those values be $\underline{\beta} = \left\{\beta_i\right\}_{i = 1}^{N}$. Now the parties locally compute
\begin{equation*}
	\sharing{v} = \sum_{i = 1}^{N} \beta_i \cdot \sharing{v_i}.
\end{equation*}
Finally, a last random shared value $\sharing{r}$ is generated, $\sharing{w} = \sharing{v} \cdot \sharing{r} = \sharing{v \cdot r}$ is computed and opened. If $w = 0$, we are guaranteed that all the triples in $\left\{\sharing{x_i}, \sharing{y_i}, \sharing{z_i}\right\}_{i = 1}^{N}$ are correct multiplication triples with probability $1 - \frac{1}{|\mathbb{F}|}$, where $\mathbb{F}$ denotes the field we are working in. The multiplication triples can either be verified during pre-processing or after the evaluation of the circuit (post-sacrifice).

\subsubsection{Batch preprocessing, preprocessing in rings, post-processing \cite{malrep}} \label{subsubsec:Eerikson_et_al}
We now present three different methods to validate Beaver triples, we will call them \emph{batch preprocessing}, \emph{batch preprocessing for rings} and \emph{postprocessing}. Those three methods were originally proposed by Eeriksson et al in \cite{malrep}. \newline
\paragraph{Batch Preprocessing.} Let $\left\{\sharing{x_i}, \sharing{y_i}, \sharing{z_i}\right\}_{i = 1}^{N} \in \mathbb{Z}_p$ be a list of multiplication triples that the computing parties have to verify and let $f$ and $g$ be polynomials of degree $N - 1$ over $\mathbb{Z}_p$ defined as $f\left(i\right) = a_i$ and $g\left(i\right) = b_i$. Now, we define another polynomial $h = f \cdot g$ which is of degree $2N - 2$. For $i \in \left[1, N\right]$, $h\left(i\right) = c_i$ and for $i \in \left[N + 1, 2N - 1\right]$, $h\left(i\right) = f\left(i\right) \cdot g\left(i\right)$. Now, if all multiplication triples are correct, then $f\left(z\right) \cdot g\left(z\right) = h\left(z\right)$ $\forall z \in \mathbb{Z}_p$. But, if some of the multiplication triples are not correct, then $f \cdot g \ne h$, and the two polynomials $f \cdot g$ and $h$ can agree on at most $2N - 2$ points. That means that for a random point $z \in \mathbb{Z}_p$
\begin{equation*}
	\mathbb{P} \left[ f\left(z\right) \cdot g\left(z\right) = h\left(z\right) | f \cdot g \ne h  \right] \le \frac{2N - 2}{|\mathbb{Z}_p|}.
\end{equation*}
\paragraph{Preprocessing on rings.} Given a single multiplication triple $\left\{\sharing{x}, \sharing{y}, \sharing{z}\right\} \in \mathbb{Z}_{2^{k + \lambda}}$ to be verified, the parties generate two random values $\left\{\sharing{a}, \sharing{b}\right\} \in \mathbb{Z}_{2^{k + \lambda}}$ and optimistically compute $\sharing{c} = \sharing{a \cdot b}$. Then, the parties jointly generate and reveal a random value $r \in \mathbb{Z}_{2^{k + \lambda}}$ and compute
\begin{equation*}
	\sharing{e} = r \cdot \sharing{x} + \sharing{a}
\end{equation*}
and 
\begin{equation*}
	\sharing{w} = r \sharing{z} + \sharing{c} - e \sharing{y}.
\end{equation*}
Finally, the computing parties verify whether $w$ is a sharing of $0$. \newline 
\paragraph{Postprocessing.} Very similar to \textbf{ABF17 preprocessing}, but all multiplications are executed optimistically and checked at the end of the evaluation of the circuit. The main difference is that $\left\{\sharing{x_i}, \sharing{y_i}, \sharing{z_i}\right\}_{i = 1}^{N}$ is defined to be the set of all performed multiplications, where $\sharing{x_i}$ and $\sharing{y_i}$ are the factors and $\sharing{z_i}$ is the product. Then $N$ instances of \textbf{Preprocessing on rings} or \textbf{triple sacrifice in fields} (\autoref{subsubsec:lindell_and_nof}) are run in parallel with $\left\{\sharing{x_i}, \sharing{y_i}, \sharing{z_i}\right\}$ being the input of the $i$-th instance. \newline

\subsubsection{Replicated secret sharing \cite{araki2016}} \label{subsubsec:araki_et_al1} In order to share an element $x \in \mathbb{Z}_{2^k}$, the dealer chooses three random values $\left\{v_i\right\}_{i = 1}^3$, then it sets the share of party $P_i$ as $\left[ \sharing{v_i}, \sharing{a_i} \right]$, where $\sharing{a_i}$ is
\begin{equation*}
	\sharing{a_i} = v_j - x,
\end{equation*}
where $j = \left[i - 1 \mod 3\right] + 1$. The Boolean protocol is similar to the one just described, with the difference that the $+$ operation is replaced by $\oplus$ and $\cdot$ is replaced by $\land$. \newline

\subsubsection{Bucket cut-and-choose \cite{araki2017optimized}} \label{subsubsec:araki_bucket_candc}
Let $N$ be the number of multiplication triples that we want to generate and let $N = \left(X - C\right)L$, $B$ be the number of buckets, $C$ the number of triples opened in each subarray and $X = \frac{N}{L} + C$ the size of each subarray. To safely generate $N$ triples, we run the following protocol:
\begin{itemize}
	\item The parties generate $2M$ sharings of random values, where $M = 2 \left(N + CL\right) \cdot \left(B - 1\right) + 2N$. Let $\left\{\sharing{a_i}, \sharing{b_i}\right\}_{i = 1}^{\frac{M}{2}}$ be the sharings they receive.
	\item The parties generate an array $\underline{D}$ of multiplication triples. Each party splits $\underline{D}$ into $B$ vectors, such that $\underline{D_1}$ contains $N$ triples and all other vectors contain $N + LC$ triples.
	\item For $k = \left[2, B\right]$ each party splits $\underline{D}_{k}$ into $L$ subarrays of equal size $X$, $\underline{D}_{k, l}$ $\forall l \in \left[1, L\right]$.
	\item $\forall k \in \left[2, B\right]$ the parties jointly generate: (1) a random permutation of the subarray $\underline{D}_{k, j}$, $\forall j \in \left[1, L\right]$ and (2) a random permutation of the set $\left[1, L\right]$ and permute the subarrays $\underline{D}_k$ accordingly. 
	\item $\forall k \in \left[2, B\right]$ and $\forall j \in \left[1, L\right]$  the parties open the first $C$ triples in $\underline{D}_{k, j}$ and remove them from the corresponding subarray. If a party rejects any of the checks it sends $\varnothing$ and outputs $\varnothing$.
	\item The remaining triples are divided into $N$ sets of triples $\left\{\underline{B}_i\right\}_{i = 1}^{N}$ such that bucket $\underline{B}_i$ contains the $i$-th triple in $\left\{\underline{D}_j\right\}_{i = 1}^{L}$.
	\item In each bucket $B - 1$ triples are used to validate the first one.
\end{itemize}
If the protocol terminates, then with high probability the remaining $N$ triples are correct multiplication triples.

\subsubsection{Secure sharing \cite{chida2018}} \label{subsubsec:Chida_et_al}
Let $\left\{x_i\right\}_{i = 1}^{N} \in \mathbb{Z}_p$ be the set of inputs. To obtain secure shares of the inputs, the parties run the following protocol
\begin{itemize}
	\item The parties generate $N$ sharings of random values $\left\{\sharing{r_i}\right\}_{i = 1}^{N}$.
	\item $\forall i \in \left[1, N\right]$ the parties send their shares of the $i$-th secret to that party $P_j$ so that party $P_j$, i.e. the owner of the $i$-th secret, receives $r_i$. If any of the parties receives $\varnothing$, then it outputs $\varnothing$ and stops.
	\item $\forall i \in \left[1, N\right]$ party $P_j$ broadcasts $w_i = x_i - r_i$.
	\item All parties broadcast $\underline{w} = \left\{w_1, ..., w_N\right\}$ or a cryptographic hash function of $\underline{w}$. Every party check the correctness of their own vectors.
	\item $\forall i \in \left[1, N\right]$, the parties compute $\sharing{v_i} = \sharing{r_i} + w_i$.
	\item The parties output $\left\{\sharing{v_i}\right\}_{i=1}^{N}$.
\end{itemize}
This protocol involves a post-processing verification of multiplication carried by generating $M$ random values, masking the shared products and factors with those values and then by opening and checking the equality to $0$.

\subsubsection{Secure sharing in rings \cite{spdzwisepow2}} \label{subsubsec:abspoel_et_al}
Let $\left\{x_i\right\}_{i = 1}^{N} \in \mathbb{Z}_{2^k}$ be the set of inputs. Assume that, even though the secrets are element of the ring $\mathbb{Z}_{2^k}$, their respective owning parties represent them as elements of $\mathbb{Z}_{2^{k + 2}}$. To share the $i$-th secret, the sharing of random value $\sharing{r_i} \in \mathbb{Z}_{2^{k + s}}$ is generated and then reconstructed to party $P_i$. Then, party $P_j$ broadcasts $x_i - r_i$ and all parties set their shares of $x_i$ to $\sharing{x_i} = \sharing{r_i} + x_i - r_i$. The broadcast operation ensures that malicious parties cannot cheat during the sharing phase of the protocol. This technique is the equivalent of \autoref{subsubsec:Chida_et_al} but for rings and not for fields.

\subsubsection{Boolean circuits without garbling \cite{tiny}} \label{subsubsec:tiny_furukawa}

From \cite{tiny} two separate building blocks are implemented in MP-SPDZ. \newline
\paragraph{Replicated secret sharing.} To share a secret bit $x$, party $P_i$ chooses three random bits $s_1$, $s_2$ and $s_3$ under the constraint that $s_1 \oplus s_2 \oplus s_3 = v$. Then, it computes $t_1 = s_3 \oplus s_1$, $t_2 = s_2 \oplus s_1$ and $t_3 = s_3 \oplus s_2$. Then, party $P_i$ sets the share of party $j$ as $\sharing{x_j} = \left\{t_j, s_j\right\}$. \newline
\paragraph{Semi-honest AND computation.} Given $\sharing{x_1} = \left\{t_j, s_j\right\}_{j= 1}^3$ $x_1$ and $\sharing{x_2} = \left\{u_j, w_j\right\}_{j= 1}^3$, the following protocols describes how to compute $x_1 \land x_2$ and it is secure in the presence of a semi-honest adversary. The protocol is the following:
\begin{itemize}
	\item Generate a sharing of $0 = \alpha_1 \oplus \alpha_2 \oplus \alpha_3$, where party $P_i$ holds $\alpha_i$.
	\item Party $P_i$ $\forall i \in \left[1,3\right]$ computes $r_i = t_iu_i \oplus s_iw_i \oplus \alpha_i$ and send $r_i$ to party $P_{i + 1}$.
	\item Party $P_i$ stores $\left\{e_i, f_i\right\}$, where $e_i = r_i \oplus r_{i - 1}$ and $f_i = r_i$.
	\item It can be easily shown that $f_1 \oplus f_2 \oplus f_3 = x_1 \land x_2$.
\end{itemize}
\paragraph{Extension to the malicious case.} The parties generate a multiplication triple $\left\{\sharing{a}, \sharing{b}, \sharing{c}\right\}$ by, first, generating two shared random values $\sharing{a}$ and $\sharing{b}$ and, then, running the semi-honest AND protocol to obtain $\sharing{c} = \sharing{a \land b}$. To verify this first multiplication triple, another triple $\left\{\sharing{x}, \sharing{y}, \sharing{z}\right\}$ is generated and sacrificed:
\begin{itemize}
	\item Parties locally compute $\sharing{\rho} = \sharing{a} \oplus \sharing{x}$ and $\sharing{\sigma} = \sharing{b} \oplus \sharing{y}$ and then reveal their shares of $\sharing{\rho}$ and $\sharing{\sigma}$. 
	\item $P_j$ sends $\left\{\rho_j, \sigma_j\right\}$ to $P_{j + 1}$, if any of the parties sees inconsistent values, the protocol is stopped. Let $\left\{t_j, s_j\right\}$ be the result of this computation held by party $P_j$.
	\item $P_{j + 1}$ computes $\sharing{z} \oplus \sharing{c} \oplus \sigma_j \land \sharing{a} \oplus \rho_j \land \sharing{b} \oplus \rho_j \land \sigma_j$ and sends it to party $P_{j + 1}$. Upon receiving $t_{j - 1}$ by party $P_{j - 1}$, party $P_j$ verifies that $t_{j - 1} = s_j$.
\end{itemize}
To generate $N$ multiplication triples the malicious triple verification is used as a low-level primitive in the context of bucket cut-and-choose, described in \ref{subsubsec:araki_bucket_candc}.

\subsubsection{Four party computation \cite{dalskov2012fantastic}} \label{subsubsec:dalskov_ff}

The protocol described in \cite{dalskov2012fantastic} assumes that there are $4$ computing parties and at most one of them is corrupted. \newline
\paragraph{Replicated secret sharing.} The secret sharing scheme described in this paper is a $3$-out-of-$4$ additive secret sharing scheme, where party $P_i$ holds $3$ out of $4$ shares of a shared value $\sharing{x}$. \newline
\paragraph{Interactive and non-interactive secret sharing.} Two protocol to share a secret are described, an interactive one, usable when a secret value $x$ is known to $2$ out of $4$ parties and a non-interactive one, faster, but usable iff a secret value $x$ is known to $3$ out of $4$ parties. The interactive protocol is as follows:
\begin{itemize}
	\item Assume that parties $P_i$ and $P_j$ know a secret value $x$ and they want to share it with parties $P_k$ and $P_g$.
	\item Let $K_g$ be a pre-shared key known to parties $P_i$, $P_j$ and $P_k$. 
	\item Parties $P_i$, $P_j$ and $P_k$ generate pseudo-random value $x_g$ using $K_g$. Then, they set $x_i = x_j = 0$ and $x_k = x - x_g $. 
	\item Finally, they verifiably send $x_k$ to $P_g$.
\end{itemize}
The non-interactive secret sharing method involves just setting $x_1 = x_2 = x_3 = 0$ and $x_4 = x$. \newline
\paragraph{Multiplication.} To multiply two secret-shared values $\sharing{x}$ and $\sharing{y}$, every pair of parties $g, h$ with $g, h \in \left[1,4\right]$ who both know $x_g$, $x_h$, $y_g$ and $y_h$ run an interactive secret sharing round with $x_h y_g + x_gy_h$ as the input. Then, each party $i$ runs the non-interactive secret sharing protocol with $x_iy_i$ as the input. The parties locally add the shares
\begin{equation*}
	\sharing{x \cdot y} = \sum_{j \ne i} \sharing{x_i y_j + x_j y_i} + \sum_{i = 1}^{4} \sharing{x_i y_i}.
\end{equation*}
This paper also provides description of $4$-parties custom protocols for truncation, edaBits generation and SPDZ-wise MAC generation on the lines of the just described multiplication protocol.

 \end{document}